\documentclass{emulateapj}
\usepackage[normalem]{ulem}
\usepackage{footnote}
\usepackage{subfigure}
\usepackage[colorlinks,urlcolor=blue,citecolor=black,linkcolor=blue]{hyperref}

\usepackage[utf8]{inputenc}

\newcommand\msun{\ensuremath{M_{\odot}}}        % solar mass
\newcommand\semi{\ensuremath{\alpha_{sc}}}		%alpha semiconvection
\newcommand\thermo{\ensuremath{\alpha_{th}}}		%thermohaline
\newcommand\overshoot{\ensuremath{f_{\rm{ov}}}}			%overshoot
\newcommand\amnu{\ensuremath{\eta_{am}}}			%am nu comeff
\newcommand\mesh{\ensuremath{\delta_{\rm{mesh}}}}		%mesh delta coeff
\newcommand\var{\ensuremath{w_t}}			%varcontrol
\newcommand\rot{\ensuremath{(\Omega/\Omega_{crit})_{i}}}		
\newcommand\msunyr{\ensuremath{\dot\msun \ {\rm yr}^{-1}}}		%accretion rate

\newcommand\review{}

\begin{document}

\title{On Carbon Burning in Super Asymptotic Giant Branch Stars}

\author{
R. Farmer\altaffilmark{1}
C.E. Fields\altaffilmark{1,2}
F.X. Timmes\altaffilmark{1,2}
}

\altaffiltext{1}{School of Earth and Space Exploration, Arizona State University, Tempe, AZ}
\altaffiltext{2}{Joint Institute for Nuclear Astrophysics}

\email{rjfarmer@asu.edu}

\begin{abstract}

We explore the detailed and broad properties of carbon burning in
Super Asymptotic Giant Branch (SAGB) stars with 2755 MESA
stellar evolution models.  The location of first carbon ignition,
quenching location of the carbon burning flames and flashes,
angular frequency of the carbon core, and carbon core mass are studied
as a function of the ZAMS mass, initial rotation rate, and mixing
parameters such as convective overshoot, semiconvection, thermohaline
and angular momentum transport.  In general terms, we find these
properties of carbon burning in SAGB models are not a strong function
of the initial rotation profile, but are a sensitive function of the
overshoot parameter.  We quasi-analytically derive an approximate
ignition density, $\rho_{ign} \approx 2.1 \times 10^6$ g cm$^{-3}$, to
predict the location of first carbon ignition in models that ignite
carbon off-center. We also find that overshoot moves the ZAMS mass
boundaries where off-center carbon ignition occurs at a nearly uniform
rate of $\Delta M_{\rm ZAMS}$/$\Delta$\overshoot $\approx$ 1.6 \msun.  For
zero overshoot, \overshoot=0.0, our models in the ZAMS mass range
\hbox{$\approx$ 8.9} to 11 \msun \ show off-center carbon ignition. For canonical
amounts of overshooting, \overshoot=0.016, the off-center carbon
ignition range shifts to $\approx$ 7.2 to 8.8 \msun. Only systems with
\overshoot $\geq 0.01$ and ZAMS mass $\approx$ 7.2-8.0 \msun \ show carbon
burning is quenched a significant distance from the center. These
results suggest a careful assessment of overshoot modeling approximations
on claims that carbon burning quenches an appreciable distance from the
center of the carbon core.

\end{abstract}


\section{Introduction}
\label{sec:introduction}

When a single star on the main sequence (MS) exhausts the supply of
hydrogen in its core, the core contracts and its temperature
increases, while the outer layers of the star expand and cool. The
star becomes a red giant
\citep[e.g.,][]{iben_1991_aa,stancliffe_2009_aa,karakas_2014_aa}. 
The subsequent onset of helium burning in the core causes the star to
populate the horizontal branch for more metal-poor stars or the red
clump for more metal-rich stars
\citep{cannon_1970_aa,faulkner_1973_aa,seidel_1987_aa,castellani_1992_aa,girardi_1999_aa}.
After the star exhausts the supply of helium in its core, the
carbon-oxygen (henceforth CO) core contracts while the envelope once again expands and
cools along a path that is aligned with its previous red-giant track. The
star becomes an asymptotic giant branch (AGB) star 
\citep[e.g.,][]{hansen_2004_aa,herwig_2005_aa,kippenhahn_2012_aa,salaris_2014_aa,fishlock_2014_aa}.

The minimum mass for carbon ignition is usually referred to as 
M$_{\rm up} \approx$ 7 \msun\ and the minimum mass for neon ignition
in the core is traditionally referred to as M$_{\rm mas} \approx$ 10 \msun\
\citep{becker_1979_aa,becker_1980_aa,garcia-berro_1997_aa}.  Stars
with zero age main-sequence (ZAMS) masses between $\approx$ 7 M$_{\odot}$
and $\approx$ 10 \msun\ are designated as super-AGB stars 
\citep[henceforth SAGB,][]{ritossa_1996_aa,ritossa_1999_aa,gil-pons_2005_aa,siess_2006_aa,
siess_2007_aa, siess_2010_aa,poelarends_2008_aa,doherty_2010_aa}.
Due to the inferred slope of the stellar initial mass
function from observations \citep[e.g.,][]{jennings_2012_aa}, single stars in this ZAMS
mass range represent the population of stars that can produce the most
massive white dwarfs, the most numerous supernovae and possibly the
least massive neutron stars \citep[e.g.,][]{doherty_2015_aa}. SAGB
stars may also make significant contributions to the Galactic
inventory of isotopes such as $^7$Li, $^{14}$N, $^{23}$Na,
$^{25-26}$Mg, $^{26-27}$Al and $^{60}$Fe
\citep{siess_2010_aa,ventura_2013_aa,doherty_2014_aa,doherty_2014_ab}.

\begin{figure}[!htb]
\centering{\includegraphics[width=3.2in]{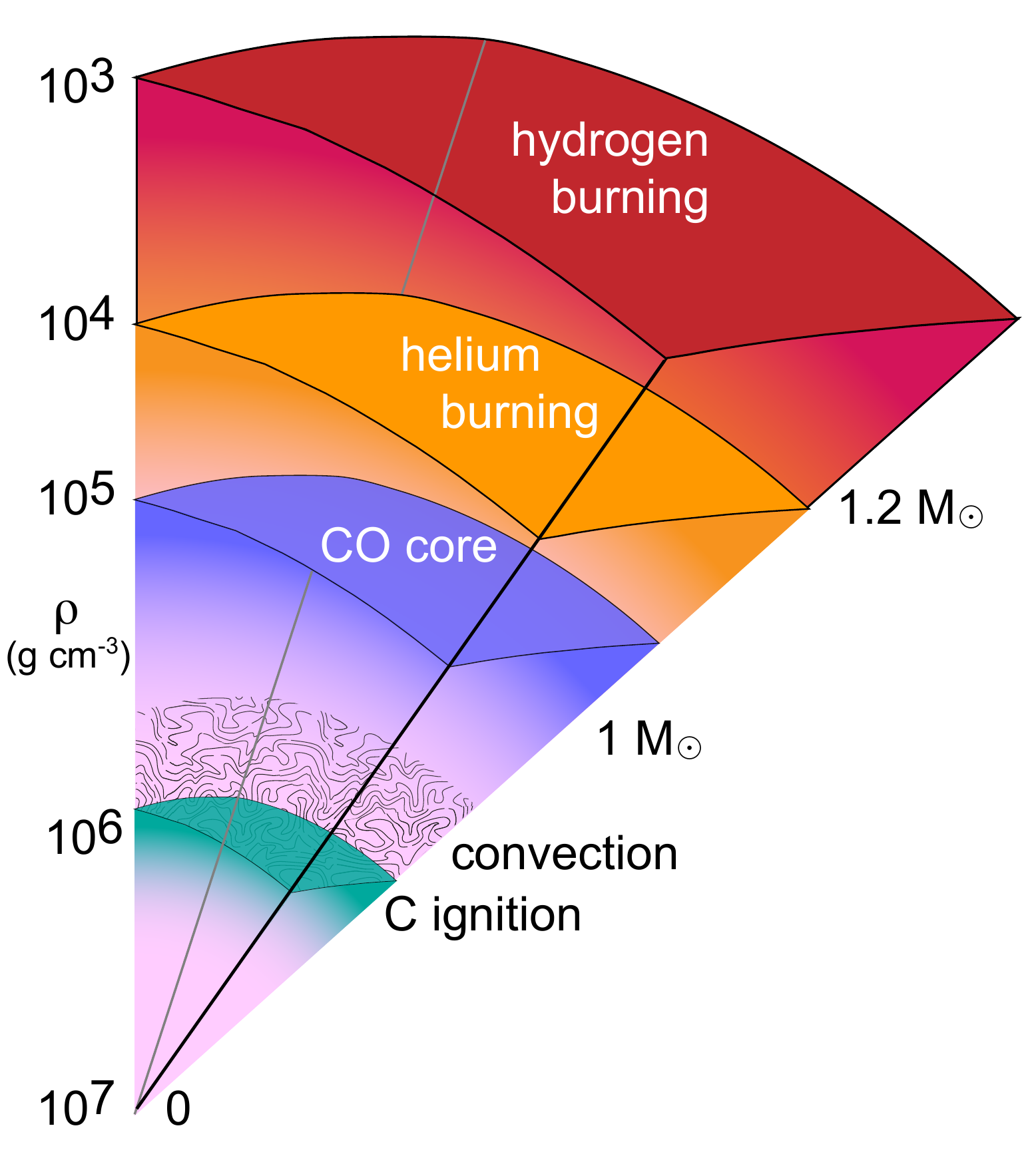}}
\caption{
Illustration of the structure of an SAGB star \review{during carbon burning}. In the center
is carbon/oxygen degenerate core, surrounded by a layer of helium which
is then surrounded by a hydrogen envelope. Ignition of carbon is followed
by a trailing convective region that drives a flash or flame towards
the center. The scale on the left is the
mass density, and the scale on the right is the enclosed mass.
}
\label{fig:illustration}
\end{figure}

After helium is exhausted in the core, \review{stars ascending the SAGB} develop partially
electron degenerate carbon–oxygen cores ranging from $\approx$ 0.9
\msun\ to $\approx$ 2.0 \msun\
\citep[pioneering studies of CO cores include][]{rakavy_1967_aa,beaudet_1969_aa,boozer_1973_aa}.
Depending primarily on the ZAMS
mass but also sensitively on the composition mixing model
\citep{poelarends_2008_aa,siess_2009_aa}, the ignition of carbon may not occur 
at all (for stars \hbox{$\lesssim$ 7 \msun}), occur at the center of the star (for
stars $\gtrsim$ 10 \msun), or occur somewhere off-center.  In the
off-center case, ignition is followed by the inward propagation of a
subsonic burning front \citep{nomoto_1985_aa,timmes_1994_aa,garcia-berro_1997_aa,saio_1998_aa}.
Trailing behind the burning front is a convective region that may
extend outward \hbox{$\approx$ 0.6 \msun}; see Figure \ref{fig:illustration} for an illustration. 
When a steady-state, convectively bounded,
subsonic, carbon burning front (henceforth a ``flame'') propagates
toward the center of the CO core, the flame leaves behind
oxygen-neon-magnesium (ONeNa) ashes. The inward propagating carbon flame 
may or may not reach the center of the star, depending on the parameters 
adopted for composition mixing beyond the convective boundary 
set by mixing-length theory 
\citep[e.g., thermohaline, overshoot, semi-convection,][]{siess_2009_aa,stancliffe_2009_aa,denissenkov_2013_ab}.
If the flame makes it to the center, then the original CO core is
converted into an ONeNa core.  Such SAGB stars can explode as electron
capture supernovae if their ONeNa core masses reach central densities
in excess of the threshold density for the $^{20}$Ne(e$^-$,$\nu$)$^{20}$F
electron capture reaction \citep{miyaji_1980_aa,nomoto_1984_aa,
  gutierrez_1996_aa,jones_2013_aa}, as may be the case for the Crab
Nebula \citep{davidson_1982_aa,nomoto_1982_aa,wanajo_2009_aa} or for
potentially explaining observations of sub-luminous type II-P
supernovae \citep[][]{smartt_2009_aa}.  If the flame does not make it
to the center, then the star is left with inner parts of the original
CO core surrounded by a layer of ONeNa. Such
hybrid white dwarfs may provide unusual Type Ia supernovae progenitors
\citep{siess_2009_aa,denissenkov_2013_ab,wang_2014_aa,chen_2014_aa}.

This paper explores the ignition and subsequent evolution of carbon
burning in SAGB stellar models as a function of the ZAMS mass, initial
rotation rate, the magnitude of various mixing parameters such as
convective overshoot, semiconvection, thermohaline and angular
momentum transport. We sample this multi-dimensional space with 2,755
MESA stellar evolution models \citep{paxton_2011_aa,paxton_2013_aa}
that are evolved from the pre-main-sequence to the end of carbon
burning. All models have $Z=0.02$ and a solar composition from \citet{grevesse_1998_aa}.
Along the way we provide quasi-analytic models for
interpreting the results.
In \hbox{\S \  \ref{sec:method}} we discuss the input physics for our
calculations, the composition mixing processes considered, and the MESA implementation of rotation and magnetic fields. 
In \hbox{\S \  \ref{sec:grids}} we define the baseline parameters used for our calculations and discuss the grids used for 
our exploration into this three dimensional parameter space. 
In  \S \  \ref{sec:cburn} we present the results of our
non-rotating models, an analytical approximation of the location of
first carbon ignition, and the evolution of carbon burning flames and flashes. 
In  \S \  \ref{sec:rotation} we present the results of the effect of rotation
and overshoot on the ignition, evolution, and death of carbon burning. 
In  \S \  \ref{sec:mixsens} we study the impact of the semiconvection, thermohaline, and
angular momentum transport coefficients on the location of first carbon ignition on our results.
In  \S \  \ref{sec:convergence} we present the results of spatial and temporal
convergence studies on our results, and 
in  \S \  \ref{sec:discussion} we discuss our results and their implications.

\section{Instrument and Methods}
\label{sec:method}
%Describe MESA

Our numerical instrument is MESA version 6794. We use the included
\texttt{sagb$\_$NeNa$\_$MgAl.net} reaction network, which follows 22
isotopes from $^{1}$H to $^{27}$Al to track hydrogen (pp chains,
CNO-, NeNa-, and MgAl-cycles), helium and carbon burning. The 51
thermonuclear reaction rates coupling these isotopes are from JINA
reaclib version V2.0 2013-04-02 \citep{cyburt_2010_aa},
energy-loss rates and their derivatives from thermal neutrinos are from
the fitting formulae of \citet{itoh_1996_aa}, and electron screening
factors for thermonuclear reactions in both the weak and strong
regimes are from \citet{dewitt_1973_aa,graboske_1973_aa} and
\citet{alastuey_1978_aa} with plasma parameters from
\citet{itoh_1979_aa}. \citet{poelarends_2008_aa} showed that increasing the mass loss rate could
decrease the mass range for systems that will become electron capture supernovae (ECSNe).
We thus use a
Reimer mass loss prescription \citep{reimer_1975_aa} with
$\eta$=0.5 on the RGB and a Bl\"ocker mass loss prescription
\citep{blocker_1995_aa} with $\eta$=0.05 on the AGB.  The MESA
inlists are publicly available\footnote{\url{http://mesastar.org/results}}.

%Describe post-analysis tools.
Analysis of a carbon burning event requires knowledge of when and where carbon
ignites. We identify the cell location of carbon burning, hence the ignition mass
coordinate ($M_{f,s}$), by three criteria. First, we require $\epsilon_{{\rm nuc}} \gg
\epsilon_{\nu}$ in a CO core. Second, we require that $^4$He is
depleted in the ignition region as some of the lowest mass stars
investigated would have a small amount of $^{12}$C burning near the
CO core, $^4$He shell boundary. Finally, we require 
X($^{20}\rm{Ne})>X(^{23}\rm{Na})>X(^{24}\rm{Mg})$, which indicates we
have vigorous $^{12}$C +$^{12}$C burning.

The end of a carbon burning event is defined when no cell within
10\% of the mass location of the flame, during the next time step, has 
$\epsilon_{{\rm nuc}} \gg \epsilon_{\nu}$. We then define the final flame
location ($M_{f,e}$), where carbon burning is quenched, as the minimum value of
the mass location taken at the end of all the carbon burning events. This is 
independent of any subsequent carbon flashes, as we record only the 
closest approach the carbon burning makes to the core.

\subsection{Mixing}
\label{sec:mix_theory}

%Mix Preamble
Treatment of convective processes within stellar interiors is essential 
for a physically accurate stellar model. We briefly discuss the 
composition mixing processes used in our calculations, how MESA 
models the mixing processes, and previous studies that
guide our choices for our baseline mixing parameters. 
Values for our baseline parameters are given in $\S$ \ref{sec:grids}.

%MLT
We use the Schwarzschild criterion for convection along with the Cox
implementation of Mixing Length Theory (MLT) \citep{cox_1968_aa}. The
Schwarzschild criterion describes that a region is stable to
convection if the gradient of a  piece of adiabatic matter
is less than that of the temperature gradient of the stellar
atmosphere:
\begin{equation}\label{eq:conv}
\left | \frac{d \ \textup{ln} \ T}{d \ \textup{ln} \ P} \right |_{ad} > \left | \frac{d \ \textup{ln} \ T}{d \ \textup{ln} \ P} \right |_{rad}.
\end{equation}
MLT has a free parameter, $\alpha_{\rm{MLT}}$, as described by
\citet{bohm_1958_aa}. Values within $1.6 \lesssim \alpha_{\rm{MLT}}
\lesssim 2.2$ have been inferred by \citet{noels_1991_aa} who compared
observations of the $\alpha$ Centauri binary star system with stellar
evolution models. \citet{trampedach_2014_aa} also suggests $1.6
\lesssim \alpha_{\rm{MLT}} \lesssim 2.05$ from calibrating 1D stellar
models to 3D radiation-coupled hydrodynamics simulations of convection
in stellar surface layers.

%OVERSHOOT

Turbulent velocity fields have been suggested 
to decay exponentially beyond the Schwarzschild convective boundary 
defined by equation \ref{eq:conv}
\citep{herwig_1997_aa,ventura_1998_aa,mazzitelli_1999_aa,blocker_2000_aa,herwig_2000_aa}
leading to a diffusive treatment of mixing as
\begin{equation}\label{eq:over}
 D_{OV}=D_{conv,0}\exp\left(-\frac{2z}{\overshoot \lambda_{\rm{P,0}}}\right)
\enskip.
\end{equation}
Here $D_{\rm{conv,0}}$ is the convective diffusion coefficient at the
convective boundary, $z$\ is the radial distance from the convective
boundary, $\lambda_{\rm{P,0}}$ is the local pressure scale height and
$\overshoot$ is an adjustable parameter. MESA offers the flexibility
of allowing \overshoot \ to be different for different convective
regions (H burning, He burning, metal burning and non
burning). However, we set \overshoot \ to the same value in all
convective regions (see \S3 for the values chosen).

%SEMICONVECTION
Semiconvection occurs when regions of the stellar interior are stable
to the Ledoux criterion and unstable to the Schwarzschild criterion 
\citep{kippenhahn_2012_aa}.
This occurs when $\nabla_{ad} < \nabla_{T} < \nabla_{L}$, where
$\nabla_{L} = \nabla_{ad} + B $ and $B$ is the Brunt composition
gradient. MESA treats semiconvection as a diffusive process 
\citep{langer_1983_aa,langer_1985_aa,heger_2000_aa,zaussinger_2013_aa} with 
a diffusion coefficient 
\begin{equation}\label{eq:semi}
 D_{\rm{sc}}=\semi \left(\frac{K}{6C_p\rho}
\right)\frac{\nabla_{T}-\nabla_{ad}}{\nabla_{L}
-\nabla_{T}},
\end{equation}
where $K$ is the radiative conductivity, $C_p$ is the specific heat at constant 
pressure and \semi \ is an adjustable dimensionless parameter describing 
the speed with which convective mixing may occur at the boundary defined
by equation \ref{eq:conv}. Ongoing efforts to calibrate such semiconvection models include 
multidimensional numerical simulations of double-diffusive convection
\citep{zaussinger_2013_aa,spruit_2013_aa} and comparing massive star 
models with observations \citep{yoon_2006_aa}.

\begin{figure*}[!htb]
\centering{\includegraphics[width=6.0in]{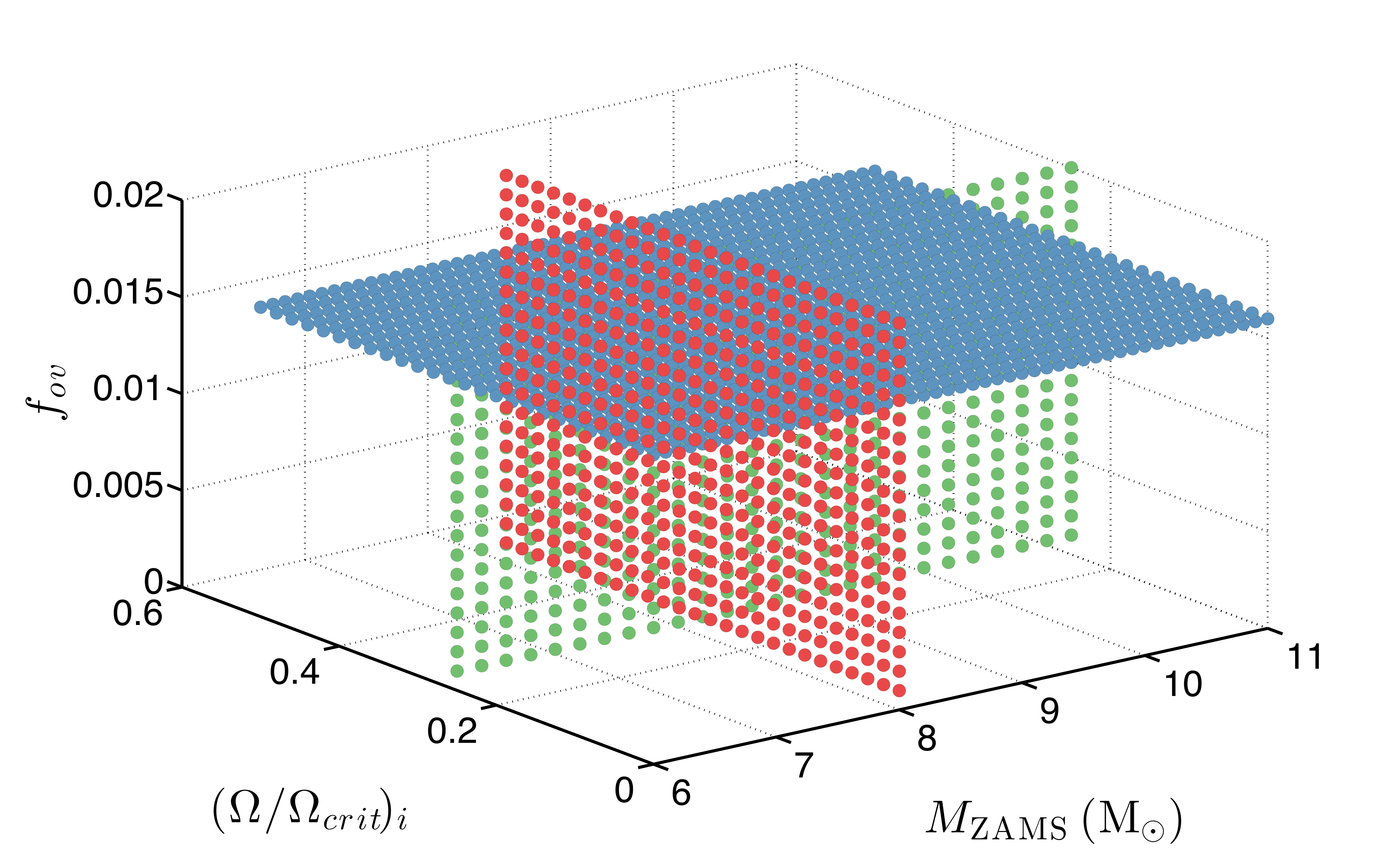}}
\caption{
Three slices explored in the (mass, rotation, overshoot) parameter
space.  We calculate 1326 models in mass-rotation rate plane (blue
slice), 546 models in the rotation rate-overshoot plane (red slice),
and 546 models in the mass-overshoot plane (green slice).
}
\label{fig:cube}
\end{figure*}

%THERMOHALINE

Thermohaline mixing occurs when $ \nabla_T-\nabla_{ad}\leq B \leq 0$.
These are regions stable against convection, according to the Ledoux
criterion, but have an inversion of the mean molecular weight
\citep{kippenhahn_1980_aa}. This type mixing forms elongated fluid parcels, 
sometimes called ``salt-fingers''. MESA treats thermohaline mixing as a diffusion process 
\citep{ulrich_1972_aa,kippenhahn_1980_aa,brown_2013_aa,zemskova_2014_aa} 
\begin{equation}
 D_{th}=\thermo \frac{3K}{2\rho C_{p}}
\frac{B}{\left(\nabla_{T} - \nabla_{ad}\right)}
\enskip ,
\end{equation}
where $\thermo$\ is a dimensionless parameter, 
related to the aspect ratio of the salt fingers.
Estimates for this parameter range from $1 \lesssim \thermo \lesssim 667$
\citep{kippenhahn_1980_aa,charbonnel_2007_aa,cantiello_2010_aa,
stancliffe_2010_aa,wachlin_2011_aa}, with some multidimensional
simulations suggesting this parameter is significantly overestimated 
\citep{denissenkov_2010_aa,traxler_2011_aa,denissenkov_2011_aa,
lattanzio_2015_aa}.

\subsection{Rotation and Magnetic Fields}
\label{sec:rot}

% Shellular Rotation
MESA implements rotation by making the assumption that the angular
velocity, $\omega$, is constant over isobars; see \citet{paxton_2013_aa} for the implementation 
of rotation into MESA.
Such an assumption is often referred to as shellular approximation
\citep{zahn_1992_aa,meynet_1997_aa}, and allows the stellar structure
equations to be solved in one dimension for a rotating star.  For this
study, rotation is initialized by imposing a solid body rotation law
at ZAMS, where the total luminosity equals the
nuclear burning luminosity.

The transport
of angular momentum and material due to rotationally induced
instabilities is followed using a diffusion approximation
\citep[e.g.,][]{endal_1978_aa, pinsonneault_1989_aa,heger_2000_aa,
maeder_2003_aa,maeder_2004_aa,heger_2005_aa,suijs_2008_aa}
for the dynamical shear instability (DSI), secular shear instability
(SSI), Eddington-Sweet circulation (ES), Goldreich-Schubert-Fricke
instability (GSF), and Spruit-Tayler dynamo (ST).
See \citet{heger_2000_aa} for a description of the physics of the
different instabilities and the calculation of the respective
diffusion coefficients.

\citet{berger_2005_aa} investigated the Ca line profiles of a sample of DA white dwarfs, concluding their rotational velocities are generally 
less than 10 km s$^{-1}$. These values are significantly less than the values determined by the rotating, nonmagnetic models of 
\citet{langer_1999_aa}. Internal magnetic torques as proposed by \citet{spruit_1998_aa} have been suggested as an effective 
mechanism to spin down the cores of these white dwarf progenitors during the giant phase. \citet{suijs_2008_aa} showed that magnetic 
torques as calculated in \citet{spruit_2002_aa} produce rotational velocities in better agreement with the observed values of 
\citet{berger_2005_aa}. We therefore include internal magnetic fields and the Spruit Tayler dynamo angular momentum mechanism 
for our rotating MESA models.

Magnetic fields are implemented in MESA using the formalism of \citet{heger_2005_aa}, where a magnetic torque due
to a dynamo \citep{spruit_2002_aa} allows angular momentum to be transported inside the star. The 
radial component, $B_r$, and the azimuthal component, $B_\phi$, of the magnetic field are modeled as
\begin{equation} 
B_\phi \sim r(4\pi\rho)^{1/2}\omega_A
\end{equation}
\begin{equation} 
B_r \sim B_\phi (rk)^{-1}
\enskip ,
\end{equation}
where $r$ is the radial coordinate, $\rho$ the density, $\omega_A$ the Alfv\'en frequency, and $k$ the wavenumber. 
These magnetic fields then provide a torque
\begin{equation} 
S = \frac{B_rB_\phi}{4\pi}
\end{equation}
which acts to slow down the stars rotation rate by decreasing the amount of differential rotation inside the star \citep{heger_2005_aa}.

The initial rotation
is normalized against the critical rotation rate for the star 
$\Omega_{crit}=\sqrt{(1 - L/L_{edd}) c M / R^3}$, where 
c the speed of light, $M$ the mass of the star,
$R$ the stellar radius, $L$ the luminosity and $L_{edd}$ the Eddington luminosity. The initial magnetic field, $B_r = B_\phi = 0$ 
for all our rotating models. Effects of rotationally induced mass loss are not included.

\section{Grids}
\label{sec:grids}

We define a set of baseline parameters and construct a number of grids
surrounding that baseline set to investigate the sensitivity of carbon burning in the SAGB models
with respect to variation in the baseline parameters. Choices in the numerical values of the
baseline parameters are based on the current understanding of the canonical values for SAGB and other stars
when using MESA. Choices in the range of values explored were designed such that we could explore 
the various competing factors involved in stellar evolution models.

Table
\ref{table:baseline} lists the baseline mixing, spatial resolution,
and temporal resolution parameters.  The parameter $f_{c}$ is the
ratio of turbulent viscosity to the diffusion coefficient \citep{heger_2000_aa}, $f_{\mu}$
is the sensitivity to $\mu$-gradients \citep{heger_2000_aa}, while $\delta_{\rm{mesh}}$
controls the spatial resolution by determining the relative magnitude
of changes between the adjacent cells \citep{paxton_2011_aa}, and $w_{t}$ controls the
temporal resolution by modulating the magnitude of the allowed changes between time steps \citep{paxton_2011_aa}. 
Baseline values for $\delta_{\rm{mesh}}$ and $w_{t}$ were based on computational requirements. However
see \S \ \ref{sec:convergence} for a discussion into their relative effects on our results.
The other baseline parameters listed in
Table \ref{table:baseline} are discussed in \S \ref{sec:mix_theory}.

\subsection{Mass-Rotation-Overshoot Grid}

We explore the (ZAMS mass, initial rotation, overshoot) parameter
space with three slices through this 3D data cube.  Table
\ref{table:cube} lists the start, stop, and step values for two of the
three quantities while holding the third quantity constant.  The
number of SAGB models is 1326 in the mass-rotation plane, 546 in the
rotation-overshoot plane, and 546 in the mass-overshoot plane for a
total of 2418 SAGB models. Figure \ref{fig:cube} illustrates these
three orthogonal slices though this 3D parameter space.

For each slice within the 3D data cube we chose one quantity to be held fixed. In the mass-rotation plane this is the 
overshoot value, and our choice of \overshoot=0.016 is based on the canonical value for this overshoot model \citep{herwig_2000_aa}.
In the rotation-overshoot plane we held the initial mass fixed at $M=8\msun$, which was selected because, based on the 
non-rotating models, we expected we could induce a range of behaviors from non-ignition, to off-center ignition to central ignition.
In the mass-overshoot plane, we held the initial rotation fixed at \rot=0.25, purely as a middle ground value between non-rotating models
and our upper bound of \rot=0.5. Our choice for the range of values covered by each grid was based 
on a requirement to have a comprehensive sample over the canonical baseline values for SAGB stars.

\begin{deluxetable}{ll}{h}
\tablecolumns{2}
%\tablewidth{0pc}
\tablewidth{0.8\linewidth}
\tablecaption{Baseline Parameters}
\tablehead{\colhead{Parameter} & \colhead{Value} }
\startdata
Mixing Length Theory ($\alpha_{\rm{MLT}}$) & $2.0000$ \\
Semiconvection ($\alpha_{\rm{sc}}$) & $0.0100$  \\
Thermohaline ($\alpha_{\rm{th}}$) & 1.0000 \\
Overshoot ($f_{\rm{ov}}$) & $0.0160$ \\
Angular Momentum (\amnu) & $1.0000$ \\
Turbulent Viscosity ($f_{\rm{c}}$) & 0.0333  \\
$\mu$-gradient Sensitivity ($f_{\rm{\mu}}$) & 0.0500 \\
Mesh Delta Coefficient ($\delta_{\rm{mesh}}$) & 0.5000 \\
Variance Control Target ($w_{\rm{t}}$) & 0.0001
\enddata
\label{table:baseline}
\end{deluxetable}

\begin{deluxetable}{lllll}{h}
\tablecolumns{5}
%\tablewidth{0pc}
\tablewidth{0.8\linewidth}
\tablecaption{Mass-Rotation-Overshoot Grid}
\tablehead{\colhead{Variable} & \colhead{Start} & \colhead{Stop} & \colhead{Step} & \colhead{Constant}}
\startdata
$M_{\rm{ZAMS}}$ & 6.0 & 11.0 & 0.1  & $f_{ov}$=0.016\\
\rot       & 0.0 & 0.5 & 0.02  &  \\
& & & \\
\rot       & 0.0 & 0.5 & 0.02  & $M_{\rm{ZAMS}}$=8.0\\
$f_{ov}$   & 0.0 & 0.02 & 0.001 & \\
& & & \\
$M_{\rm{ZAMS}}$ & 6.0 & 11.0 & 0.2   & \rot = 0.25 \\
$f_{ov}$   & 0.0 & 0.02  & 0.001 & 
\enddata
\label{table:cube}
\end{deluxetable}

\begin{deluxetable}{lllll}{h}
\tablecolumns{5}
%\tablewidth{0pc}
\tablewidth{0.8\linewidth}
\tablecaption{Mixing Coefficients Grid}
\tablehead{\colhead{Variable} &   \multicolumn{4}{c}{Values}}
\startdata
\semi      &  0.0 & 10$^{-3}$ & 10$^{-2}$ & 10$^{-1}$ \\
\thermo    & 0.0 & 0.1 & 1.0 & 10.0 \\
\overshoot & 0.000 & 0.001 & 0.016 & 0.020 \\
\amnu      & 0.0 & 0.5 & 1.0 & 1.5 
\enddata
\label{table:mix}
\end{deluxetable}

\begin{deluxetable}{llll}{h}
\tablecolumns{4}
%\tablewidth{0pc}
\tablewidth{0.8\linewidth}
\tablecaption{Spatial and Temporal Convergence Grid}
\tablehead{\colhead{Variable} &   \multicolumn{3}{c}{Values}}
\startdata
$M_{\rm{ZAMS}}$  &  7 &  8 &  9 \\
\rot             & 0.0 & 0.25 & 0.5 \\
\mesh            & 0.1 & 0.5 & 1.0 \\
\var             & 10$^{-5}$ & 10$^{-4}$ & 10$^{-3}$
\enddata
\label{table:converge}
\end{deluxetable}

\subsection{Mixing Coefficients Grid}

We investigate the (semiconvection, thermohaline, overshoot, angular
momentum diffusion) parameter space on the location of the first
carbon ignition in a \hbox{8 \msun} ZAMS, \rot=0.25 model with
selected points in this 4D data cube.  Table \ref{table:mix} lists
these quantities and their selected values.  
We choose to limit the range of \thermo in this grid to span only 
the lower values discussed in the literature. \amnu\ is a scale factor that alters the strength of
angular momentum diffusion in each cell; see  \citet{paxton_2013_aa} for details.
The total number of SAGB
models in this grid is 4$^4$=256.  This grid permits assessment of the
relative strengths of each mixing process.

\subsection{Spatial and Temporal Convergence Grid}

Finally, we examine the spatial and temporal convergence properties of
a subset of our SAGB models. The MESA parameter \mesh \ broadly
controls the spatial resolution and \var broadly relates to the
temporal resolution. For \mesh=0.5\ there are $\approx$ 5,000$-$10,000
spatial points center to surface. Spatial resolutions necessary
to capture carbon burning flames and flashes are discussed in
\S \ref{sec:convergence}.  For $\var=10^{-4}$ a temporal resolution of
$\approx$ 10 yrs is achieved.  Table \ref{table:converge} lists these
quantities and their values.  The total number of SAGB models in this
resolution sensitivity grid is 3$^4$=81.

\section{Results From Non-Rotating Models with Baseline Mixing Parameters}
\label{sec:cburn}

\begin{figure*}[ht]
         \centering
         %ratio is 1200x944
        \begin{subfigure}{
                \includegraphics[width=3.4in,height=2.75in]{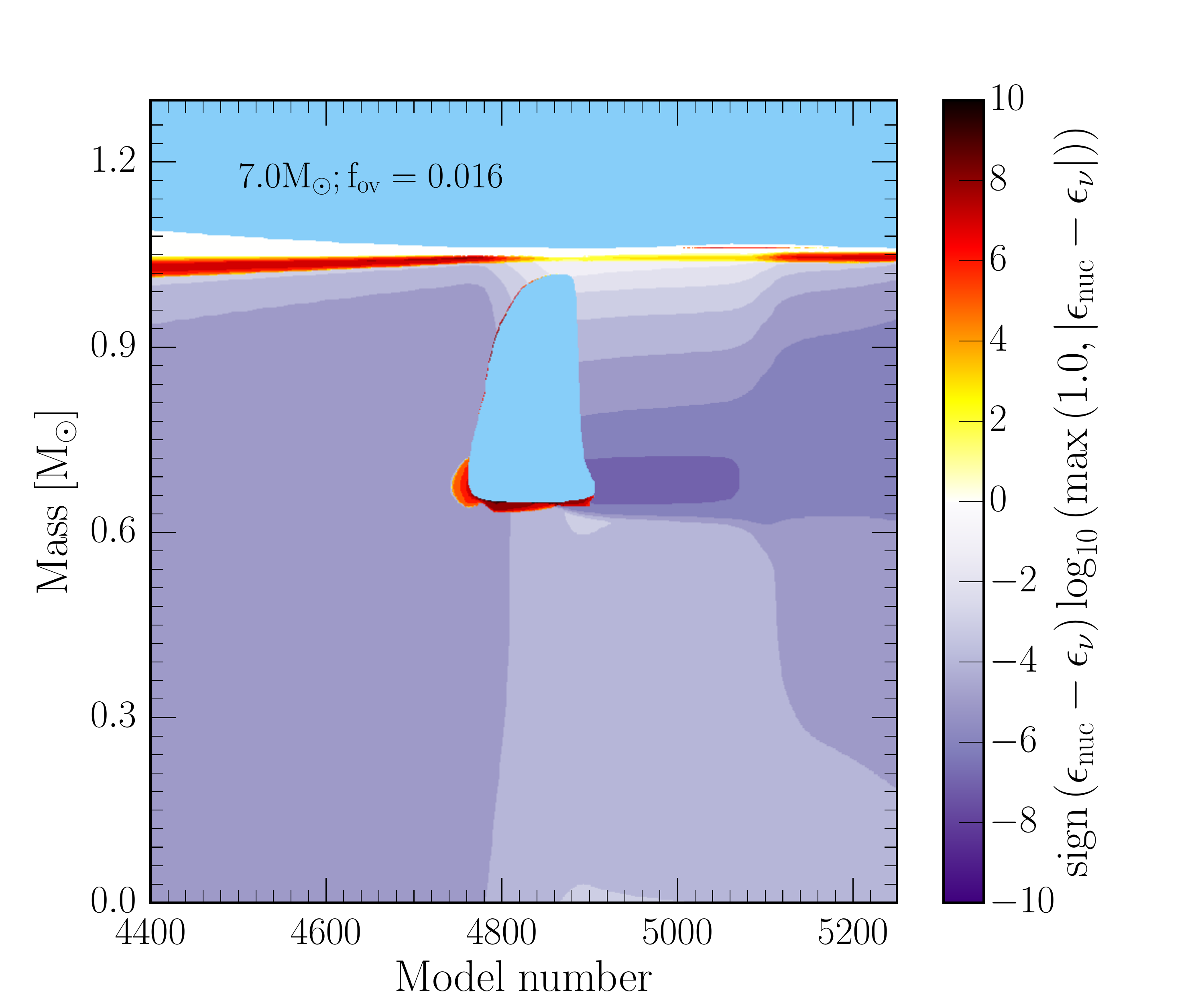}}
%                  \caption{7.0 \msun}
                \label{fig:kip7}
        \end{subfigure}%
        \begin{subfigure}{
                \includegraphics[width=3.4in,height=2.75in]{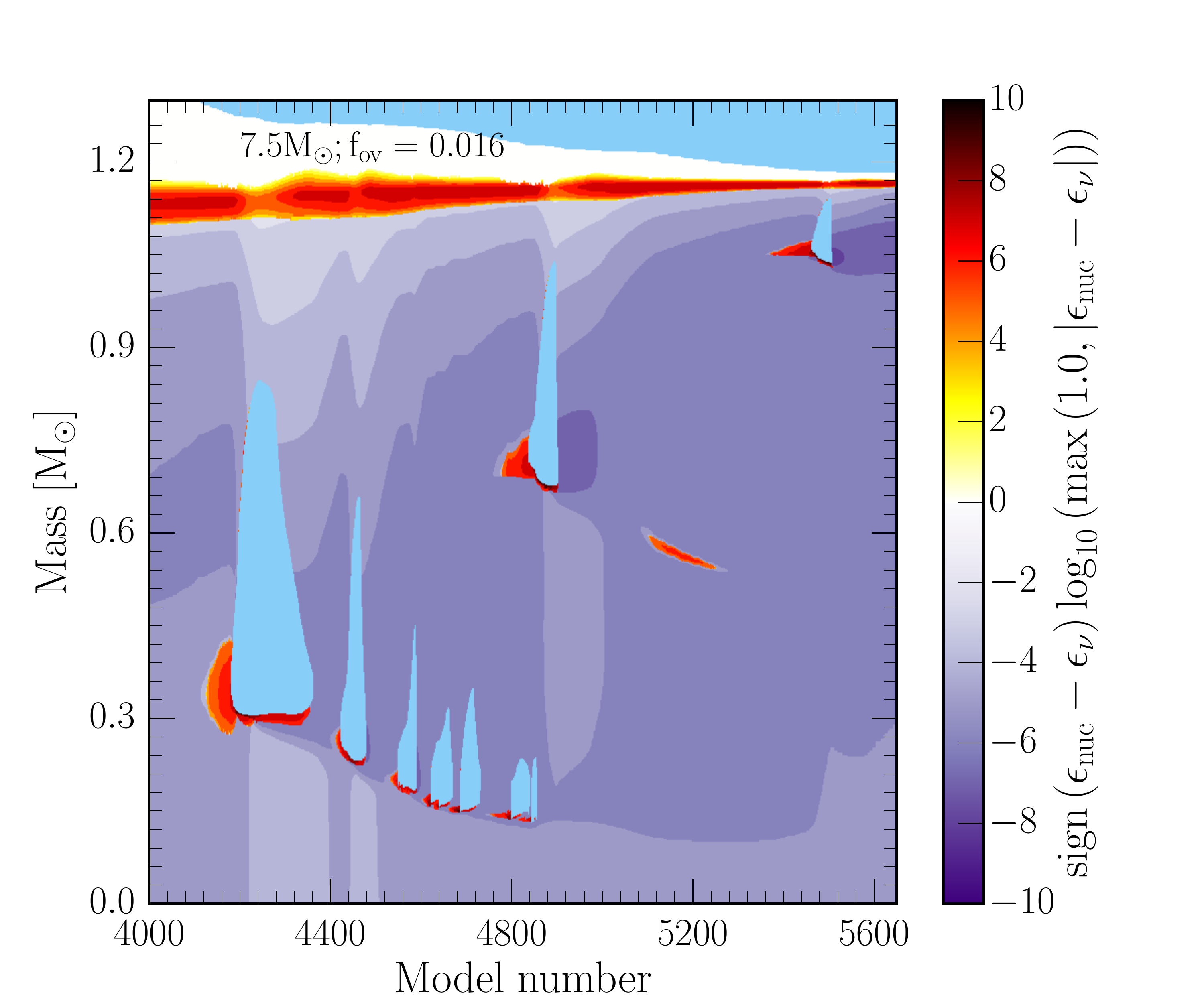}}
%                  \caption{7.5 \msun}
                \label{fig:kip75}
        \end{subfigure}
        
        \begin{subfigure}{
                \includegraphics[width=3.4in,height=2.75in]{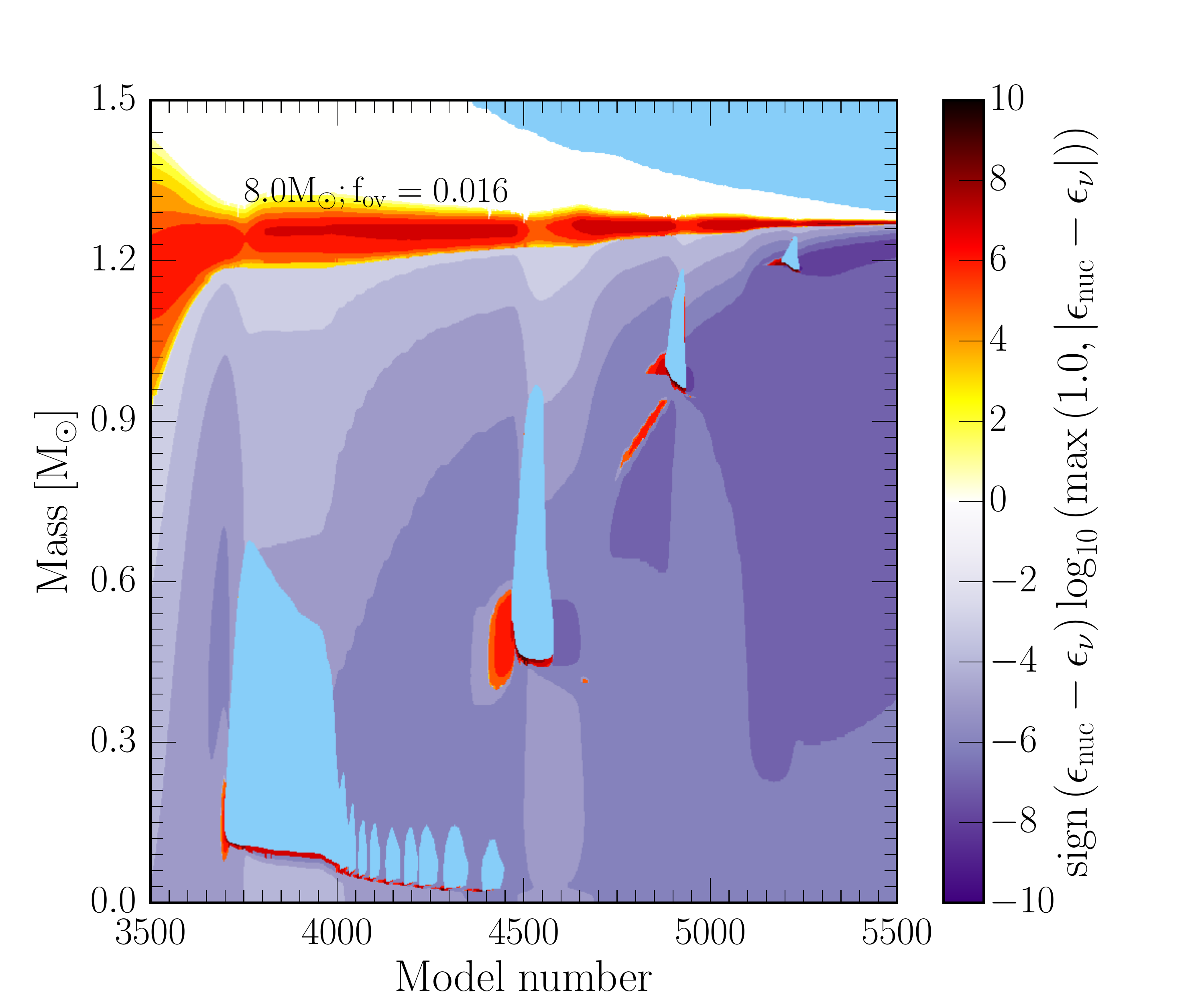}}
%                  \caption{8.0 \msun}
                \label{fig:kip8}
        \end{subfigure}%
%        ~
        \begin{subfigure}{
                \includegraphics[width=3.4in,height=2.75in]{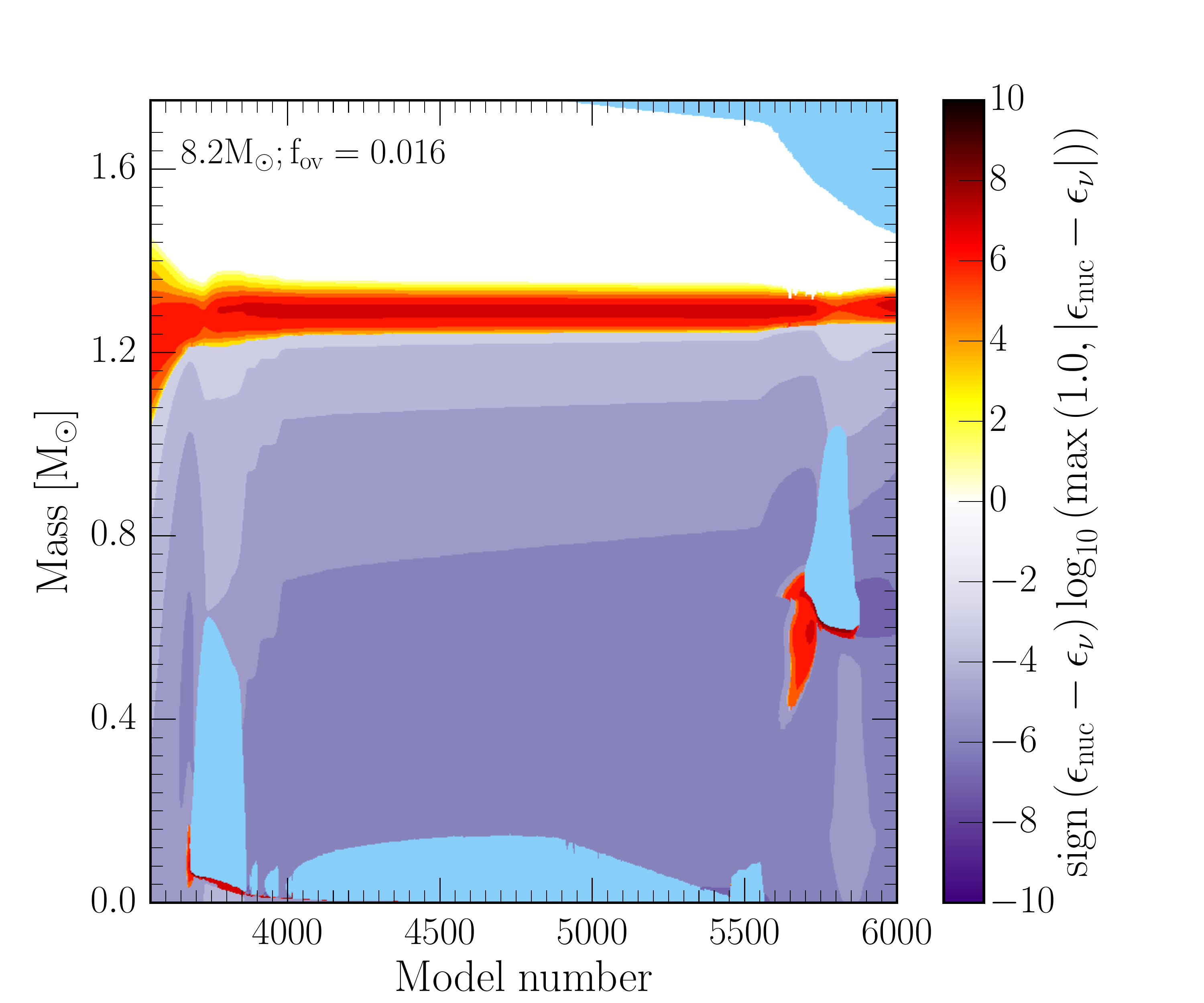}}
%                  \caption{9.0 \msun}
                \label{fig:kip82}
        \end{subfigure}      
        
        \begin{subfigure}{
                \includegraphics[width=3.4in,height=2.75in]{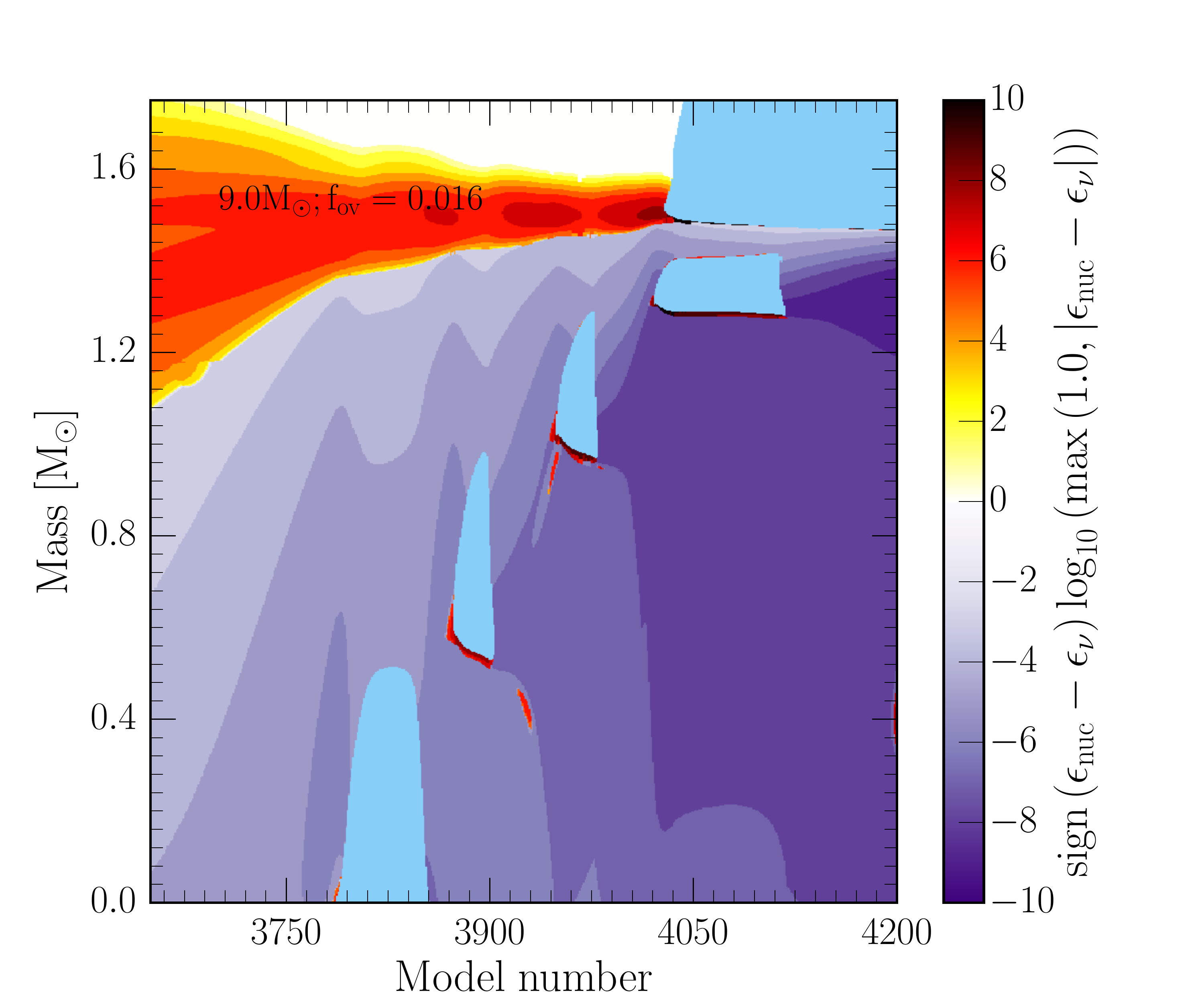}}
%                  \caption{9.0 \msun}
                \label{fig:kip9}
        \end{subfigure}
        ~
        \begin{subfigure}{
                \includegraphics[width=3.4in,height=2.75in]{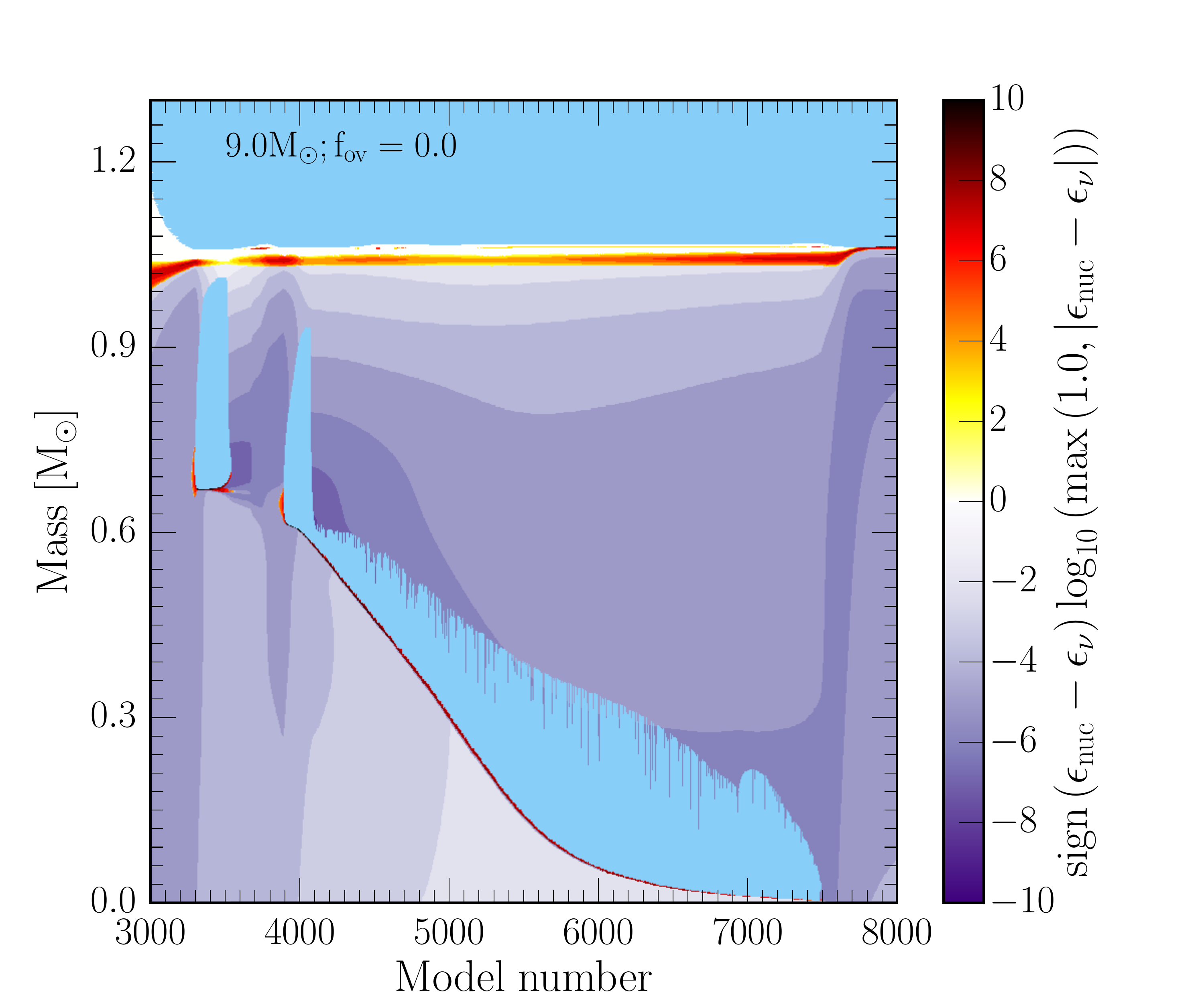}}
%                  \caption{9.0 \msun}
                \label{fig:kip9noover}
        \end{subfigure}
        \caption{Kippenhahn diagrams of flashes and steady state flames of 7.0 \msun\ (top left),
        7.5 \msun\ (top right),
8.0 \msun\ (middle left), 8.2 \msun\ (middle right) and 9.0 \msun\ (bottom left) \ ZAMS model stars, for non-rotating
\overshoot=0.016. Bottom right a 9.0 \msun\ non-rotating, \overshoot=0.0 model. 
Dark purple regions
indicate regions of cooling, primarily from thermal neutrino losses,
with the darker shades of purple representing a logarithmic increase in
the cooling rate.  Red regions indicate significant nuclear burning,
light blue regions indicate convection. For clarity the regions
undergoing other types of mixing are not shown.}\label{fig:f5}
\end{figure*}

We begin by considering a series of non-rotating stellar models using our
baseline mixing parameters as described in \S \ \ref{sec:method}. Figure
\ref{fig:f5} shows the Kippenhahn plots of these non-rotating stars as
representative samples for all the stellar models. In the 7 \msun \ case, 
carbon ignites off-center at $\approx$ 0.6 \msun. However, it rapidly
quenches and does not propagate towards the core. The 7.5 \msun \ case undergoes 
a series of carbon flashes, with each flash igniting closer to the core but it
is unable to form a steady state flame. In the 8 \msun \ case, 
an off-center ignition occurs at $\approx$ 0.15 \msun. A flame propagates
inwards and through a series of distinct flashes as the flame
approaches the core, and almost reaches the center. For the 9 \msun \ case,
carbon ignites at the center. In both the 8 and 9 \msun \ models,
secondary flashes at $\approx$ 0.5 \msun\ are due to off-center carbon burning.
For all cases the core is undergoing significant cooling, primarily
through photo-neutrino and plasma neutrino losses prior to the
first ignition of carbon \citep{nomoto_1984_aa,nomoto_1987_aa,ritossa_1996_aa,ritossa_1999_aa}.
The 7.5, 8.0 and 9.0 \msun \ cases the stars undergo a series of subsequent 
carbon flashes that travel away from the core. For the 9.0\msun \ case with no overshoot
(bottom right in Figure \ref{fig:f5}) the flame ignites off-center, contrary to the center ignition of the  9 \msun \ model with \overshoot.
The star undergoes a flash and then a flame, which propagates all the way to the center. The model with closest morphology is the 8 \msun \ case 
which has a flash and flame, but carbon burning does not reach the center. The difference between models with overshoot and without will
be discussed further in \S \ref{sec:mixsens}.

%NON - ROT KIPP - FIGURE 5
% \begin{figure}[htb]
% \centering{\includegraphics[width=3in]{figs_kip_7}}
% \centering{\includegraphics[width=3in]{figs_kip_8}}
% \centering{\includegraphics[width=3in]{figs_kip_9}}
% \caption{
% Kippenhahn diagrams of flashes and steady state flames of 7.0 \msun,
% 8.0 \msun, and 9.0 \msun \ ZAMS model stars.  Dark blue regions
% indicate regions of cooling, primarily from thermal neutrino losses,
% with the darker shades of blue representing a logarithmic increase in
% the cooling rate.  Red regions indicate significant nuclear burning,
% while light blue regions indicate convection. For clarity the regions
% undergoing other types of mixing are not shown.}
% \label{fig:f5}
% \end{figure}

%\subsection{Helium and Carbon Core Masses}
%\label{sec:core_masses}

\begin{figure}[htb]
\centering{\includegraphics[width=3.5in]{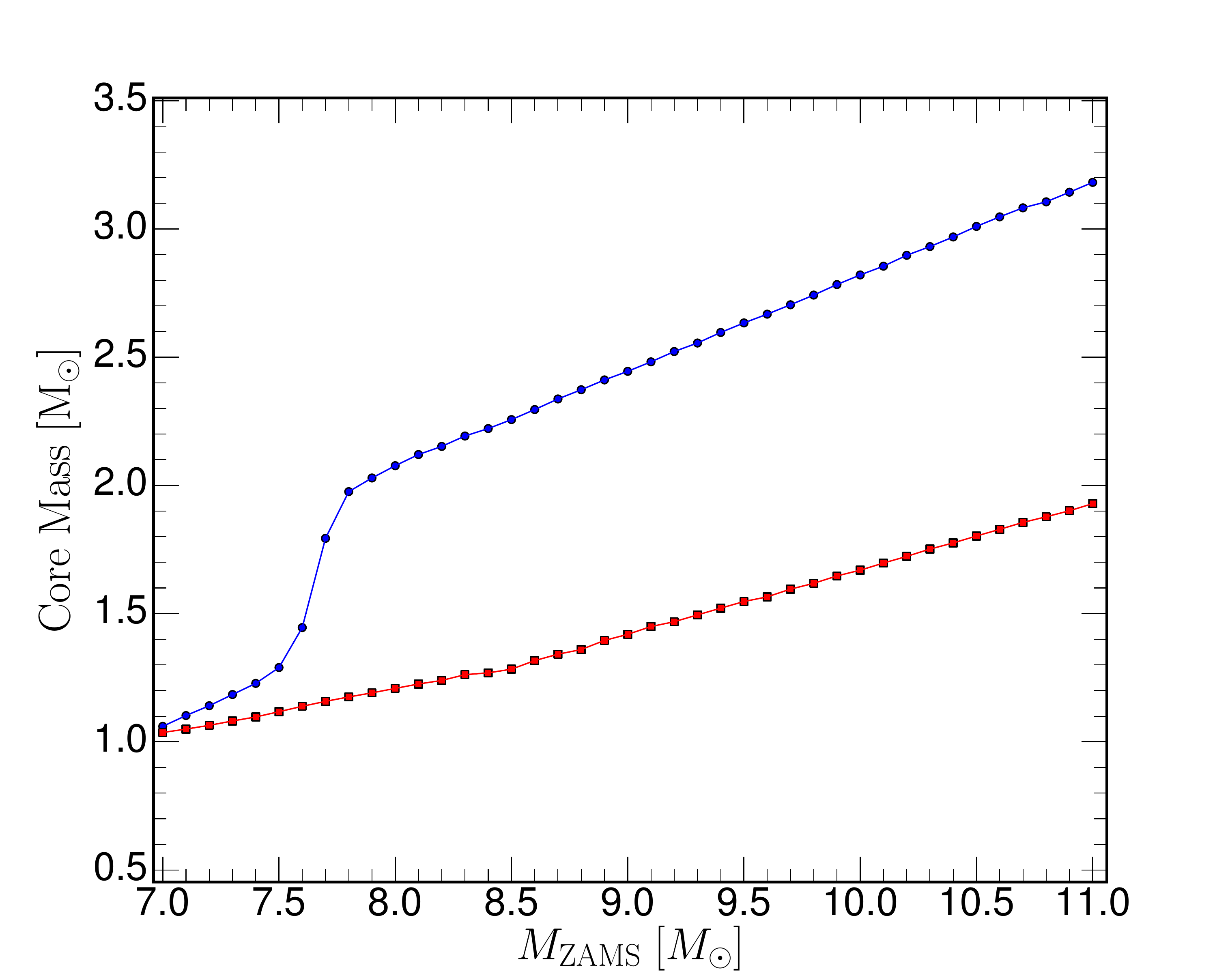}}
\caption{He and CO core mass as a function of ZAMS mass at the first 
ignition of carbon \review{, for our baseline mixing parameters}. In blue is the $^4$He core mass, while in red is the CO core mass.
Stars with M$_{\rm{ZAMS}}$ $<7.0 \ \msun$ are not shown as they do not have a carbon ignition point.
SAGB models with 7 \msun $\le$ M$_{{\rm ZAMS}}$ $\le$ 7.5 \msun \ have 
thin helium envelopes while models with M$_{{\rm ZAMS}}$ $> 7.5 \ \msun$ have thick helium envelopes.
}
\label{fig:f6}
\end{figure}

The He and CO core masses for the non-rotating models are shown in
Figure \ref{fig:f6} as a function of the ZAMS mass.  Stellar models
with M$_{\rm{ZAMS}}$ $<$ 7 \msun \ do not ignite carbon
\citep{becker_1979_aa,becker_1980_aa} and are not shown.  The carbon
core mass increases linearly with ZAMS mass over the range shown.  For
M$_{{\rm ZAMS}} \lesssim$ 7.5 \msun, the He forms a radiative,
geometrically thin, burning shell accreting material onto the CO core at a
rate of $\approx$ 10$^{-6}\msunyr$.  Between $\approx$ 7.5 \msun \ and
$\approx$ 8.0 \msun, Figure \ref{fig:f6} shows the He shell transitions
from an geometrically thin shell to an geometrically thick shell.  
For M$_{{\rm ZAMS}} \lesssim$8.0 \msun, the
geometrically thick He shell mass grows linearly with ZAMS mass over the
range shown, accreting material onto the CO core at $\approx$ 3$\times
10^{-6}\msunyr$ at 8.0 \msun, and increasing to $\approx$ 2.0$\times
10^{-5}\msunyr$ for the $11$ \msun \ stellar models. 

As the star evolves, the He shell grows in mass reaching $\approx$ 1.7
\msun \ for the 7 \msun \ stars and up to 3.2 for the 11 \msun
\ stars. Once the He shell reaches its peak size, the $^{4}$He
depletes quickly leaving a CO core.  Shortly afterwards the $^{4}$He shell
begins rapidly accreting material onto the CO core, eventually
depleting itself as well.  At the lowest ZAMS masses ($\lesssim 7\msun$), ignition occurs
after the $^{4}$He shell has accreted onto the CO core and the star has finished
its second dredge up (2DU) \citep{becker_1979_aa}. In the transition region where we have a series of flashes ($\approx 7-7.8\msun$),
the stars are undergoing their 2DU while igniting carbon. In the higher mass systems, $\approx$ 7.9$-$8.2 \msun,
where we have steady state flames or central carbon ignition, the star ignites 
carbon before the 2DU and before any significant accretion on to the CO core can occur.
Above 8.3 \msun \ we have dredge out events \citep{ritossa_1999_aa}, where the $^4$He shell grows an 
an outwardly moving convection zone which merges with the inwardly moving convective envelope.

At the base of the flame, we can ask whether a packet of convective material can 
penetrate into the region ahead of the flame transferring energy which will
decrease the flames lifetime. 
A simplified derivation (L. Bildsten, private communication) assumes 
the length scale $l$ of the flame front (the distance over which the temperature decreases
from a peak inside the flame to the background value) is much less than the local pressure scale height, $l \ll H$.
This implies the pressure is constant across the subsonic flame front. A convective packet will move from a region
of high temperature to a lower temperature region, at constant pressure. Assuming adiabatic motion, a 
fluid packet keeps its original temperature. Thus, the buoyancy felt by this convective packet is:
\begin{equation}
a=g \frac{\rm{d} \ln T}{\rm{d}r} \delta r
\enskip .
\end{equation}
Where $a$ is the buoyancy acceleration, $g$ is the local acceleration due to gravity, 
and $\delta r$ is the distance the packet moves ahead of the flame.
Simplifying,
\begin{equation}
a \approx -\frac{g}{l} \delta r
\enskip ,
\end{equation}
and solving this harmonic oscillator equation we find
\begin{equation}
 \delta r \approx \frac{v_c}{N}
\enskip ,
\end{equation}
where $v_c$ is the convective velocity and $N$ is the Brunt–V\"{a}is\"{a}l\"{a} frequency. Using
the local scale height $H=c_s^2/g$, where $c_s$ is the local sound speed:
\begin{equation}
 \frac{\delta r}{l}=\frac{v_c}{c_s} \left(\frac{H}{l}\right)^{1/2}
\enskip .
\end{equation}
In a typical carbon flame we have $v_c/c_s\approx 10^{-4}$ and $H/l \approx 10$, thus $\delta r \ll l$ and a convective fluid packet
cannot penetrate the flame front.

%NON - ROT HE/C CORE - FIGURE 6

%NON - ROT Mass/Density IGN - FIGURE 7
\begin{figure}[htb]
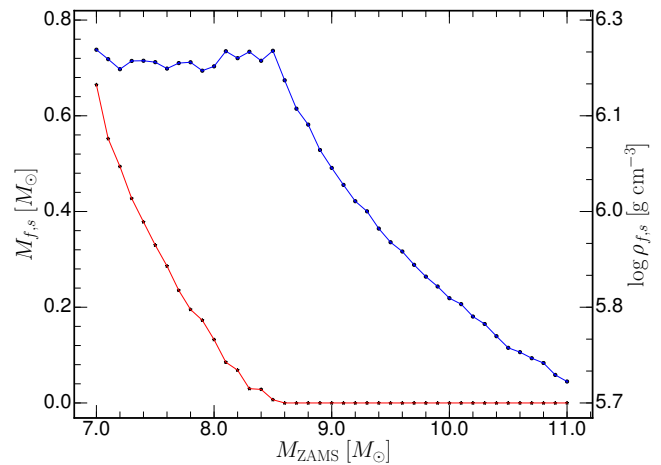

\centering{\includegraphics[width=3.5in]{{{figs_non_rot_ignMassDen}}}}
\caption{
Mass location of first carbon ignition (red) and local density at first carbon ignition (blue)
as a function of the ZAMS mass for non-rotating\review{, baseline mixing parameters models}.
}
\label{fig:f7}
\end{figure}

\subsection{First Ignition of Carbon}
\label{sec:first_ignition}

%Where inside a stellar model does carbon first ignite? Why does
%carbon first ignite at specific mass locations?  Answering these
%questions is the main subject of this subsection.

Figure \ref{fig:f7} shows the mass location of the first ignition as a
function of the stellar model ZAMS mass.  For our choice of baseline
mixing parameters, the lowest mass star to ignite carbon is a
7$ \ \msun$\ model and ignition occurs off-center at a mass coordinate of
$\approx$ 0.65 \msun.  As the ZAMS mass increases, the location of the off-center
first ignition location moves steadily inwards in mass
\citep{siess_2007_aa}. For ZAMS masses larger than 8.4 \msun, carbon
ignites at the center.  Figure \ref{fig:f7} also shows the local
density at the location of first ignition.  All stellar models that
ignite carbon off-center, 7 \msun $\le$ M$_{{\rm ZAMS}}$ $\le$
8.4 \msun, do so at a nearly constant density of $\log \rho \approx$ 6.2, or 
\hbox{$\rho \approx$ 1.5$\times$10$^6$} g cm$^{-3}$.  For stellar models that
undergo central carbon ignition, 8.4 \msun $<$ M$_{{\rm ZAMS}}$ $\le$
11 \msun, the density at ignition monotonically decreases.
Models which do not ignite carbon will eventually form a CO WD. Those models that undergo off-center carbon ignition
but where the burning does not reach the center
will form hybrid CO+ONe WD. Model stars which ignite carbon at the center will 
eventually form an ONeNa WD \review{which may explode as} an ECSNe.

To a first approximation, at ignition the nuclear burning timescale
and thermal diffusion timescales \review{are equal \citep{timmes_1992_aa}}
\begin{equation}
\tau_{\rm diff} \sim \frac{\sigma}{\rho  \ C_{p}}
\hskip 0.5in
\tau_{\rm burn} \sim \frac{E}{\epsilon_{{\rm nuc}}}
\enskip ,
\label{eq:timescales}
\end{equation}
where $\sigma$ is the thermal conductivity, 
$C_{p}$ is the specific heat capacity at constant pressure, $E$ is 
the local
thermal energy, and $\epsilon_{{\rm nuc}}$ is the
screened nuclear energy generation rate during carbon burning
including energy losses due to neutrino cooling. 
For a given temperature, density, and composition both $C_{p}$ and $E$
may be calculated from an equation of state.
For carbon burning,
$\epsilon_{{\rm nuc}}$ takes the form \citep[e.g.,][]{woosley_2004_aa}:
\begin{equation}
\epsilon_{{\rm nuc}} = 
6.7 \times 10^{23} \ X^2(^{12}{\rm C}) \ \rho_6 \ f_{{\rm screen}} \ \lambda_{12,12}  \ - \ \epsilon_{\nu}
\enskip ,
\label{eq:sdot}
\end{equation}
where X($^{12}$C) is the carbon mass fraction, $\rho_6$ is the density
divided by 10$^6$ g cm$^{-3}$, $\lambda_{12,12}$ is the unscreened
nuclear reaction rate for carbon burning and $f_{\rm{screen}}$ is the
electron screening factor.  Using a MESA 501 isotope reaction network that includes
neutrino losses with an initial
initial composition of X($^{12}$C) = 0.3 and X($^{16}$O) = 0.7 to calculate 
$\epsilon_{{\rm nuc}}$ over the relevant range in the $\rho$-T plane, we find
positive values of $\epsilon_{{\rm nuc}}$ may be approximated by the power law
\begin{equation}\label{eqn:c_burn_nuc}
\epsilon_{{\rm nuc}} \approx 
1.6 \times 10^{7} \left (\frac{T}{7 \times 10^{8}} \right )^{23} \left ( \frac{\rho}{2 \times 10^{6}}\right )^{1.2} 
\enskip .
\label{eq:sdot_fit}
\end{equation}
Results for carbon ignition 
\review{for any large reaction network, including the 501 isotope network used here,} are 
expected to be similar to that of the smaller 22 isotope net used in the SAGB models
because both networks have the same set of \review{key} isotopes and reaction rates crucial for carbon burning.
\review{\citep[e.g.,][]{timmes_2000_ab}.}

%NON - ROT S DOT - FIGURE 9
\begin{figure}[!htb]
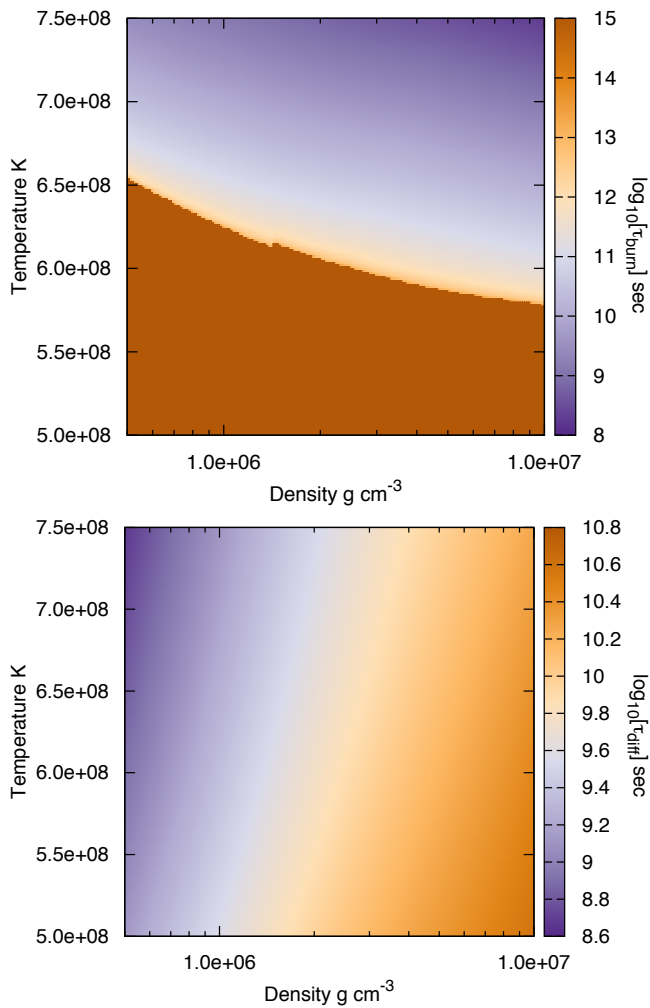

\includegraphics[width=3.4in]{{{figs_tau_burn-crop}}}
\includegraphics[width=3.4in]{{{figs_tau_diffu-crop}}}
\caption{
Nuclear burning (top) and thermal diffusion (bottom) timescale in the $\rho$-T plane
for X($^{12}$C) = 0.3 and X($^{16}$O) = 0.7.
}
\label{fig:f9}
\end{figure}

Fitting the other quantities in equation \ref{eq:timescales} in a
similar manner, we find the following expressions for the nuclear
burning timescale and thermal diffusion timescales
\begin{equation}
\tau_{{\rm burn}}  = 
5.1 \times 10^{9} \left(\frac{T}{7 \times 10^{8}} \right)^{-32} \left( \frac{\rho}{2 \times 10^{6}}\right)^{-0.8}
\label{eq:tau_burn}
\end{equation}
\begin{equation}
\tau_{{\rm diff}}  = 
4.0 \times 10^{9} \left(\frac{T}{7 \times 10^{8}} \right)^{-2.4} \left( \frac{\rho}{2 \times 10^{6}}\right)^{-1} 
\enskip .
\label{eq:tau_diff}
\end{equation}
These two timescales are shown in Figure \ref{fig:f9}.
Equating the two timescales gives
\begin{equation}
 \left (\frac{T}{7 \times 10^{8}} \right )^{29.6} \left ( \frac{\rho}{2 \times 10^{6}}\right )^{-0.2}  = 1.3
\enskip .
\end{equation}
At the threshold of vigorous carbon burning, \hbox{$T\approx$ 7$\times$10$^{8}$ K}, this expression 
gives a unique and constant ignition density
\begin{equation}
\rho_{ign} \approx 2.1 \times 10^6 \quad {\rm g \ cm}^{-3}
\enskip ,
\label{eq:rho_ign}
\end{equation}
which is consistent with the constant ignition density of 
\hbox{$\rho \approx$ 1.5$\times$10$^6$ g cm$^{-3}$} found in the MESA SAGB models that
ignite carbon off-center, 7 \msun $\le$ M$_{{\rm ZAMS}}$ $\le$
8.4 \msun, of Figure \ref{fig:f7}. We find this result also holds for our
rotating SAGB models.

\begin{figure}[!htb]
\centering{\includegraphics[width=3.7in]{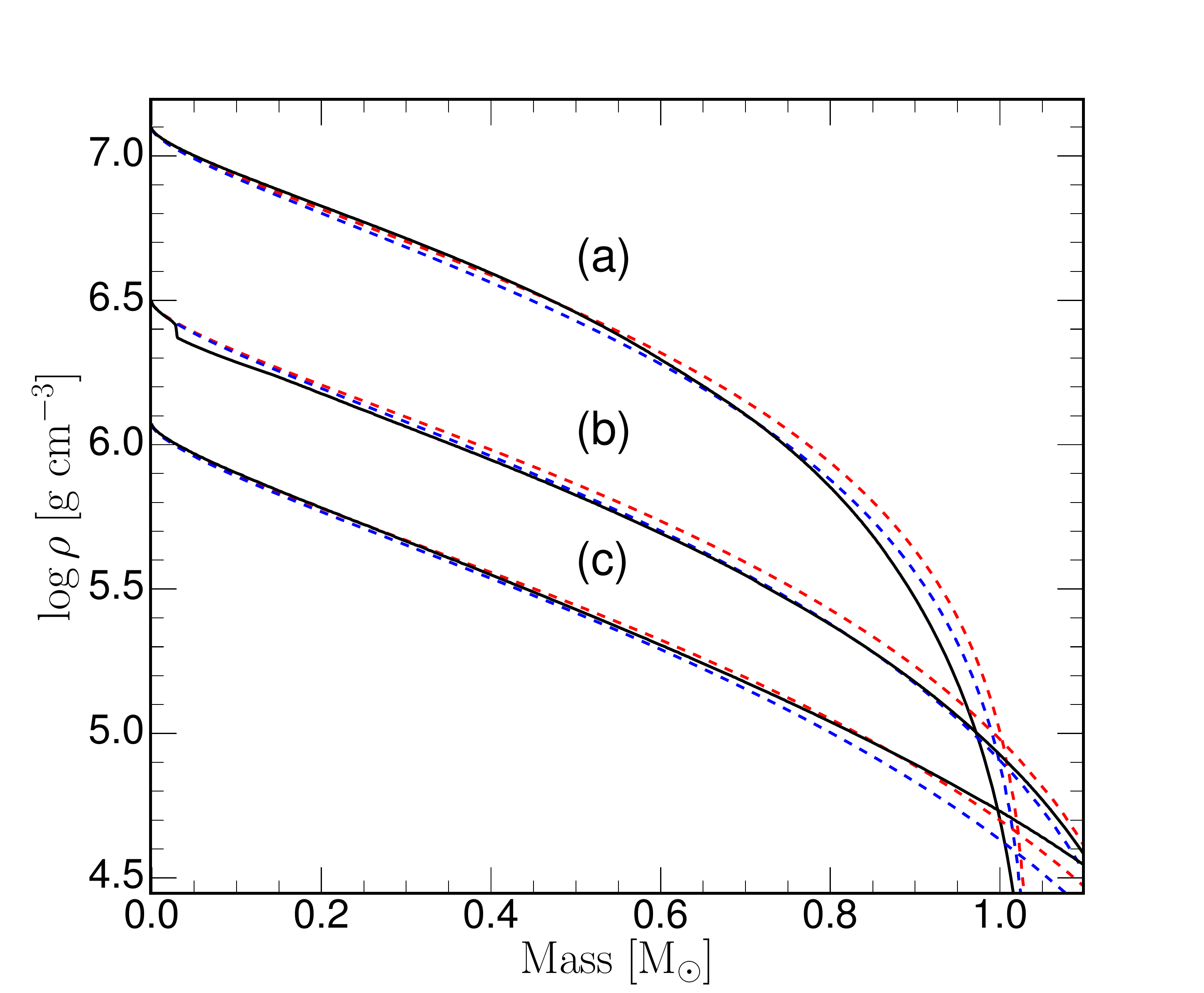}}
\caption{Polytrope fits to the MESA carbon core density structure for the 
non-rotating (a) 7.0 \msun,(b) 8.0 \msun, and (c) 9.0 \msun \ ZAMS model stars
from top to bottom, respectively. In each case, the MESA model is bounded by 
two polytropic fits; (a) red n=2.9, blue n=3.1; (b) red n=3.6, blue n=3.8;
(c) red n=4.2, blue n=4.4.
}
\label{fig:poly}
\end{figure}

\begin{figure}[htb]
\centering{\includegraphics[width=3.4in]{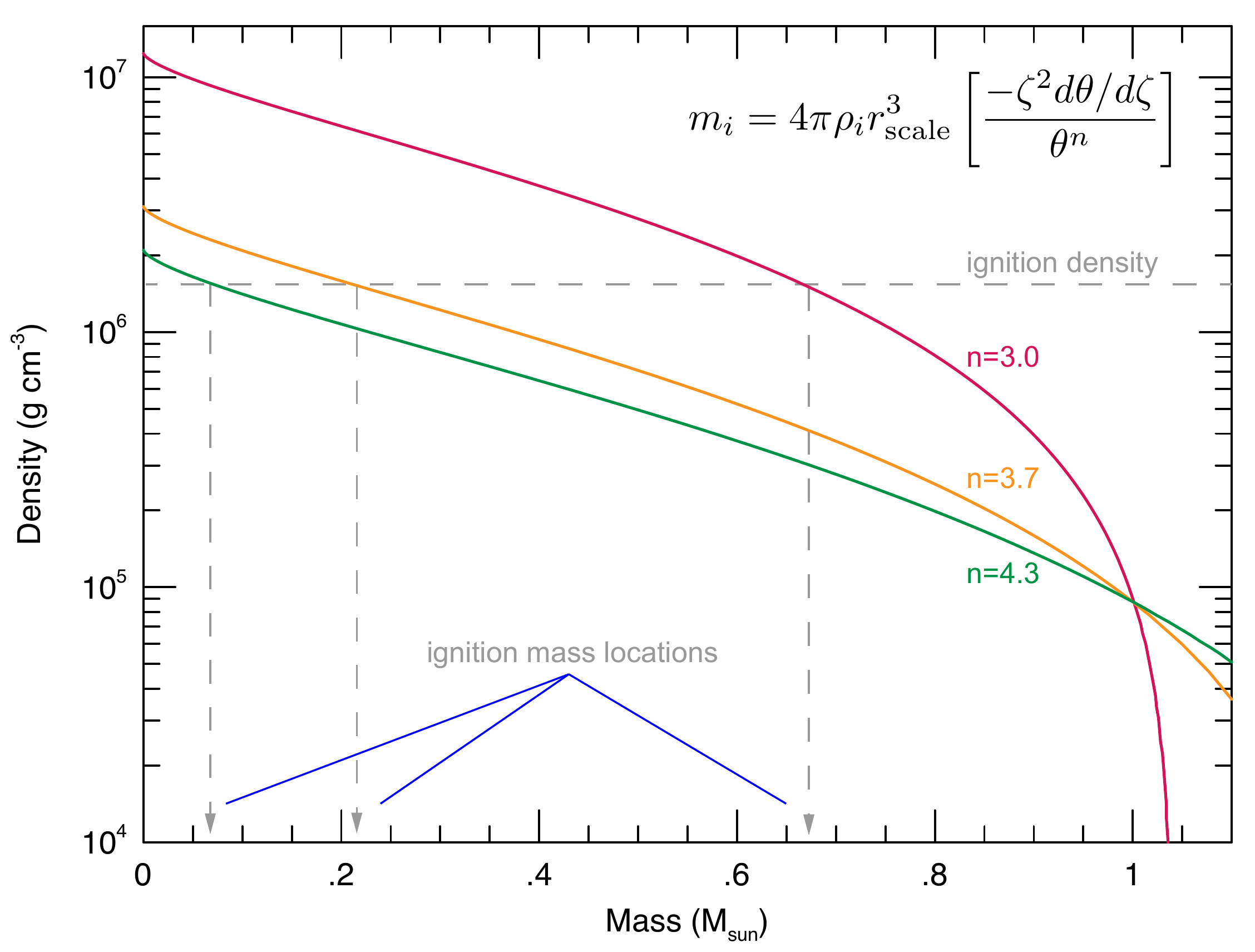}}
\caption{
Mass locations of first carbon ignition from the ignition curve and polytropic models
in the density$-$mass plane.
}
\label{fig:f8}
\end{figure}

Figure \ref{fig:poly} (a) shows the carbon core of a
non-rotating \hbox{7 \msun} ZAMS star can be well approximated by a 
$n \approx$ 3.0 polytropic model, (b) a 8 \msun\ ZAMS star by a 
$n \approx$ 3.7 polytrope, and (c) a 9 \msun\ ZAMS star by a 
$n \approx$ 4.3 polytrope.  The density structures for these polytropic
models were calculated using an open-source
tool\footnote{\url{http://cococubed.asu.edu/code\_pages/polytrope.shtml}}.
In addition, the polytropic relations offer an explanation for why 
the location of the first ignition moves steadily inwards in mass for the off-center ignition cases
(see Figure \ref{fig:f7}).
The density
structure of the $n$ = 3.0, 3.7, and 4.3 polytrope models
are shown in Figure \ref{fig:f8}. The mass
locations for a fixed ignition density (dashed horizontal line),
where ignition occurs (dash vertical lines) moves
monotonically inwards as the polytropic index increases, as the carbon
core mass increases, as the ZAMS mass increases.

%NON - ROT POLYTROPE - FIGURE 8

We now turn to the decrease in the central density for
those SAGB models in Figure \ref{fig:f6} that centrally ignite carbon.
Homology relations between the central density $\rho_c$ and
the mass $M$ \citep[e.g.,][]{hansen_2004_aa,kippenhahn_2012_aa}
for a chemically homogeneous star characterized by a mean
molecular weight $\mu$, constant opacity, ideal gas equation of state
and power-law energy generation rate $\epsilon_{{\rm nuc}} \propto T^{\nu}$ indicate
\begin{equation}
\rho_c \propto \mu^{\frac{3(4-\nu)}{\nu+3}} M^{\frac{2(3-\nu)}{\nu+3}}.
\label{eq:homology}
\end{equation}
For carbon burning near ignition, equation 
\ref{eq:sdot_fit} shows \hbox{$\nu \approx$ 23}, and equation \ref{eq:homology} then gives
$\rho_c \propto$ M$^{-1.5}$. The negative exponent shows that the density at 
first ignition, for those \hbox{8.4 \msun $<$ M$_{{\rm ZAMS}}$ $\le$ 11 \msun} 
models that undergo central carbon ignition, monotonically
decreases as the mass of the carbon core increases 
with a slope that is consistent with the rate of
decline shown in Figure \ref{fig:f7}.

\subsection{Carbon Burning flashes and Transition to a Steady State Flame}
\label{sec:flashes}

\begin{figure}[htb]
\centering{\includegraphics[trim = 2.1in 1.8in .1in .1in, clip, width=4.9in]{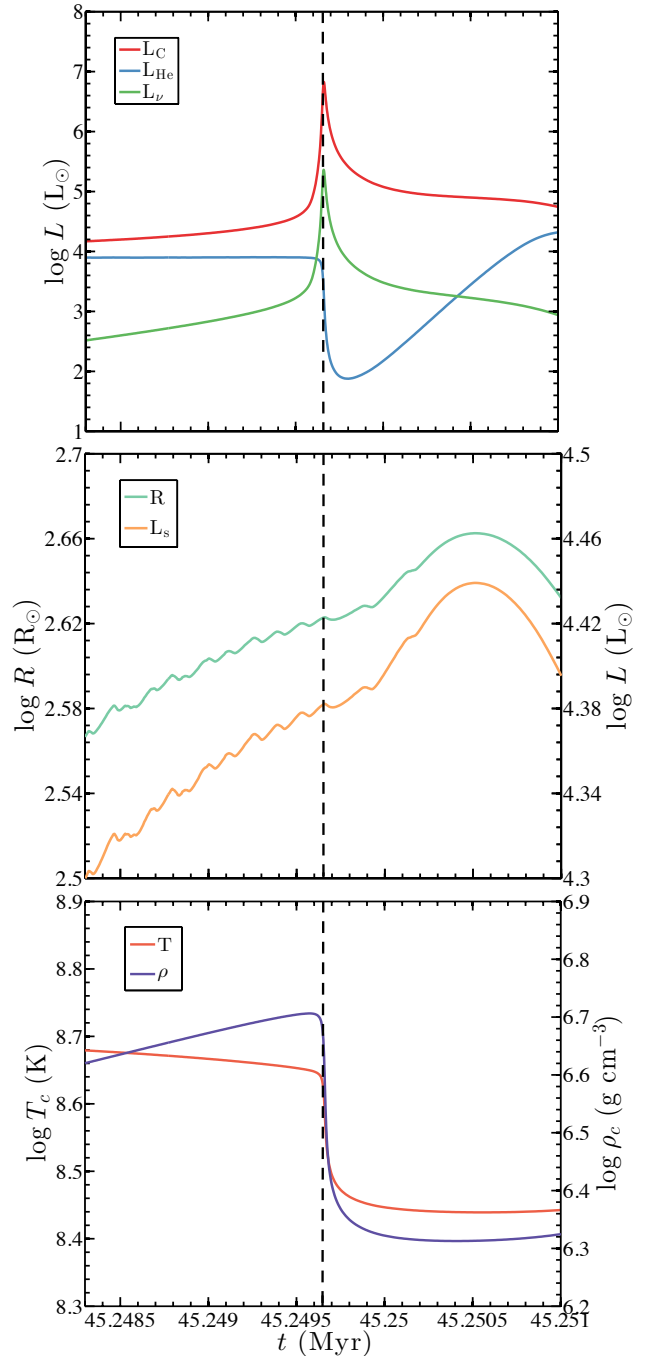}}
\caption{
Local and global characteristics before, during, and after the first carbon
burning flash in an 7.5$ \ \msun$ model. $L_{C}$ is the total luminosity due to \review{carbon burning}, $L_{He}$ the luminosity due to nuclear
 helium burning, $L_{\nu}$ the luminosity due to thermal neutrino losses, $T_c$ the central temperature, 
$\rho_c$ central density, $R$ the surface radius, and $L_s$ the surface luminosity.
The black dotted line marks first ignition of carbon.}
\label{fig:flash}
\end{figure}

Local characteristics before, during, and after the first off-center
carbon flash in a 7.5 \msun \ model is shown Figure \ref{fig:flash}.  
Before the first ignition of carbon, the CO core is in its most
compact, most electron degenerate configuration.  The first off-center
carbon burning flash is thus the most powerful; any subsequent flashes
or steady-state flames take place under more extended, less degenerate
configurations.  In addition, the energy released during the first
ignition decreases as the ZAMS mass increases because the CO core is
not as compact and not as degenerate.  Furthermore, the first carbon
flash impacts the base of the convective envelope more strongly in
lower mass SAGB models than in higher mass SAGB models
\citep{garcia-berro_1997_aa,siess_2006_aa}, partly because of their
more compact configuration but also because the first flash occurs
farther from the center in lower mass models (see Figure
\ref{fig:f7}).

At the start of carbon burning at the ignition density of $\rho_{ign}
\approx$ 1.5$\times$10$^6$ g cm$^{-3}$, marked by the vertical dotted line
in Figure \ref{fig:flash}, the energy generation rate rapidly rises.
Figure \ref{fig:flash} shows that $\approx$ 10\% of the energy produced by
nuclear reactions is lost to neutrinos, with the balance of the
thermal energy transported by convection \citep{ritossa_1996_aa,saio_1998_aa}. 
This is not the $\approx$ 50\% expected for steady state burning, as the 7.5\msun \ model
is undergoing a time-dependent flash and not a steady state flame.
Were it not for carbon burning, the surface luminosity and radius of
these SAGB models would continue to increase in step with an
increasing helium-burning luminosity, similar to lower mass, M$_{\rm{ZAMS}}$ $\lesssim$
7 \msun \ AGB stars that do not experience carbon burning
\citep{garcia-berro_1997_aa,karakas_2014_aa}.  These dramatic events
mainly impact the core and have only a modest effect on the structure
of the outer envelope and surface luminosity.

Figure \ref{fig:flash} shows the rapid injection of energy causes the
entire core to expand and cool \citep{ritossa_1999_aa}. A convective
region also develops above the ignition location in response to the
injection of energy from carbon burning.  The overall expansion of the
core decreases the electron degeneracy parameter ($\eta = \mu_e/kT
\rightarrow -\infty$ means non-degenerate where $ \mu_e$ is the
electron chemical potential, $\eta \rightarrow \infty$ means perfect
degeneracy). The expansion extinguishes the flash partially because
the nuclear energy generation rate drops below the critical luminosity
\citep[e.g.,][]{siess_2006_aa} and partially because the ignition density
$\rho_{ign}$ is pushed deeper into the stellar model. The first flash
thus quenches and does not become a steady-state flame that propagates
inwards toward the center. 

After the first flash is quenched and the deposition of energy from
nuclear reactions ceases, the core again contracts but to a less
compact, less degenerate configuration. This contraction leads to the
second ignition of carbon. The second flash (and any subsequent
flashes) occurs at a deeper mass coordinate because the accretion of C
from the He burning shell slightly increases the core mass which moves
the location of the critical density $\rho_{ign}$ inwards where there
is also fresh, previously unburned fuel. Similar evolution pathways
can occur for other fuels and other masses, for example, 
neon burning in more massive models \citep[e.g.,][]{jones_2013_aa}.

The evolution of the second and any subsequent flashes is notably
different than the first flash as shown in Figure \ref{fig:f5}.  When
these later flashes develop, their convective region grows into regions
previously occupied by previous flashes where carbon had been
depleted. Thus the nuclear energy production rate, which from
equation \ref{eq:sdot} is proportional to X$^2$($^{12}$C), 
is reduced. Furthermore, the subsequent core expansion and induced
thermodynamic changes are significantly smaller. The first
flash lasts the longest and is the most vigorous, while subsequent flashes have shorter
lifespans, release less energy, and the time interval between flashes becomes shorter
Models which have a geometrically thin helium envelope show more flashes
than models with a  geometrically thick helium envelope (see Figure \ref{fig:f6}).

Each flash releases less energy, expands the core by a smaller amount,
and moves the mass location of the critical density inward at a slower
rate.  This allows a flash to transition to a steady-state flame.
Thus, after one or more flashes a steady-state may be achieved.
Combustion in steady-state flames is incomplete; only a small portion
of the carbon burns. A condition of balanced power is set up where the
rate of energy emitted as neutrinos from the base of the convective
region equals the power available from the unburned fuel that crosses
the flame front. The inward propagation of the flame by thermal
conduction is limited by the temperature at the base of the convective
shell, which cannot greatly exceed the adiabatic value. These two
local conditions give a unique speed for the flame, with typical
values of \hbox{$\approx$ 0.1 cm s$^{-1}$}. We verified the flame speed at
several locations in the SAGB models and it is commensurate with previous
local studies, with speeds of order $10^{-3}-10^{-2}$cm s$^{-1}$\citep[e.g.,][]{timmes_1994_aa, ritossa_1996_aa,
  garcia-berro_1997_aa,siess_2006_aa,denissenkov_2013_ab}. The flame lives $\approx$ 20,000 years on its
journey towards the center.
 
The convective nature of the material behind the flame has two key
consequences for its journey to the core.  First, the temperature
behind the flame front (i.e., towards the surface) is bounded. The
ONeNa ashes of the burning are not allowed to assume an arbitrary
value of temperature; rather convection fixes the temperature behind
the flame front. The second feature of the convective material is the
lack of abundance gradients behind the flame. That is, convection
uniformly mixes the ashes of the partial burning. 

For our standard mixing parameter settings, models with ZAMS masses in
the 7$-$7.6 \msun \ range do not achieve a steady-state flame.
Instead, they undergo a series of flashes, where each flash occurs
closer to the core \citep[][and see Figure \ref{fig:f5}]{siess_2009_aa,denissenkov_2013_ab}.  
The number of flashes increases as the ZAMS mass increases, until the
ZAMS mass exceeds 7.6 \msun \ when the first flash transitions into
steady-state flame.  For stars between 7.7$-$8.4 \msun \ the off-center
steady-state flame begins closer to the center.  The dependence of
this mass range on the composition mixing parameters is discussed in
\S\ref{sec:rotation}.

\subsection{Does the Burning Reach the Center?}
\label{sec:center}

Whether off-center carbon burning, either as a steady-state flame or
as a series of time-dependent flashes, reaches the center in these
models depends on the ZAMS mass and the adopted mixing parameters.
When convective mixing operates within the Schwarzschild boundaries, a
flame \review{will propagate all the way toward} \review{the center}
\citep{nomoto_1985_aa,garcia-berro_1997_aa,saio_1998_aa, siess_2009_aa,doherty_2010_aa}.
\review{Convective overshoot is not strictly required for carbon flames
to be quenched away from the center. Additional mixing processes such as thermohaline
mixing have been shown to effectively quench the carbon flame away from the center
while convection is operating within only the strict Schwarzschild boundaries. 
  \citep{siess_2009_aa,denissenkov_2013_ab}  }
Mixing processes that, by design, extend beyond the MLT convective
boundary take unburned carbon fuel ahead of the flame front and mix
this fresh fuel with the ashes of the convective region behind the
flame front.  This starves the flame of fuel, with the nuclear energy
production rate proportional to the square of the carbon abundance
(see equation \ref{eq:sdot}).  For instance, instead of a fresh carbon
mass fraction of 0.3 the carbon mass fraction near the ignition point
may be depleted to 0.1 and polluted with enhanced abundances of
$^{16}$O, $^{20}$Ne, $^{23}$Na and $^{24}$Mg.  Examples of such
convective boundary mixing processes include thermohaline and
overshoot.  For large enough mixing parameters, the flame either
disintegrates and sputters in a series of fuel starved flashes moving
towards the center, or is extinguished before reaching the center
\citep{siess_2009_aa,denissenkov_2013_ab}.  Where the flame is
extinguished, if it is extinguished, can have repercussions for the
composition of the subsequent white dwarf that is formed, and from
there, possible consequences for supernova Type Ia models.

%NON - ROT FLAME TRAVEL - FIGURE 11
\begin{figure}[htb]
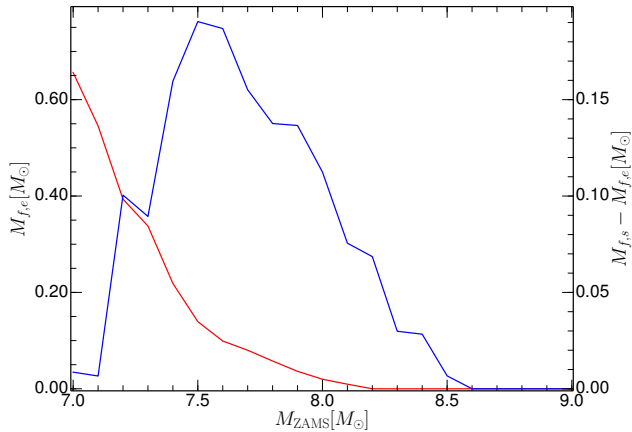

\centering{\includegraphics[trim = 0.4in .1in .0in .4in, clip,width=3.3in]{{{figs_non_rot_finalStatemass}}}}
\caption{
Carbon burning extinction location in mass coordinate (red curve) and mass
traversed (blue curve) for the non-rotating models with baseline mixing
parameters.  Stars with $M>9$\ \msun \ are not shown as they all ignite at
the center.
}
\label{fig:f11}
\end{figure}

Figure \ref{fig:f11} shows the location where carbon burning is
extinguished (red curve, left y-axis) for the non-rotating models with
baseline mixing parameters.  The distance (in mass coordinates) that
the flame traveled from birth to death is shown by the blue curve and the 
right y-axis.  For 7.0 $\le$ M$_{\rm{ZAMS}}$ $\le$ 8.2, carbon burning does not
reach the center.  As the ZAMS mass increases, the flame or flashes get
closer to the center, eventually reaching the center at 8.2 \msun.  In
terms of the mass traversed, the flame (or flashes) increases its
travel distance from 7.0 to 7.5 \msun \ and then decreases for higher
mass models.  This transition occurs as stars with masses between 7.0
and 7.5 \msun \ undergo a one or more of flashes, where each flash does
not travel, but each subsequent flashes ignites closer to the core. For
higher mass models a steady-state flame reaches the center.  The
dependence of these results on the chosen mixing parameters is
discussed \S \ \ref{sec:rotation}.

\begin{figure}[htb]
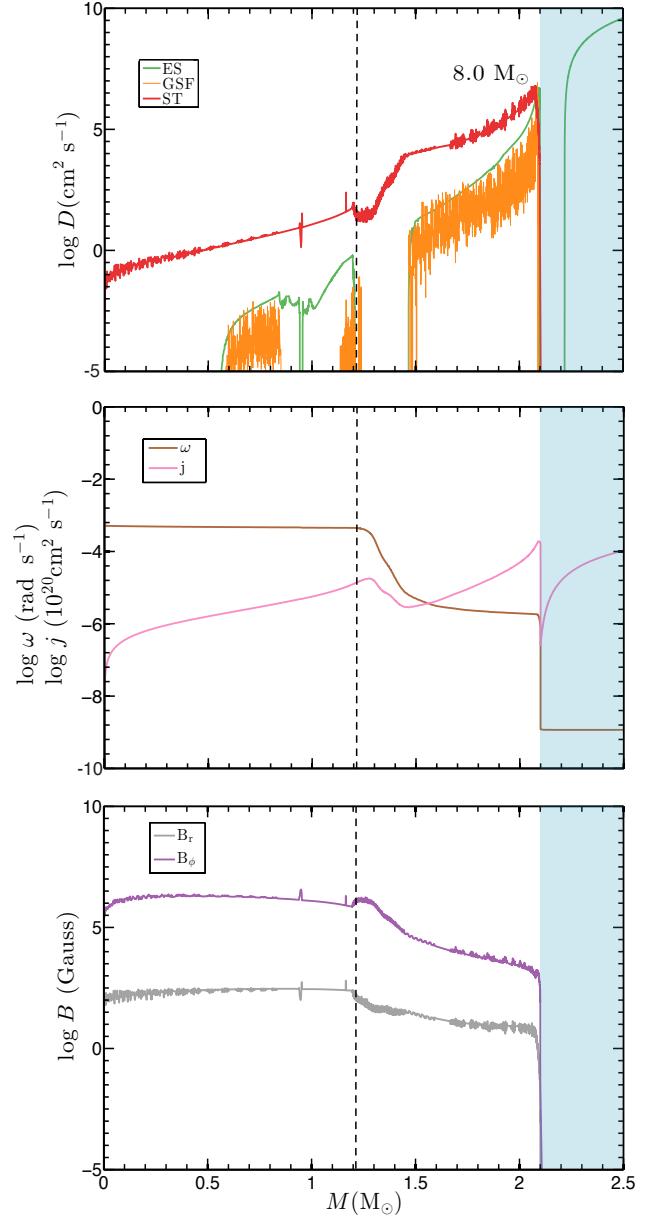

\flushleft{\includegraphics[trim = 2.2in 1.7in 1.4in 1.2in, clip, width=4in]{{{figs_rot_intro_stacked_rotation_rev1}}}}
\caption{
Angular momentum diffusion coefficients (top panel); specific angular
momentum $j$, angular frequency $\omega$ (middle panel), and radial magnetic field $B_r$, and azimuthal magnetic field $B_\phi$,
components (bottom panel), of an 8.0M$_\odot$ ZAMS model with an
initial rotation at ZAMS of \rot=0.2 \review{, baseline mixing parameters,} at the onset of carbon
ignition. The blue shaded region indicates a convective region, and
the dashed (black) line shows the boundary of the CO core.  The
angular momentum diffusion coefficients shown are the Eddington-Sweet
circulation (ES), Goldreich-Schubert-Fricke instability (GSF), and
Spruit-Tayler dynamo (ST).
}
\label{fig:f1}
\end{figure}

For the models between 7.8 \msun\ and 8.2 \msun, which do not reach
the center, they transition from a steady state flame into a series of
flashes, as seen in Figure \ref{fig:f5} for the 8 \msun\ case. These
flashes ignite in regions where the X($^{12}$C) abundance has dropped
due to the mixing. A subsequent flash thus requires a higher critical density
(see equation \ref{eq:rho_ign}) to ignite in the presence of a lower
abundance of carbon.  These flashes are able to drive a local expansion
of the core and the critical density to move inwards into the core,
analogous to the flashes seen in the $<7.8 \ \msun$\ models.  The
flashes eventually stop forming once the core can no longer reach the
critical density leaving unburned carbon in the core.

\vskip 0.5in

\begin{figure}
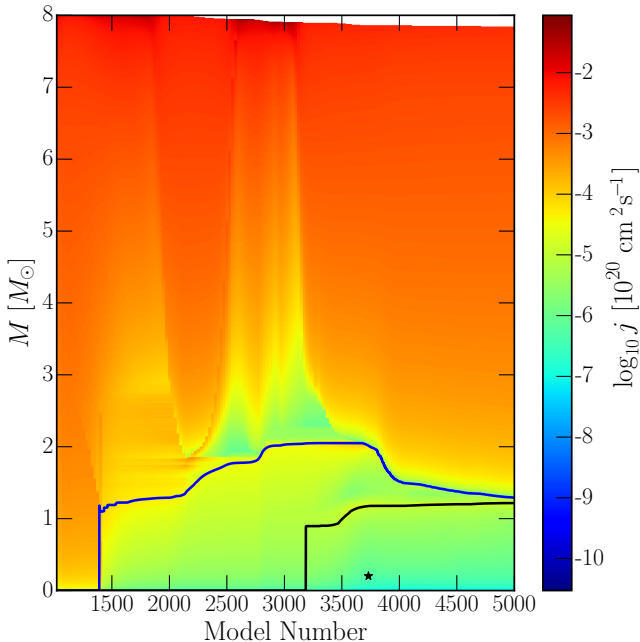

 \includegraphics[width=1\linewidth]{{{figs_rot_intro_am_evol}}}
 \caption{Angular momentum evolution of a 8 \msun \ star, with initial rotation \rot=0.25 and
 overshoot \overshoot=0.016. The blue line shows the extent of the $^4$He core, the black line the extent
 of the CO core and the black star marks the location of the first ignition. The white region at the top is where
 the star has lost mass.}
 \label{fig:am_evol}
\end{figure}

\begin{figure}[htb]
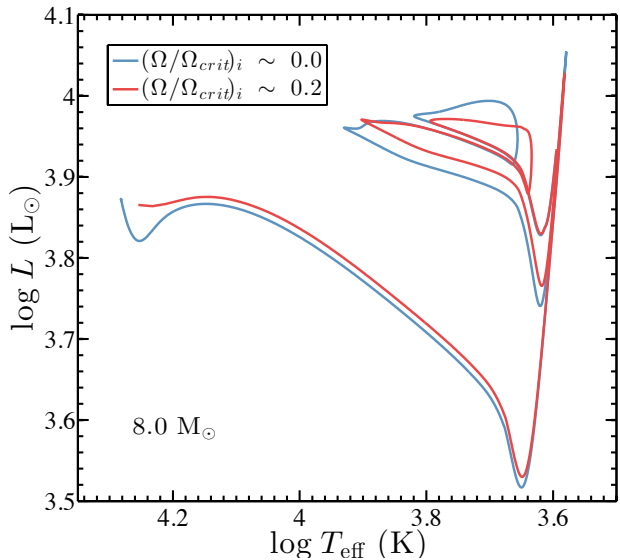

\centering{\includegraphics[trim = 1.7in 0.9in 1in 0.5in, clip, width=3.6in]{{{figs_rot_grid_8m_he_depletion_rev1}}}}
\caption{HR diagram of two 8 \msun \ models: one non-rotating (blue line) 
and one with  $\rot $ $\sim$ 0.2 (red) \review{, for baseline mixing parameters}. 
The diagram shows the evolution from H depletion to He depletion for both models.
}
\label{fig:HR}
\end{figure}

\section{Results From The Grid Of Rotating and Overshooting Models}
\label{sec:rotation}

In this section we investigate the impact of rotation and convective
overshoot on carbon burning in the SAGB models using the range of
values in Table \ref{table:cube} and baseline mixing parameters listed
in Table \ref{table:baseline}.  As an example of the rotation
characteristics in the carbon core, Figure \ref{fig:f1} shows the
angular momentum diffusion coefficients, specific angular momentum,
angular frequency, and magnetic field profiles at first carbon
ignition for an 8.0 \msun\ ZAMS model with \rot=0.2. Rotation is
initialized by imposing a solid body rotation law at ZAMS. For
\rot=0.2 this corresponds to an initial angular frequency of
$\omega$=1.522$\times$10$^{-5}$ rad s$^{-1}$ and a total angular
momentum of $L$=1.90$\times$10$^{51}$ erg s.  At first carbon
ignition, Figure \ref{fig:f1} (middle panel) shows the carbon core has
spun up a factor of $\approx$ 30 to $\omega \approx$ 6$\times$10$^{-4}$ rad
s$^{-1}$ and rotates as a solid body.  The largest angular momentum
diffusion coefficient, by several orders of magnitude, at first carbon
ignition is due to the Spruit-Tayler dynamo. The implied $\approx$ 1 MG
radial component of the magnetic field is shown in the bottom panel. 
DSI and SSI are not
shown due to their negligible contributions in this model.

Figure \ref{fig:am_evol} illustrates the evolution of the specific
angular momentum from ZAMS to first carbon ignition for the
8.0 \msun\ ZAMS model initialized at ZAMS with \rot=0.25 star and
\overshoot=0.016. During the main-sequence phase (model numbers
1000$-$1400) the specific angular momentum is uniformly distributed
throughout the model. As the star ascends the RGB and then the AGB
(model numbers 1400$-$3200) the specific angular momentum is extracted
from the core and redistributed into the envelope, decreasing the
specific angular momentum in the core by a factor of $\approx$ 100.
Boundaries between convective and non-convective regions are
distinguished by sharp transitions in the specific angular momentum,
with non-convective regions having the least specific angular
momentum. The first ignition of carbon, off-center in this case, 
occurs around model 3700 and is marked.

While \citet{poelarends_2008_aa} claims that mass loss will strongly 
effect the final outcome of SAGB, we note that in our models (with a only 
one mass loss rate used) that mass loss seems to have a minimal effect on the star 
\review{up to carbon ignition}.
In figure \ref{fig:am_evol} the effect of mass loss is visible as the white region at the top
of the figure and only significant near the end of the 
core helium burning phase. While mass loss will extract angular momentum 
from the star, the core's angular momentum is unaffected.

Rotation during the main sequence supplies a prolonged source of
hydrogen fuel that builds a slightly more massive helium core than non-rotating
models \citep[e.g.,][]{maeder_2000_aa,heger_2000_aa,lagarde_2012_aa}.
The increase in core mass ($\approx 0.05 - 0.2$ \msun), shifts the
effective temperature and luminosity. However, once core helium is
depleted the two tracks converge on the AGB and there is little
difference in the CO core masses between rotating and non-rotating stars.  
Figure \ref{fig:HR}
shows a portion of the Hertzsprung-Russel diagram of two 8 \msun \ ZAMS
models from H depletion to He depletion: one with \rot=0.0 (blue
curve), and one with \rot=0.2 (red curve).

\subsection{Evolution of Core Rotation}\label{sec:rot_evol}

\begin{figure*}[!ht]
         \centering
         %ratio is 1200x944
        \begin{subfigure}{
                \includegraphics[width=3.4in,height=2.75in]{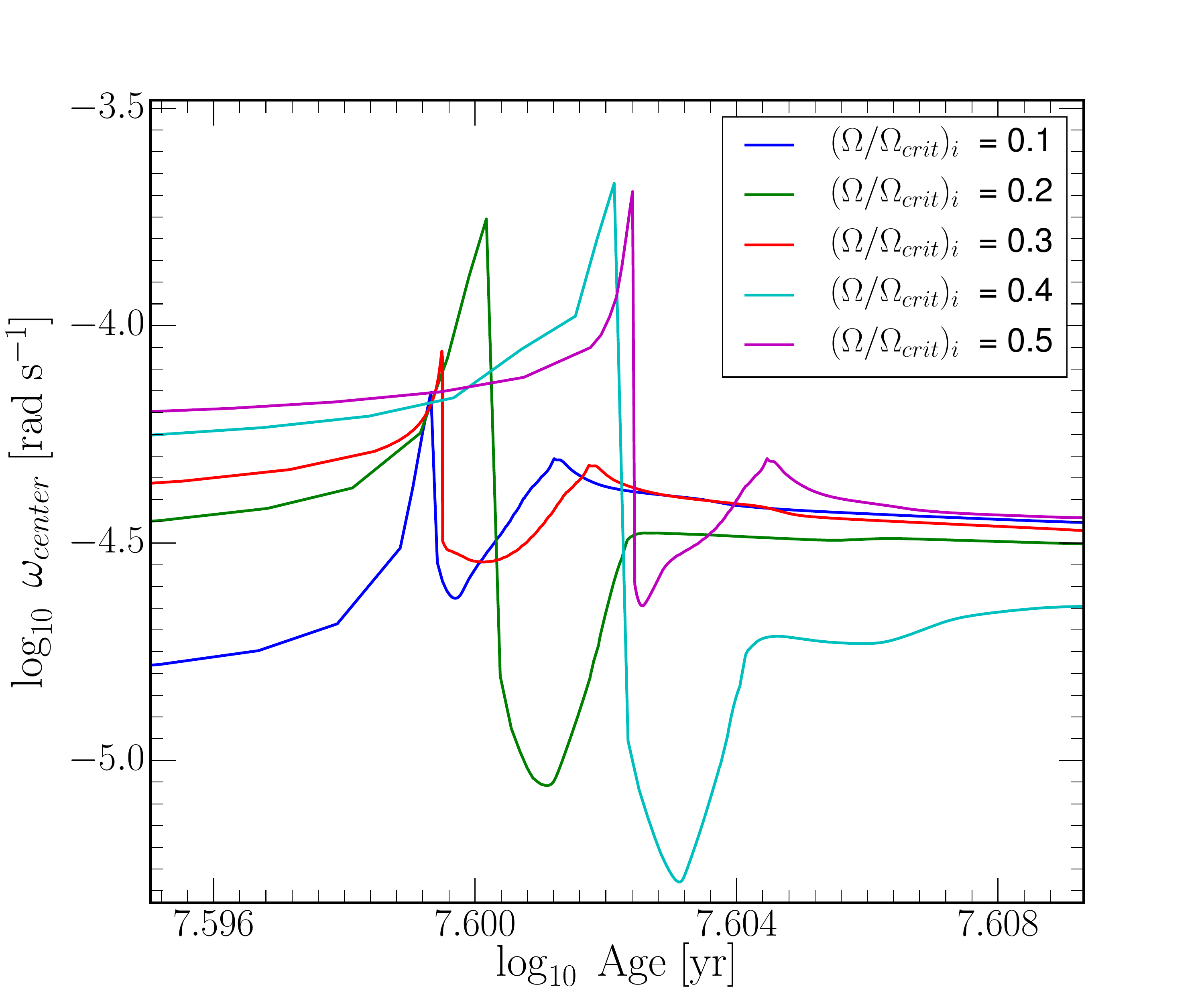}}
%                  \caption{7.0 \msun}
                \label{fig:rot_evol_ch}
        \end{subfigure}%
        ~
        \begin{subfigure}{
                \includegraphics[width=3.4in,height=2.75in]{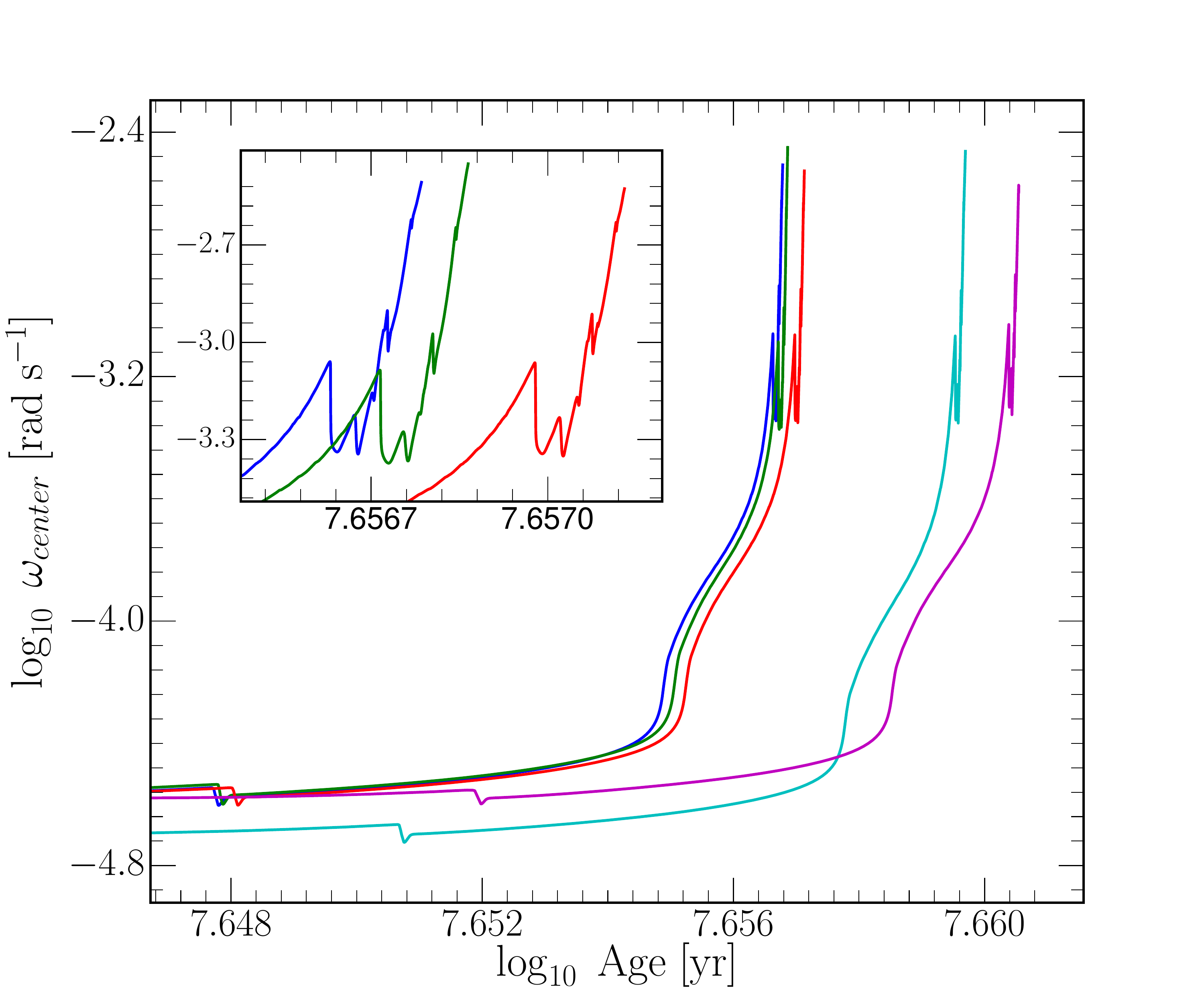}}
%                  \caption{7.5 \msun}
                \label{fig:rot_evol_cburn}
        \end{subfigure}
        
        \caption{The evolution of the central rotation rate $\omega_{{\rm center}}$, for a 7.5\msun\ at \rot=0.1,0.2,0.3,0.4,0.5.
        Left panel: Evolution from the TAMS to the ignition of core Helium burning. Right panel:
        Evolution during formation of the CO core and carbon ignition. Right panel insert: Zoom in on 
        \rot=0.1,0.2,0.3 during the carbon ignition}\label{fig:rot_evol}
\end{figure*}

Figure \ref{fig:rot_evol} (left panel)  shows the evolution of the central rotation rate, $\omega_{{\rm center}}$, for
different initial rotation rates for the 7.5 \msun \ ZAMS mass model.  During the main-sequence phase
there is little evolution in the central rotation rate, spinning down by $\approx 20\%$. Figure \ref{fig:rot_evol} (left panel) shows the transition 
region between core hydrogen burning to shell hydrogen burning and then to core helium burning.
As a star leaves the core hydrogen burning phase ($\log_{10} \rm{Age}\sim7.596$ for the \rot=0.1 model), the nuclear energy in the core decreases.
This causes a contraction of the core and subsequently a spin up of the core \citep{palacios_2006_aa}. The convective core
spins at a constant rate throughout the convective region, and spins faster than the outer-non
convective layers. As the nuclear energy 
generated decreases, the convective region inside the core shrinks towards the center \citep{sills_2000_aa}. As it 
shrinks, magnetic fields in the outer radiative layers propagate inwards into regions
which where previously convective. These fields act to slow down the core, removing the rotation
differential between the core and the envelope.

Eventually the convective region 
recedes entirely allowing the magnetic fields to propagate through to the center, which causes
the rapid spin down of $\omega_{center}$. At this point the core is still contracting, thus the 
core begins to spin up again.

There are now two possible outcomes, seen in Figure \ref{fig:rot_evol} (left panel), either the 
core has a second peak in core rotation (\rot=0.1,0.3,0.5) or the core rotation plateaus (\rot=0.2,0.4), the
outcome of which depends on the sign of $\rm{d}\omega/\rm{dM}$. Stars with $\rm{d}\omega/\rm{dM} < 0$ form
a second peak in the core rotation profile. When convection restarts in the core, due to helium burning, the convective region expands outwards.
As it does so it engulfs slower rotating material, which slows down the convective center of the star. Hence
the second peak in core rotation occurs when convection restarts. For stars with $\rm{d}\omega/\rm{dM} >0$, a plateau is reached
in the core rotation profile. Here, when the convection restarts and expands outwards it engulfs faster rotating material and thus spins up.
However this forces the convective core to expand, thus as more material is engulfed in to the convective 
region, it will begin to slow the core down. Hence the maximum $\omega_{center}$ occurs after convection has started. Typically, we find
\review{that the convective core will grow to $\approx 0.05 \msun$ before the spin up ceases}. While we believe these qualitative aspects hold, we caution against inferring 
quantitative predictions from this, as we have \review{found that the sign of $\rm{d}\omega/\rm{dM}$} to be model resolution dependent.

Figure \ref{fig:rot_evol} (right panel) show the evolution of $\omega_{\rm{center}}$ at the end of the core helium burning 
phase up to the start of carbon burning. Note the change in the scale of $\omega_{\rm{center}}$; the centers have spun down by $\approx 25\%$, 
during the core helium burning phase. First, we see a glitch in the rotation rate at the end 
of the core helium phase. This is due to the same process that occurs at the end
of core hydrogen burning. The convective core shrinks, allowing the magnetic fields to propagate
inwards slowing the core down. This is countered by the core contracting and spinning up. We can also see that the initially faster rotating 
stars are evolving slower due to their ability to mix fresh fuel into the core \citep{maeder_2000_aa,heger_2000_aa}.

We see that as the star forms its CO core the core spins up, due to the core contraction. Carbon ignition occurs
at the glitches seen at $\log_{10}\; \omega_{\rm{center}}\approx -3.12$ rad s$^{-1}$, when all stars have the same core rotation rate.
These glitches occur due to the core expansion due to the carbon burning events. Subsequent 
episodes of carbon burning can be seen as smaller rotation glitches in figure \ref{fig:rot_evol} (right panel insert).
We can also see that the faster the star initially rotates, the later the ignition occurs, due to their slower evolution.

\subsection{Mass-Rotation Plane}

\begin{figure*}[!ht]
         \centering
         %ratio is 1200x944
        \begin{subfigure}{
                \includegraphics[width=3.4in,height=2.75in]{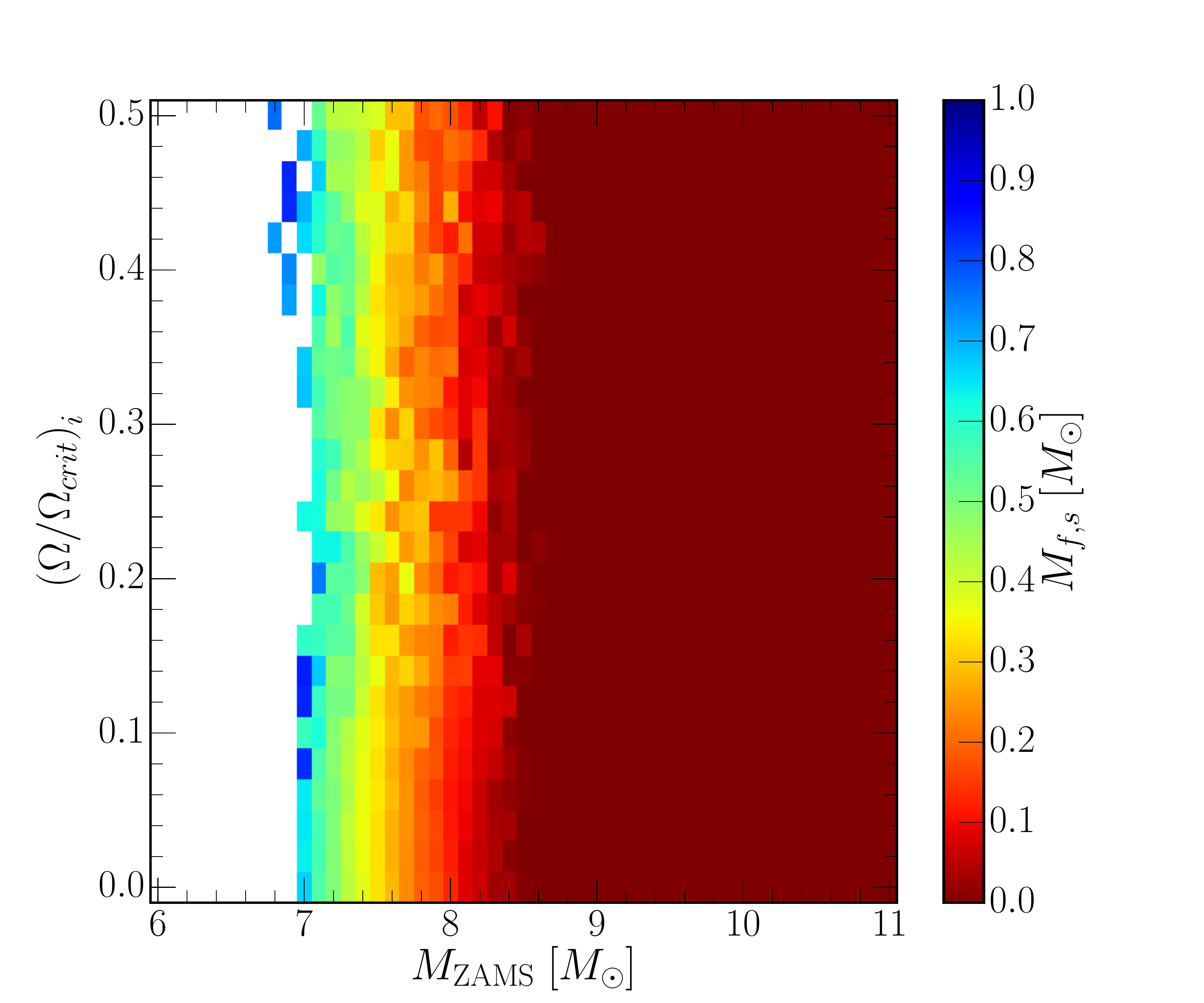}}
%                  \caption{7.0 \msun}
                \label{fig:mass_rot_ign}
        \end{subfigure}%
        ~
        \begin{subfigure}{
                \includegraphics[width=3.4in,height=2.75in]{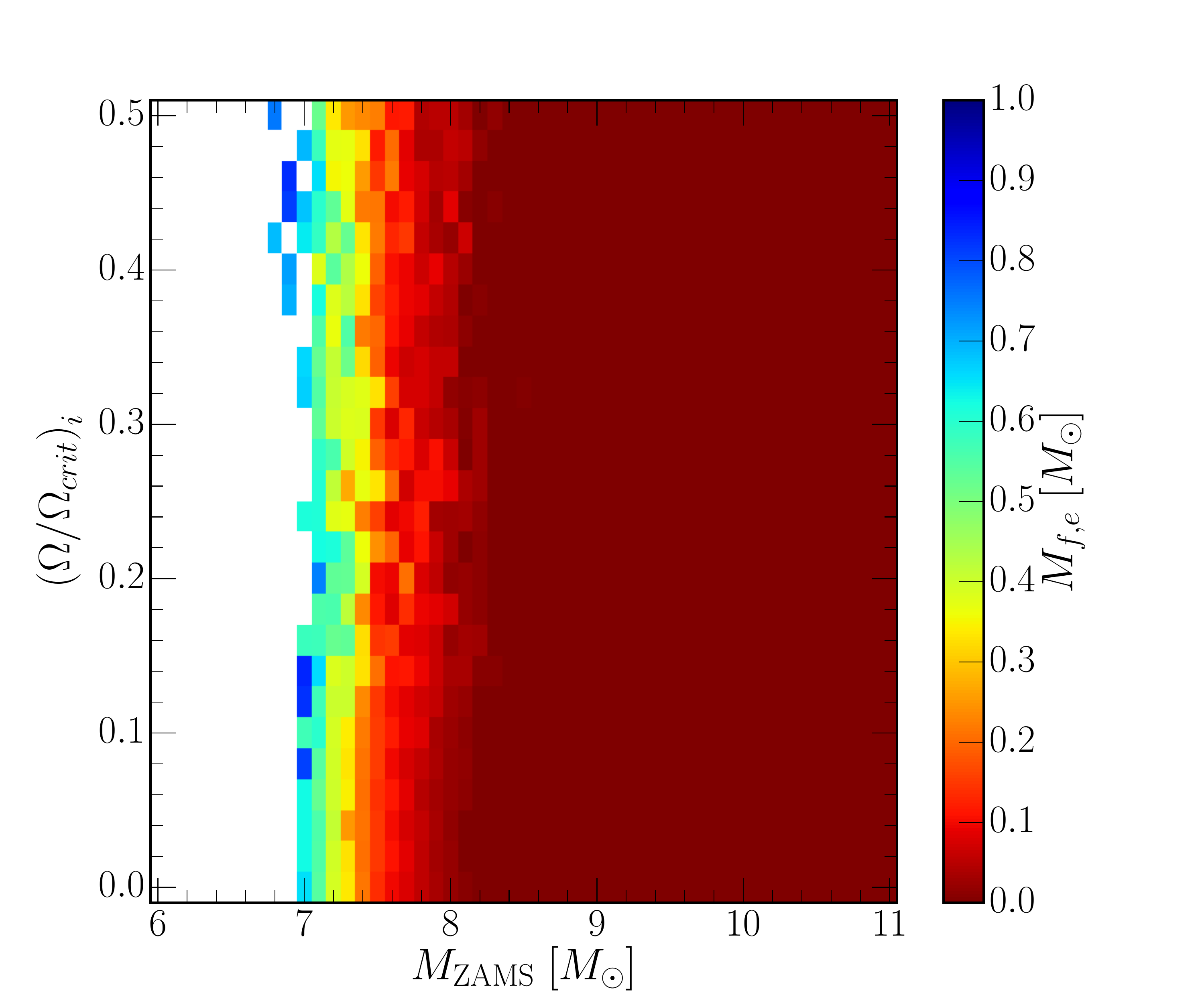}}
%                  \caption{7.5 \msun}
                \label{fig:mass_rot_fin}
        \end{subfigure}
        
        \begin{subfigure}{
                \includegraphics[width=3.4in,height=2.75in]{{{figs_rot_grid_centerOmega}}}}
%                  \caption{8.0 \msun}
                \label{fig:mass_rot_rot}
        \end{subfigure}%
        ~
        \begin{subfigure}{
                \includegraphics[width=3.4in,height=2.75in]{{{figs_rot_grid_cCoreMass.ign}}}}
%                  \caption{9.0 \msun}
                \label{fig:mass_rot_ccore}
        \end{subfigure}
        \caption{The ignition mass location (top left), minimum distance the carbon burning reaches to the core (top right),  
                 rotation of the center at ignition (bottom left) and CO core mass at ignition (bottom right)
                 as a function of the initial ZAMS mass and initial rotation \rot \ at a fixed \overshoot=0.016.
                 White regions are models that do not ignite carbon. }\label{fig:mass_rot}
\end{figure*}

\begin{figure*}[!ht]
         \centering
         %ratio is 1200x944
        \begin{subfigure}{
                \includegraphics[width=3.4in,height=2.75in]{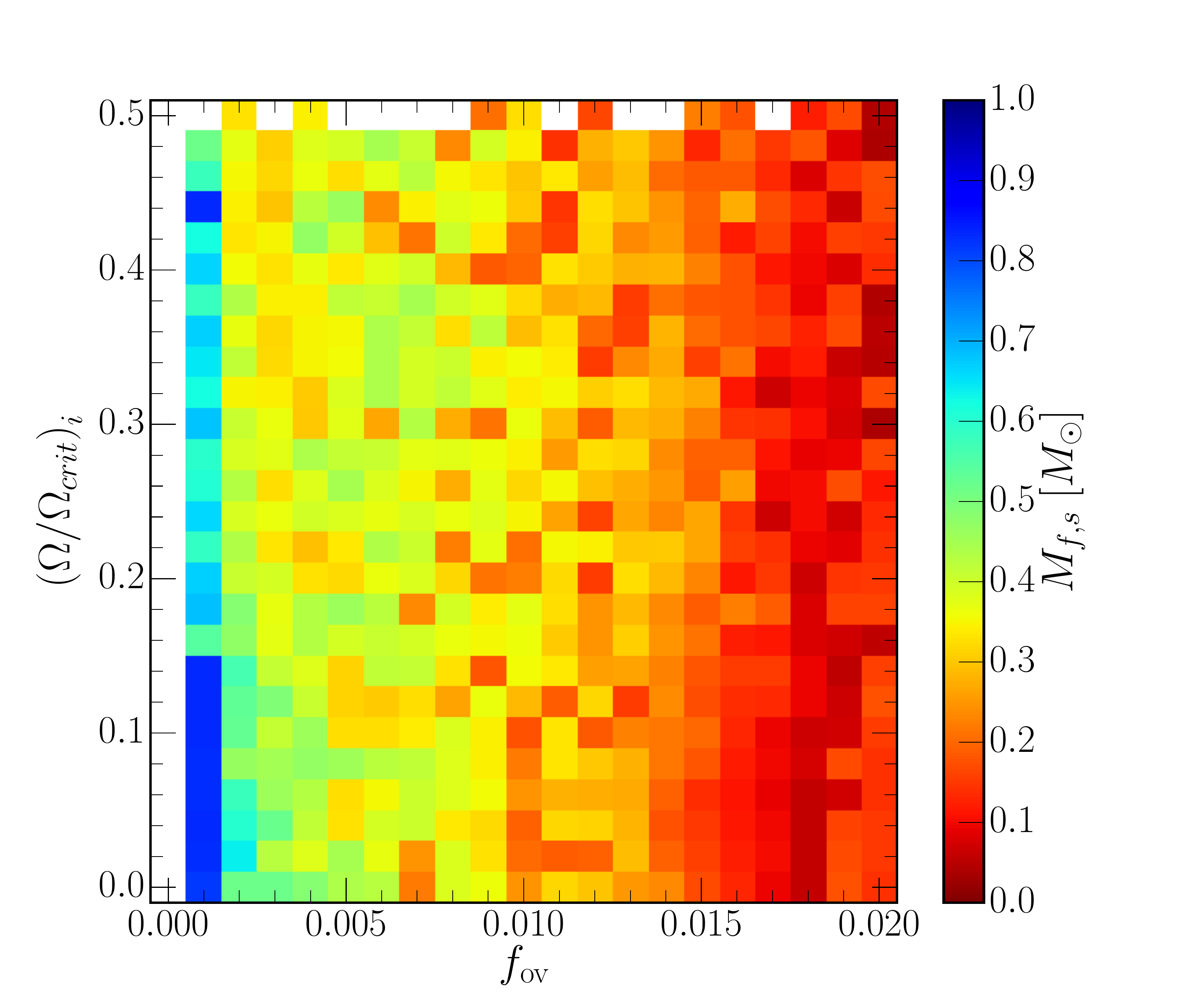}}
%                  \caption{7.0 \msun}
                \label{fig:over_rot_ign}
        \end{subfigure}%
        ~
        \begin{subfigure}{
                \includegraphics[width=3.4in,height=2.75in]{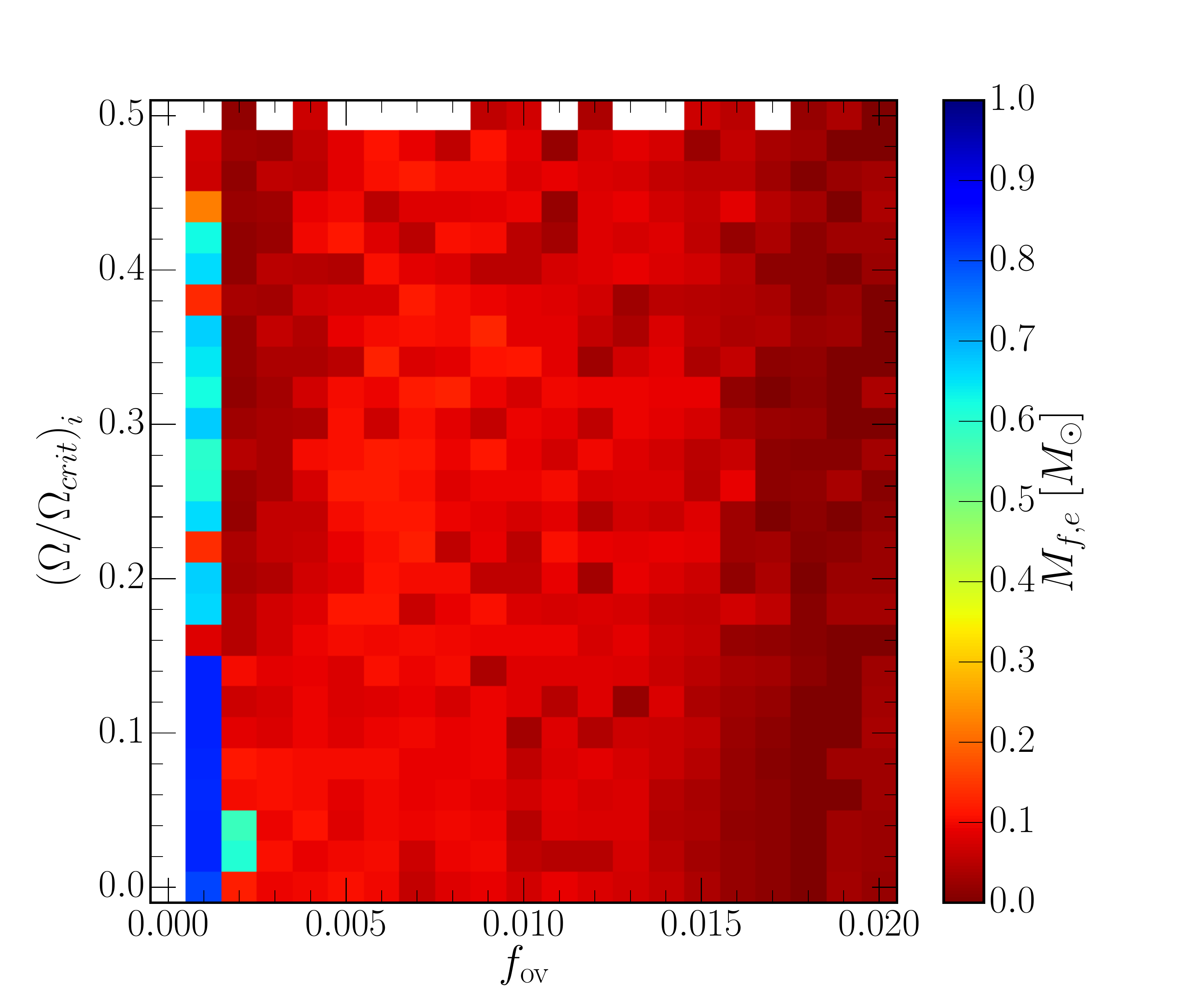}}
%                  \caption{7.5 \msun}
                \label{fig:over_rot_fin}
        \end{subfigure}
        
        \begin{subfigure}{
                \includegraphics[width=3.4in,height=2.75in]{{{figs_over_grid_centerOmega}}}}
%                  \caption{8.0 \msun}
                \label{fig:over_rot_rot}
        \end{subfigure}%
        ~
        \begin{subfigure}{
                \includegraphics[width=3.4in,height=2.75in]{{{figs_over_grid_cCoreMass}}}}
%                  \caption{9.0 \msun}
                \label{fig:over_rot_ccore}
        \end{subfigure}
        \caption{The ignition mass (top left) as a function of the overshoot (\overshoot) and
        initial rotation at a fixed mass of 8 \msun, minimum distance the flame reaches to the core (top right),  
        rotation of of the center at ignition (bottom left) and finally the CO core mass
        at ignition (bottom right). White regions are models that do not ignite. Note the scale on 
        the CO core mass is different to that of figures \ref{fig:mass_rot} \& \ref{fig:mass_over}}\label{fig:over_rot}
\end{figure*}

\begin{figure*}[!ht]
         \centering
         %ratio is 1200x944
        \begin{subfigure}{
                \includegraphics[width=3.4in,height=2.75in]{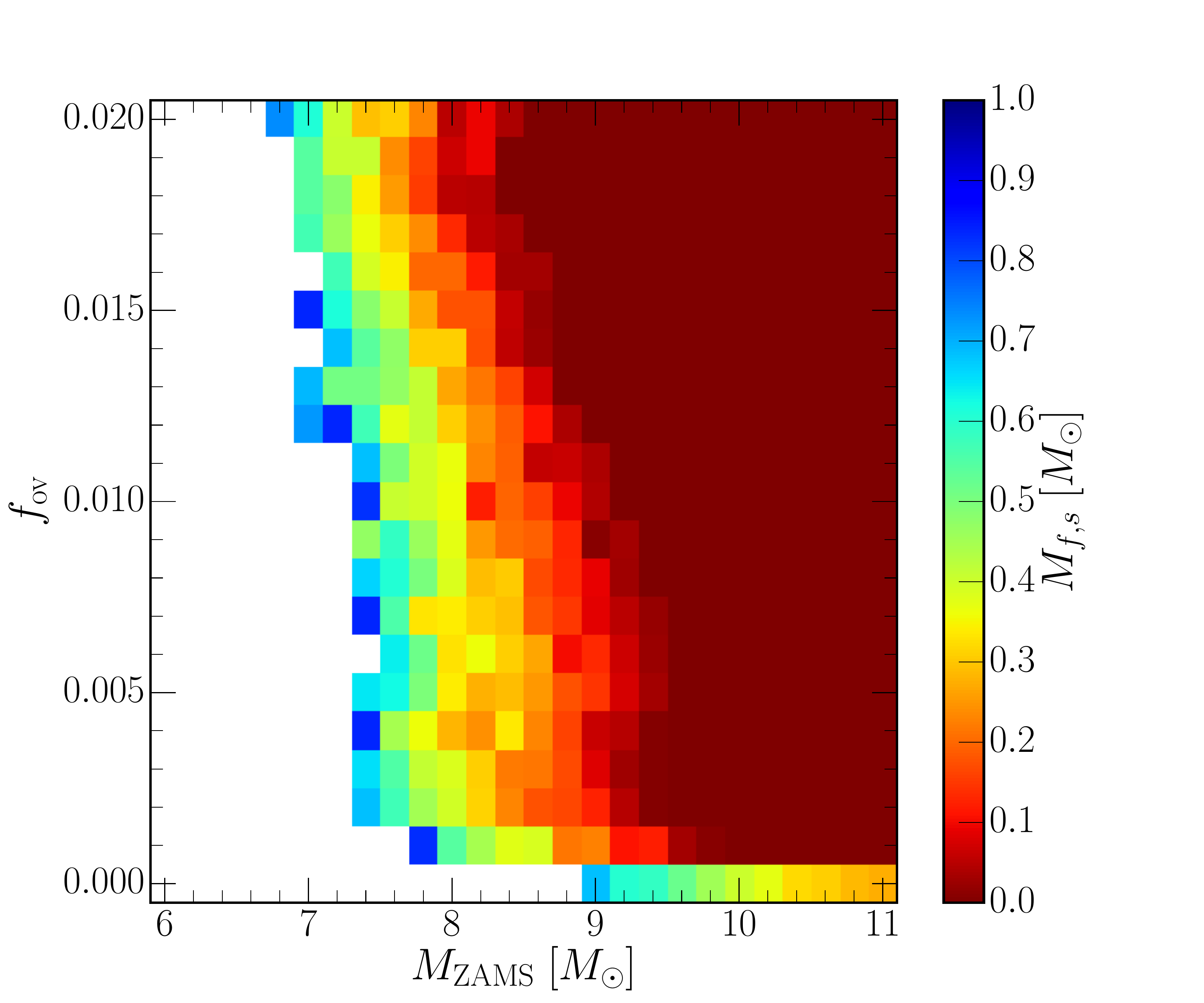}}
%                  \caption{7.0 \msun}
                \label{fig:mass_over_ign}
        \end{subfigure}%
        ~
        \begin{subfigure}{
                \includegraphics[width=3.4in,height=2.75in]{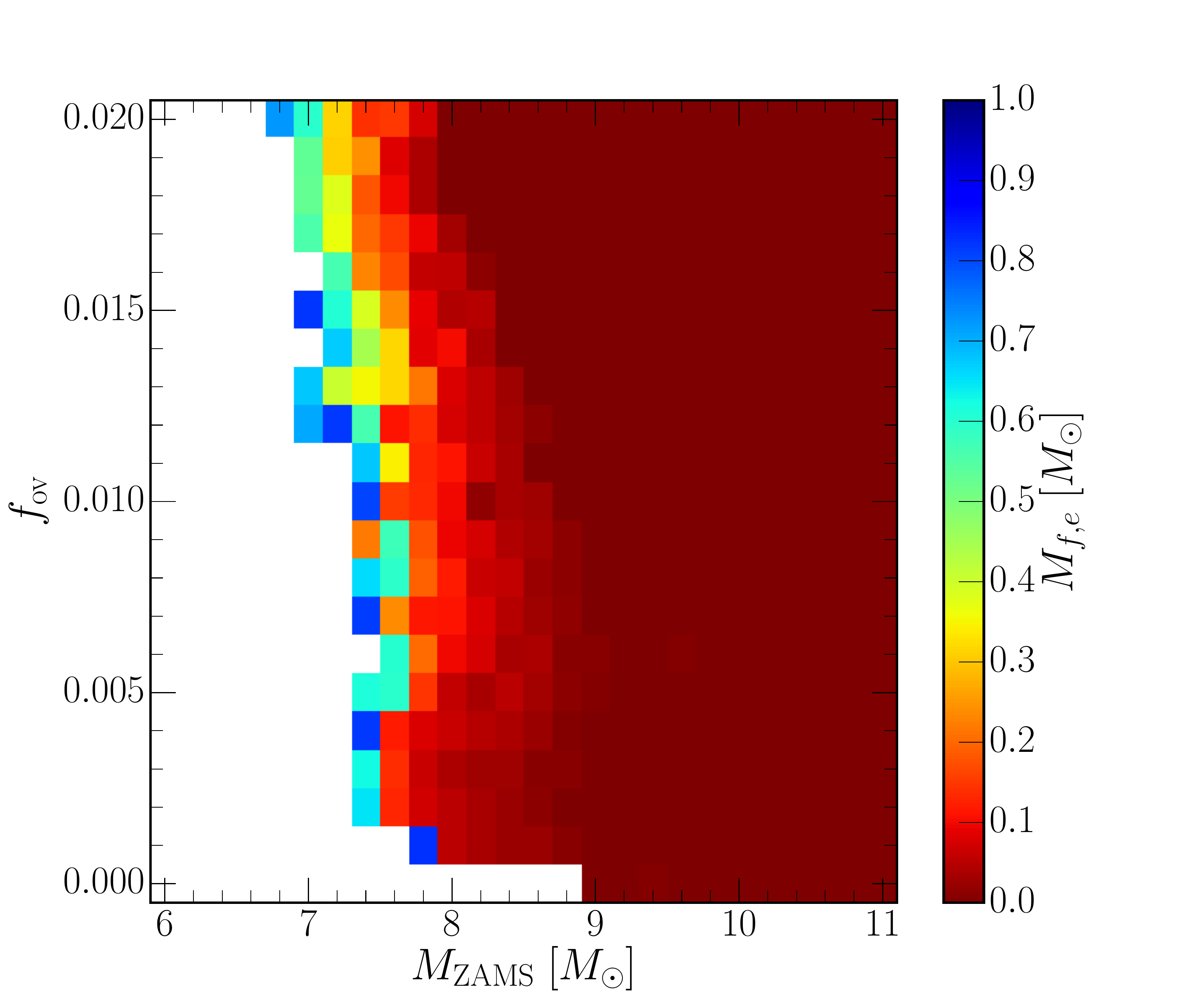}}
%                  \caption{7.5 \msun}
                \label{fig:mass_over_fin}
        \end{subfigure}
        
        \begin{subfigure}{
                \includegraphics[width=3.4in,height=2.75in]{{{figs_mass_over_grid_centerOmega}}}}
%                  \caption{8.0 \msun}
                \label{fig:mass_over_rot}
        \end{subfigure}%
        ~
        \begin{subfigure}{
                \includegraphics[width=3.4in,height=2.75in]{{{figs_mass_over_grid_cCoreMass}}}}
%                  \caption{9.0 \msun}
                \label{fig:mass_over_ccore}
        \end{subfigure}
        \caption{The ignition mass (top left) as a function of the initial ZAMS mass and
        overshoot (\overshoot) at a fixed \rot = 0.25, minimum distance the flame reaches to the core (top right),  
        rotation of the center at ignition (bottom left) and finally the CO core mass
        at ignition (bottom right). White regions are models that do not ignite.}\label{fig:mass_over}
\end{figure*}

Figure \ref{fig:mass_rot} (top left) shows the location of the first
carbon ignition in the mass-rotation plane for a fixed
\overshoot=0.016. Only models with 7 \msun $\lesssim M_{\rm ZAMS}
\lesssim$ 8 \msun \ feature off-center ignition, as models with
$M_{ZAMS} \lesssim 7 \ \msun$ do not ignite carbon, while models with
$M_{ZAMS} \gtrsim 8 \ \msun$ feature central carbon ignition. Figure
\ref{fig:mass_rot} (top right) shows the location where carbon flames
and flashes are quenched. Only models having off-center ignition within
the relatively narrow range 8.0 \msun $\lesssim M_{\rm ZAMS} \lesssim$
8.2 \msun \ does the flames or flashes reach the center.

For a fixed ZAMS mass the ignition and quenching locations are mostly
independent of \rot \ values between 0.0 and 0.5.  This occurs because
the transport of angular momentum transport from the core to the
overlying layers is efficient during the giant branch phases of
evolution.  Thus, regardless of the initial rotation rate, by the time
the carbon core forms the central regions are rotating as a solid body
with similar angular frequencies for a fixed ZAMS mass.  Figure
\ref{fig:mass_rot} (bottom left) shows the center angular frequency,
$\omega_{\rm center}$, in the mass-rotation plane at the fixed
baseline overshoot value. Note $\omega_{\rm center}$ only spans a
factor of $\approx$ 2 over the entire plane; all models rotate with
similar angular frequencies at a fixed ZAMS mass.  We can
quantitatively explain, to first order, the rate at which the
carbon core spins up between formation of the carbon core and first
ignition of carbon from angular momentum conservation and the
mass-radius relationship of polytropes.  When the rotating carbon core
forms, its total angular momentum is
\begin{equation}
L_i = I_i \omega_i \sim c_i M_i R_i^2 \omega_i
\enskip ,
\end{equation}
where $I$ is the moment of inertia, $M_i$ is the mass of the carbon core,
$R_i$ is the radius of the carbon core, and $c_i$ is a constant that depends 
on the density structure. At first ignition of carbon, the angular momentum
of the more massive contracting \review{CO} core is 
\begin{equation}
L_f = I_f \omega_f \sim c_f M_f R_f^2 \omega_f
\enskip .
\end{equation}
Conserving angular momentum over this phase of evolution gives
\begin{equation}
\frac{\omega_f}{\omega_i}  \sim \frac{M_i R_i^2}{M_f R_f^2}
\label{eq:omega_conserv}
\enskip .
\end{equation}
Assuming the non-rotating polytropic mass-radius relation 
\begin{equation}
R \sim M^{(1-n)/(3-n)}
\end{equation}
applies at the first order, substitution into eq. \ref{eq:omega_conserv} gives
\begin{equation}\label{eq:spinup}
\frac{\omega_f}{\omega_i}  \sim \frac{M_i^{(5-3n_i)/(3-n_i)}}{M_f^{(5-3n_f)/(3-n_f)}}
\label{eq:omega_ratio}
\enskip ,
\end{equation}
where $n_i$ is the polytropic index at the formation of the carbon
core and $n_f$ is the polytropic index at first carbon ignition. For 
the angular frequency of the core to increase, $\omega_f > \omega_i$, 
the polytropic index is restricted to be in the range 1 $< n <$ 3. For
example, for an 8 \msun \ ZAMS model with \rot=0.2 at formation of the
carbon core, we find $M_i \approx$ 0.92 \msun \ and $n_i \approx$ 1.5.  Using a
least squares fitting program to generate the polytropic index $n$ for
a sequence of MESA profiles between formation of the carbon core and
first carbon ignition (see Figure \ref{fig:poly} for an example), we
find the left-hand side and the right-hand side of
eq. \ref{eq:omega_ratio} agree to within a factor of $\approx$ 2 for the
7 \msun, 8 \msun, and 9 \msun \ models shown in Figure
\ref{fig:mass_rot}.  The center spins up,
on average, by a factor of $\approx$ 40 between formation of the carbon
core and first carbon ignition.  Since the carbon core at first
ignition rotates as a solid body with similar angular frequencies for
a fixed ZAMS mass, the carbon core masses are nearly in the same state
independent of the initial rotation rate, as shown in Figure
\ref{fig:mass_rot} (bottom right).

We find the carbon cores are rotating with periods between 0.1$-$1.0
days at first carbon ignition on the AGB.  During the RGB phase of
evolution we find the helium cores have periods of $\approx$ 2.5 days,
again independent of the initial rotation rate. \citet{mosser_2012_aa} 
measured rotational splittings in a sample of about 300 red giants
observed during more than two years with {\it Kepler}.  They found
these splittings are dominated by core rotation.  Periods range
between 10$-$100 days with larger periods for red clump stars compared
to RGB stars. They inferred a ZAMS mass range of 1.2$-$1.5 \msun,
less massive than our rotating SAGB models.

Stars with masses $<7 \msun$ will eventually form CO WDs after removing their
outer envelopes. Between \hbox{7 \msun} and $\lesssim$ 8 \msun\, where the carbon burning
does not reach the core, our models suggest these stars will form CO+ONe hybrid WDs. 
Stars with masses $\gtrsim8$\msun\
form ONe WDs as the carbon flames will burn away the $^{12}$C. Electron capture supernovae
are expected for stars with masses $>$9 \msun, due to the CO core mass being greater than the Chandrasekhar
mass \review{\citep{eldridge_2004_aa}.}

\subsection{Overshoot-Rotation Plane}

Figure \ref{fig:over_rot} (top left) shows the location of the first
carbon ignition in the overshoot-rotation plane for a fixed ZAMS mass
of 8 \msun.  For this case, overshoot is a dominant factor in setting
the location of the first, off-center, carbon ignition.  This first
ignition of carbon can be made to occur at almost any mass coordinate
within the carbon core of the 8 \msun \ model by varying the overshoot
parameter. The \overshoot=0.0 case, where convective mixing operates
only within the Schwarzschild boundaries and does not extend beyond
the MLT convective boundary, does not ignite carbon for any of the
initial rotations rates.  The smallest non-zero overshoot parameter in
our grid, \overshoot=0.001, gives mass locations for the first
ignition of carbon that are furthest from the center, closest to the
outer boundary of the carbon core.  Progressively larger values of the
overshoot parameter generally move the location of the off-center
ignition location closer to the center.

Figure \ref{fig:over_rot} (top right) shows the quenching location,
where the flame and flashes die, in the overshoot-rotation plane for a
fixed ZAMS mass of 8 \msun.  The flame and flashes approaches the
center for nearly all the models; only models with $\overshoot\le
0.02$ does the burning become quenched relatively far from the center
\citep{denissenkov_2013_ab}.
Similar to our analysis the mass-rotation plane, Figure
\ref{fig:over_rot} (bottom left) shows carbon core rotation rate is
approximately constant, to within a factor of $\approx$ 2, regardless of
the initial rotation rate. There is evidence for a weak dependence on
the rotation rate to the overshoot parameter.  Figure
\ref{fig:over_rot} (bottom right) shows the carbon core mass increases
with increasing values of \overshoot, again nearly independent of the
initial \rot.  A larger core mass, in turn, leads to the first carbon
ignition occurring deeper in the star.

Comparing the results of the overshoot-rotation grid with the
mass-rotation grid, we find the carbon core mass is the quantity that
most strongly determines the structure of the flame.  For example, the
boundary between cases that ignite off center and those that do not
ignite carbon (ZAMS masses $\approx$ 7 \msun) depends on whether the star
can form a carbon core of $\approx$ 1.05 \msun, which is necessary to
reach the critical density in equation \ref{eq:rho_ign}.

\subsection{Mass-Overshoot Plane}\label{sec:mass_over}

Figure \ref{fig:mass_over} (top left) shows the mass location of
carbon ignition in the mass-overshoot plane at a fixed ZAMS rotation
of \rot=0.25.  For \overshoot=0.0 the minimum mass needed to ignite
carbon is 9 \msun, $\approx$ 2 \msun\ greater than the baseline case, and
models up to 11 \msun\ ignite carbon off-center.  While no overshoot
may be unphysical, even a small amount of overshoot moves the minimum
ZAMS mass for ignition considerably, down to \hbox{7.8 \msun}. Increasing
\overshoot \ decreases the required ZAMS mass to off-center carbon
ignition and decreases the minimum ZAMS mass needed for central
ignition of carbon.

The width of the ZAMS mass range where the model stars ignite carbon
off-center is approximately constant with respect to overshoot,
$\Delta M_{\rm ZAMS}$/$\Delta$\overshoot$\approx$ 1.6 \msun.
That is, overshoot uniformly moves the ZAMS mass boundaries where 
off-center carbon ignition occurs.  For example, the sloped contours in Figure
\ref{fig:mass_over} (top left) show that when \overshoot=0.0, models
in the mass range $\approx$ 8.9 to 11 \msun \ have off-center ignition
\citep[as found by e.g.][]{siess_2006_aa,siess_2007_aa} When
\overshoot=0.008 this mass range shifts to $\approx$ 7.4 to 9.4 \msun, and
when \overshoot=0.016 the off-center carbon ignition range shifts to
$\approx$ 7.2 to 8.8 \msun.

Figure \ref{fig:mass_over} (top right) shows the final fate of the
carbon burning flames and flashes, the quenching location in the
mass-overshoot plane.  With \overshoot=0.0, all flames and flashes
reach the core (e.g., bottom right plot of Figure \ref{fig:f5}).  As the overshoot parameter increases, the carbon
burning is less likely to reach the center.  For $\overshoot<$0.01,
the flame either has a single flash (similar to Figure \ref{fig:f5}
top left panel for the 7 \msun \ model) or undergoes a single flash then
a steady state flame (similar to Figure \ref{fig:f5} middle right
panel for the 8.2 \msun \ model).  Only when the overshoot parameter is
large, $\overshoot>0.01$, does an intermediate evolution of a flash
and then a steady state flame that does not reach the core (similar to
Figure \ref{fig:f5} middle left panel for the \hbox{8.0 \msun} \ model).

Figure \ref{fig:mass_over} (bottom left) shows the central angular
frequency at the point of first ignition in the mass-overshoot plane
with a fixed ZAMS rotation of \rot=0.25. We find
that models which undergo central carbon ignition have lower
angular frequencies than the off-center igniting stars. This is due 
to the carbon cores being spun up during the cooling phase, due to the
contraction of the carbon core (see equation \ref{eq:spinup}). As the heavier stars ignite carbon earlier
they have less time in which to be spun up compared to the lower mass stars have at first ignition. As before,
only models with carbon core masses greater than $\approx$ 1.05 \msun
\ ignite carbon, as shown in in Figure \ref{fig:mass_over} (bottom
right). As before, the maximum carbon core mass that ignites
off-center carbon burning is $\approx$ 1.2 \msun, similar to that shown in
Figures \ref{fig:mass_rot} and \ref{fig:over_rot}.

\begin{figure}[htb]
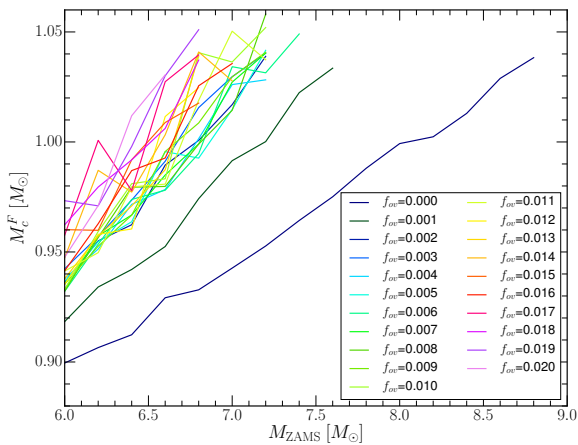

\centering{\includegraphics[width=1\linewidth]{{{figs_mass_over_grid_co_int_fin_mass}}}}
\caption{ZAMS mass and final C/O core masses for non-igniting models in the 
mass-overshoot plane, measured once the $^4$He shell has been depleted.
}
\label{fig:co_fin_mass}
\end{figure}

Figure \ref{fig:co_fin_mass} shows the final CO core mass for the
non-igniting models (white region in Figure \ref{fig:mass_over}). The
maximum mass for a CO core is 1.05 \msun; stars with
heavier CO cores ignite carbon burning. We also find a trend for
increasing overshoot to increase the final CO core mass, as noted
previously.  \citet{doherty_2015_aa} reported
a grid of models with the Monash stellar evolution code, over a range of metallicities, to
investigate the fate of AGB and SAGB stars.  Comparing our results
with their solar metalicity results (their Fig. 6), we find that for
a given ZAMS mass our rotating CO core masses are 0.05$-$0.1\msun \ 
larger. Increasing the overshoot can mimic decreasing the metalicity
in terms of the final CO WD mass.

\begin{deluxetable*}{llllllllllllllllll}{h}
\tablecolumns{18}
\tablewidth{0pc}
\tablecaption{Ignition locations in solar masses for the mixing grid.}
\tablehead{ \colhead{} & \colhead{} & \multicolumn{16}{c}{\amnu}}
% \tablehead{\colhead{\thermo} & \colhead{\thermo} 
% & \multicolumn{4}{c}{\overshoot}& \multicolumn{4}{c}{\overshoot}
% & \multicolumn{4}{c}{\overshoot} & \multicolumn{4}{c}{\shoot}}

 \startdata
 & & \multicolumn{4}{c}{0.0} & \multicolumn{4}{c}{0.5} & \multicolumn{4}{c}{1.0} & \multicolumn{4}{c}{1.5} \\
\colhead{\semi} & \colhead{\thermo} & \multicolumn{4}{c}{\overshoot}& \multicolumn{4}{c}{\overshoot} & \multicolumn{4}{c}{\overshoot} & \multicolumn{4}{c}{\overshoot} \\
& & 0.000 & 0.001 & 0.016 & 0.020 & 0.000 & 0.001 & 0.016 & 0.020 & 0.000 & 0.001 & 0.016 & 0.020 & 0.000 & 0.001 & 0.016 & 0.020 \\
%   & & \multicolumn{16}{c}{} \\
  \hline \\

0.000 &  0.00 & $\dots$\footnote{Ellipses represents models with no ignition} & $\dots$ & 0.65 & 0.28 & $\dots$ & 0.49 & 0.15 & 0.18 & $\dots$ & 0.43 & 0.17 & 0.16 & $\dots$ & 0.49 & 0.17 & 0.13  \\
0.000 &  0.10 & 0.50 & 0.71 & $\dots$ & $\dots$ & $\dots$ & 0.46 & 0.17 & 0.05 & 0.43 & 0.42 & 0.14 & 0.05 & $\dots$ & 0.47 & 0.18 & 0.05  \\
0.000 &  1.00 & $\dots$ & 0.63 & 0.35 & 0.38 & 0.39 & 0.46 & 0.11 & 0.18 & $\dots$ & 0.40 & 0.16 & 0.16 & $\dots$ & 0.48 & 0.15 & 0.16  \\
0.000 & 10.0 & 0.50 & 0.65 & $\dots$ & 0.54 & $\dots$ & 0.49 & 0.16 & 0.15 & $\dots$ & 0.43 & 0.18 & 0.05 & $\dots$ & 0.43 & 0.14 & 0.16  \\

0.001 &  0.00 & 0.81 & $\dots$ & 0.65 & $\dots$ & $\dots$ & 0.59 & 0.21 & 0.17 & $\dots$ & 0.83 & 0.16 & 0.14 & $\dots$ & 0.85 & 0.23 & 0.04  \\
0.001 &  0.10 & 0.79 & 0.48 & 0.28 & 0.37 & $\dots$ & 0.83 & 0.20 & 0.05 & $\dots$ & 0.85 & 0.14 & 0.18 & $\dots$ & 0.69 & 0.12 & 0.05  \\
0.001 &  1.00 & 0.83 & $\dots$ & 0.58 & 0.71 & $\dots$ & 0.84 & 0.15 & 0.17 & $\dots$ & 0.83 & 0.18 & 0.05 & $\dots$ & 0.71 & 0.17 & 0.15  \\
0.001 & 10.0 & 0.33 & 0.69 & 0.32 & 0.49 & $\dots$ & 0.57 & 0.18 & 0.16 & $\dots$ & 0.62 & 0.08 & 0.05 & $\dots$ & 0.65 & 0.11 & 0.16  \\

0.010 &  0.00 & 0.19 & 0.57 & 0.45 & 0.75 & $\dots$ & 0.63 & 0.21 & 0.16 & $\dots$ & 0.65 & 0.17 & 0.14 & $\dots$ & 0.63 & 0.16 & 0.05  \\
0.010 &  0.10 & $\dots$ & $\dots$ & 0.30 & 0.39 & $\dots$ & 0.69 & 0.20 & 0.05 & $\dots$ & 0.62 & 0.13 & 0.04 & $\dots$ & 0.68 & 0.12 & 0.05  \\
0.010 &  1.00 & 0.73 & $\dots$ & $\dots$ & 0.73 & $\dots$ & 0.66 & 0.17 & 0.18 & $\dots$ & 0.55 & \textbf{0.20}\footnote{Baseline model} & 0.05 & $\dots$ & 0.85 & 0.16 & 0.14  \\
0.010 & 10.0 & 0.49 & 0.57 & 0.66 & 0.36 & $\dots$ & 0.56 & 0.18 & 0.04 & $\dots$ & 0.63 & 0.08 & 0.17 & $\dots$ & 0.64 & 0.10 & 0.05  \\

0.100 &  0.00 & 0.28 & $\dots$ & 0.39 & 0.76 & $\dots$ & 0.66 & 0.12 & 0.17 & $\dots$ & 0.84 & 0.15 & 0.14 & $\dots$ & 0.83 & 0.16 & 0.06  \\
0.100 &  0.10 & 0.54 & $\dots$ & 0.28 & 0.37 & $\dots$ & 0.52 & 0.20 & 0.08 & $\dots$ & 0.58 & 0.14 & 0.17 & $\dots$ & 0.83 & 0.11 & 0.05  \\
0.100 &  1.00 & 0.51 & $\dots$ & 0.57 & 0.73 & $\dots$ & 0.68 & 0.15 & 0.18 & $\dots$ & 0.83 & 0.20 & 0.06 & $\dots$ & 0.55 & 0.17 & 0.13  \\
0.100 & 10.0 & 0.69 & $\dots$ & 0.52 & 0.55 & $\dots$ & 0.49 & 0.18 & 0.04 & $\dots$ & 0.58 & 0.08 & 0.15 & $\dots$ & 0.60 & 0.12 & 0.16  \\

 \enddata
\label{table:mixing_grid}
\end{deluxetable*}

%ZONING PLOT
\begin{figure}[!htb]
%\centering{\includegraphics[width=1\linewidth]{figs_flame_zoning}}
\centering{\includegraphics[trim = 1.3in .4in .05in .8in, clip,width=3.6in]{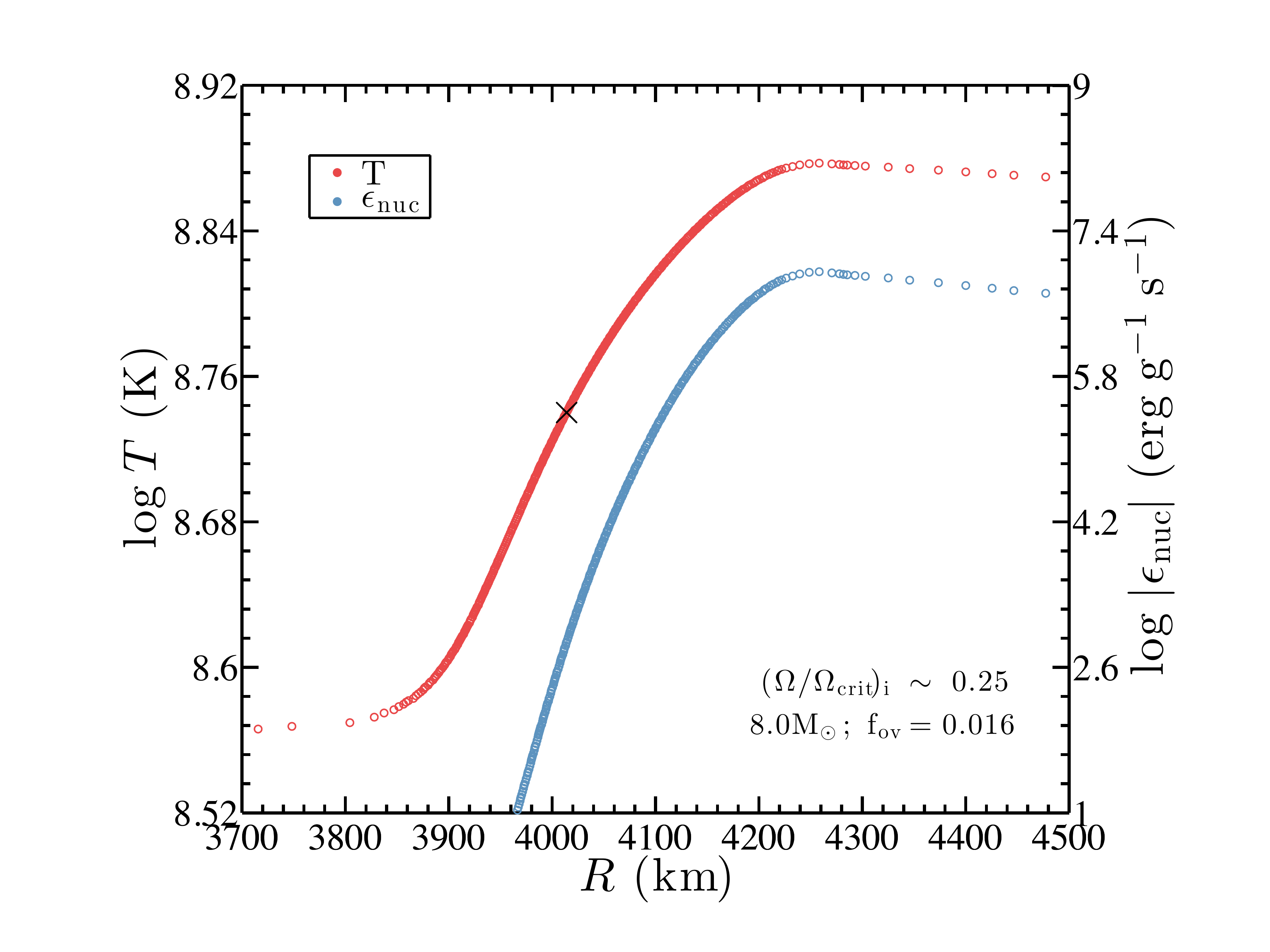}}
\caption{Temperature and nuclear energy generation rate profile of a carbon flame front within an 8 \msun \
ZAMS model with \rot=0.25 and \overshoot=0.016. Red open circles
show the mesh locations of the temperature while blue open circles 
show the absolute value of the nuclear energy generation rate. The distance between mesh locations within
the body of the flame is $\lesssim$1 km, which is sufficient to accurately
capture the nuclear burning and thermal transport processes.
}
\label{fig:f3}
\end{figure}

%TIMESTEP PLOT
\begin{figure}[!htb]
\centering{\includegraphics[width=1\linewidth]{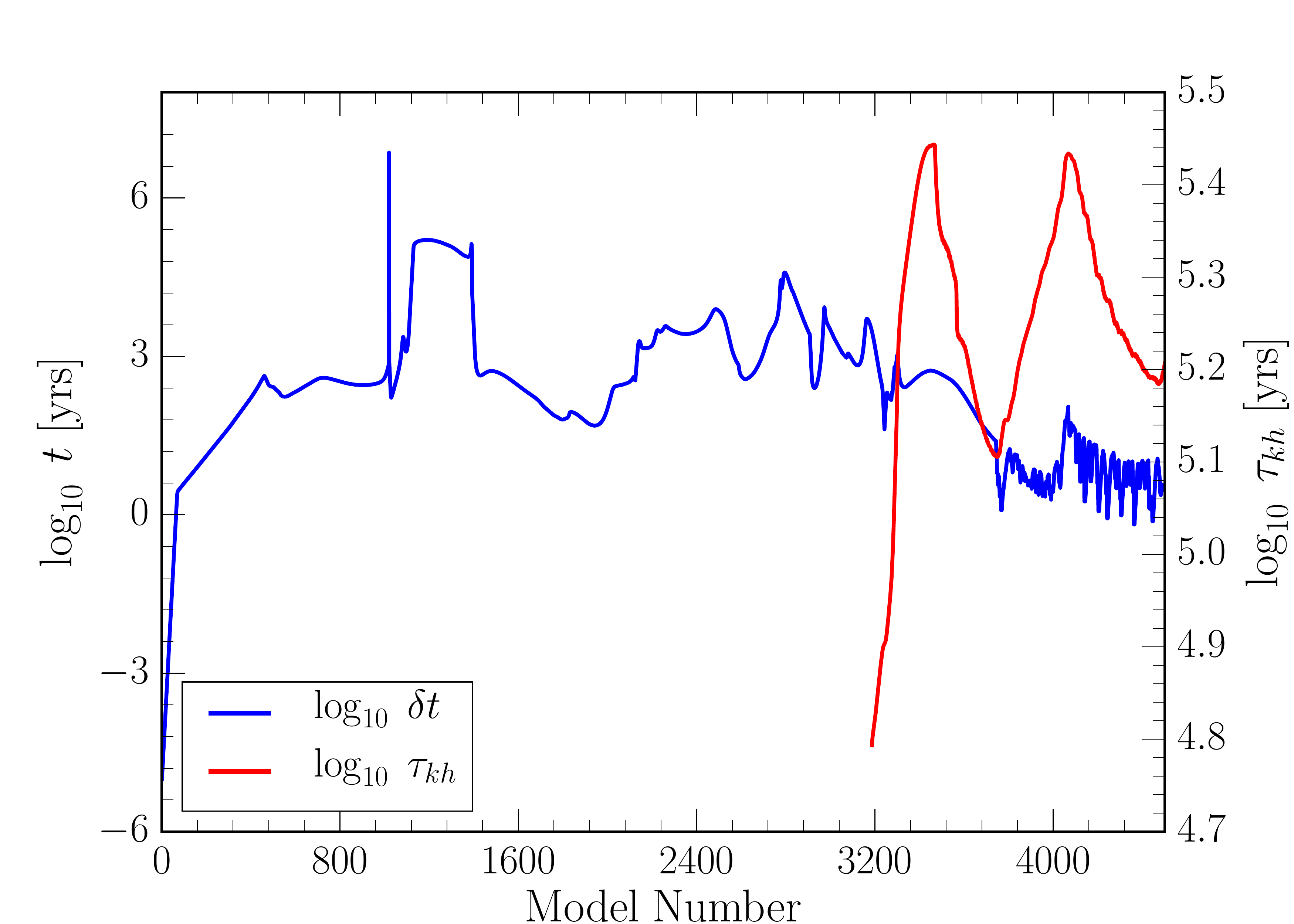}}
\caption{
Evolution of the timestep $\delta$t (left y-axis) and Kelvin-Helmholtz thermal timescale $\tau_{kh}$ of the carbon core
(right y-axis) of an 8 \msun \ ZAMS model with \rot=0.25
and \overshoot=0.016.  At model number $\approx$ 2800 the timestep begins
to decrease due to the increase in nuclear burning caused during core
helium depletion. At carbon ignition, which occurs when the thermal timescale is in a local minimum at model number $\approx$ 3700,
the timestep is $\approx$ 10 years and decreases to $\approx$ 1 year during the carbon flame and flashes.
}
\label{fig:fig4}
\end{figure}

%ROT CONVERGE - FIGURE 12
\begin{figure*}[ht]
         \centering
         %ratio is 1200x944
        \begin{subfigure}{
                \includegraphics[width=3.4in,height=2.75in]{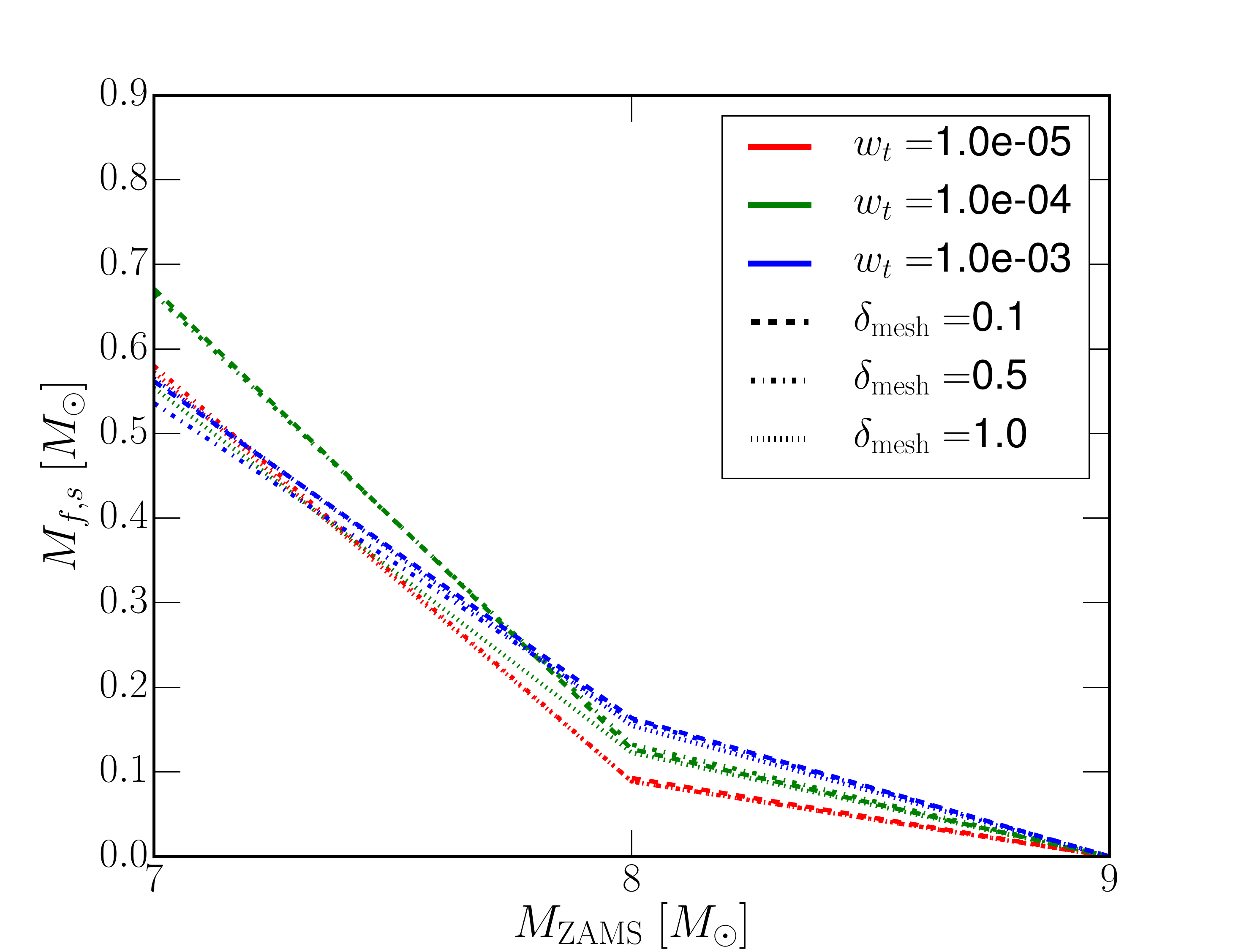}}
%                  \caption{7.0 \msun}
                \label{fig:converge_rot_00}
        \end{subfigure}%
        ~
        \begin{subfigure}{
                \includegraphics[width=3.4in,height=2.75in]{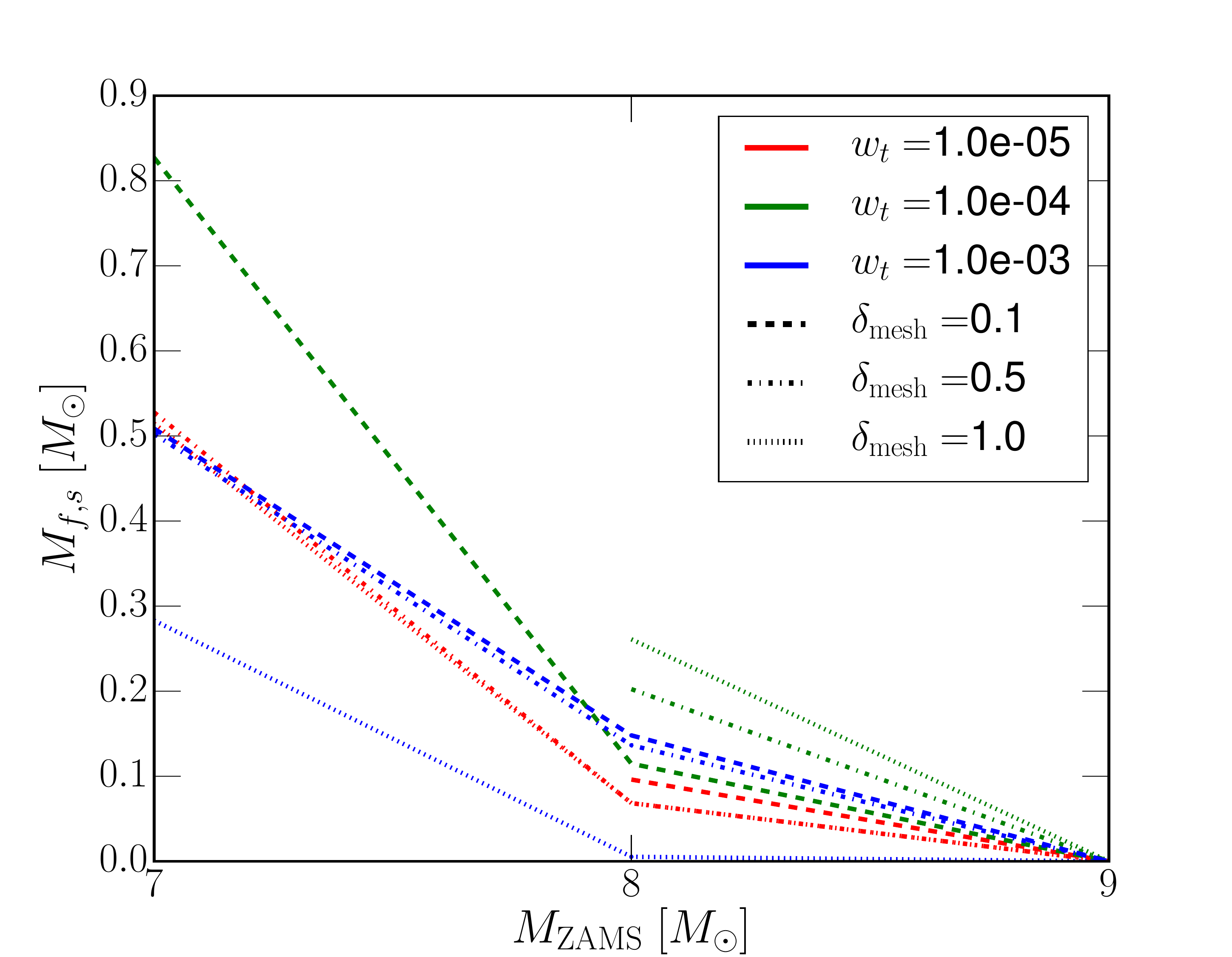}}
%                  \caption{7.5 \msun}
                \label{fig:converge_rot_025}
        \end{subfigure}
        
        \begin{subfigure}{
                \includegraphics[width=3.4in,height=2.75in]{{{figs_converge_grid_rot05}}}}
%                  \caption{8.0 \msun}
                \label{fig:converge_rot_05}
        \end{subfigure}%
        \caption{Location of the ignition mass in the 7 \msun, 8 \msun, and 9 \msun \ models with a fixed \overshoot=0.016. 
        Different values of spatial resolution \mesh (line style) and temporal resolution \var (line color) are show.
        Top left \rot=0.0, top right \rot=0.25 and bottom \rot=0.5. Color/style combinations not shown
        do not ignite.}\label{fig:converge_grid}
\end{figure*}

\section{Results From The Mixing Coefficient Grid Studies}
\label{sec:mixsens}

Table \ref{table:mixing_grid} shows overshoot has the most significant
effect, on the location of the initial flame, with no ignition for no
overshoot \review{for the 8\msun, \rot=0.25 model}, as long as the scale factor for the strength of angular momentum diffusion $\amnu>0$. 
As overshoot increases, the flame ignition occurs deeper in
the star. This is due to changes in the $^4$He and CO core masses during the stars evolution.  For instance, comparing
Figure \ref{fig:f5} bottom left and bottom right, we can see the effect of overshoot for the 9 \msun \ model. Primarily, the
model with overshoot ignites at the center, while the no overshoot model ignites off center.
Without overshoot the CO core mass, the size of the helium shell and the ignition location
are comparable to the 7 \msun \ model with overshoot.

There are two distinct populations in the thermohaline models, those with small values of \thermo, which ignite a flame at $M\approx$ 0.5 \msun \ 
and those with large \thermo \ values which ignite a flash at $M\approx$ 0.8 \msun \  (though only if \overshoot=0.001). 
Those models that ignite at $M\approx$ 0.5 \msun, thermohaline mixing has little impact
on the flame ignition location. However there is some variation due to thermohaline mixing before the flame ignites. Before the flame ignites, 
when $\epsilon_{{\rm nuc}}>>\epsilon_{\nu}$, there is a region in the vicinity of the ignition point that
undergoes weak $^{12}$C+$^{12}$C burning (with $\epsilon_{{\rm nuc}}<<\epsilon_{\nu}$). This weak burning is able 
to drive a region of weak thermohaline mixing as a precursor to the vigorous carbon burning. Those that ignite at $M\approx0.8$ ignite a flash,
but under different conditions to those that we predict for the other flashes. Here, there is thermohaline mixing between the CO and helium shell which allows 
us to form a small region ($M\approx0.05\msun$) where
$X(^{12}$C$)\sim X(^{16}$O$)$. This higher fraction of carbon allows the ignition to occur at a lower density.
This flash then prevents an ignition occurring deeper in the CO core where $X(^{12}$C$)\approx0.3$, which we assume in equation \ref{eqn:c_burn_nuc}.

Thermohaline mixing also effects the flame once burning has commenced, 
the mixing pulls $^{12}$C material from below the flame \citep{siess_2009_aa}.
As the strength of thermohaline mixing increases, the sub flashes, seen 
in figure \ref{fig:f5}, middle left panel, merge into one continuous flame, due to the
increased carbon abundance. However as \citet{siess_2009_aa} showed this mixing eventually
starves the flame of fuel preventing it from reaching the core.

The effect of semiconvection is almost negligible, over the range of values considered here. 
Semiconvection acts near regions of convection, however it only acts for short periods of time in our
models, thus has limited ability to change the composition of models before the formation of the CO core. In can however act
during the carbon burning once the convective region has formed, mixing the burnt material
with unburnt CO. Again this effect is small and plays a limited role the evolution of the flame.
In figure \ref{fig:f5}, where we have secondary carbon flashes (top left, middle left and middle right panels),
those flashes that occur near the $^4$He shell can form brief semiconvective regions across the shell and into the
convective envelope. This may provide a way to detect the product of the flashes in the surface abundances,
though these flashes (and the semi convective regions) are short lived, which will limit the material transferred.

Changes in \amnu, which is a global scale factor on the strength of angular momentum mixing, primarily acts
by changing the strength of  thermohaline mixing. As \amnu increases the amount of thermohaline mixing increases as well, which allows 
the mixing of material from the core into the flame \citep{siess_2009_aa} to increase, though again this effect is small.

\citet{zaussinger_2013_aa} found for a 15\msun \ model on the MS that the 
semiconvection mixing timescale is long ($10^{10}$yrs) 
which explains why the semiconvection has little effect on these systems.
\citet{siess_2009_aa} showed that thermohaline mixing has limited effect in the evolution up to 
the carbon ignition, due to the lack of the $^3$He($^3$He,2p)$^4$He reaction which is necessary to 
set  up the mean molecular weight inversion needed for thermohaline mixing.
\citet{brown_2013_aa} propose a model for mixing by fingering convection in the 
parameter regime relevant for stellar (and planetary) interiors that is supported by 
three-dimensional direct numerical simulations.

The angular momentum diffusion term has limited impact due to its implementation as a 
global scale factor on the angular momentum mixing process \citep{paxton_2013_aa}, thus the 
value itself is not physically motivated however we have varied to test whether missing
sources of angular momentum, like internal gravity waves \citep{kumar_1997_aa}, would have an
impact. Given however, that in Figure \ref{fig:f1} we have shown that the individual diffusion 
coefficients can vary by 10 orders of magnitude, a change of $\approx$ 50\% in \amnu\ is
insignificant. For additional sources of angular momentum mixing to have an impact, they must be able to effect
size of the CO core, like overshoot does, to have a detectable impact. Compositional changes (thermohaline and semiconvection) are
too weak to have appreciable impact on the ignition location due to their limited ability to change the CO core mass.

\section{Results From The Spatial And Temporal Convergence Studies}
\label{sec:convergence}

Accurately capturing the nuclear burning and thermal transport within
a steady-state, convectively bounded, carbon burning front, or within
the time-dependent carbon burning flashes, requires spatial resolutions
$\lesssim 2$ km \citep{timmes_1994_aa,ritossa_1996_aa,garcia-berro_1997_aa}.
\citet{siess_2006_aa} use as many as $\approx$ 50 grid points to describe
the precursor flame between the convection region and the minimum in
the luminosity profile below the convective region.
\citet{denissenkov_2013_ab} and \citet{chen_2014_aa} use more than 100
mass zones.  Figure \ref{fig:f3} shows the temperature 
and nuclear energy generation rate profiles
of a
carbon burning flame in our \hbox{8 \msun}\ ZAMS model with \rot=0.25
and \overshoot=0.016. Distances $\lesssim$ 3700 km lie ahead of the
flame front, and distances between $\approx$ 3700 km and $\approx$ 3900 km
contain the region where thermal conduction dominates nuclear burning.
Distances between $\approx$ 3900 km and $\approx$ 4200 km contains the body
of the flame front which reaches a peak temperature of $\approx$ 7.5$\times$10$^8$ K
and peak energy generation rates of $\approx$ 8.9$\times$10$^6$ erg g$^{-1}$ s$^{-1}$. 
Distances $\gtrsim$ 4200 km contain the
convectively bounded region of the flame. The critical temperature
at which the heating due to nuclear reactions equals the energy diffused 
away by neutrino and conductive processes in the steady state is about 
$T_{\rm{crit}} \sim$ 5.5$\times$10$^8$ K. The location of this critical temperature is marked in Figure~\ref{fig:f3} with a black cross.
 The profiles shown in Figure \ref{fig:f3} also capture the flame structure with 1$-$2 km resolution 
with $\approx$ 400 mesh points. The flame structure propagates
inward toward the center at speeds of $\approx$ 0.1 cm s$^{-1}$,
consistent with the values reported in \citet{timmes_1994_aa}.

Figure \ref{fig:fig4} shows the timestep and 
Kelvin-Helmholtz thermal timescale, $\tau_{kh}=GM_c^2/R_cL$, for the CO core
of the 8 \msun \ ZAMS model with \rot=0.25 and
\overshoot=0.016 model.  As helium is depleted in the core at
model number $\approx$ 2800, the timestep begins to decrease from $\approx$ 10,000 
years due to the increase in nuclear burning. At first carbon
ignition, model $\approx$ 3700, the timestep is $\approx$ 10 years and decreases to $\approx$ 1
year as the flame and flashes propagate towards the center. At model
number $\approx$ 4500 the flashes have reached their closest approach to the center.  
The thermal timescale increases as the
core  increases in mass until the first ignition, where it then rapidly decreases
due to the increased luminosity. The thermal timescale then peaks again shortly before the 
next ignition at model $\approx$ 4400. This time however the flame
generates less energy and the thermal timescale is reduced by a smaller amount, 
compared to the first ignition. On average the flame lifetime is $\approx$ 10\% that of the thermal
timescale of the core.

The location of first carbon ignition in the 7 \msun, 8 \msun, and 9
\msun \ models as a function of spatial and temporal resolution is
shown in Figure \ref{fig:converge_grid}.  Each model has the baseline
\overshoot=0.016. Spatial resolution in MESA is generally controlled
by \mesh, with smaller values providing an increase in the number of
cells.  Temporal resolution is loosely controlled by \var, the
allowed change in the size of variables during a timestep, with smaller values
decreasing the size of the timesteps taken. See \citep{paxton_2011_aa} for
a detailed discussion of these two MESA control parameters.

For zero rotation, Figure \ref{fig:converge_grid} (top left) shows all
values of \mesh \ and \var \ give the same location of first carbon
ignition, suggesting convergence has been attained. Increasing the
spatial resolution has little impact on the location of the flame,
while increasing the temporal resolution shows a slight decrease in
the ignition location.

For \rot=0.25 (Figure \ref{fig:converge_grid} top right), the location of first carbon ignition
depends on the values of \mesh \ and \var.  For the 7\msun \ case
the highest resolution model (red-dashed), $\var=10^{-5}$ and
\mesh=0.1, agree with our baseline model (green-dot-dashed),
$\var=10^{-4}$ and \mesh=0.5, that there is no ignition occurs. At
$\var=10^{-3}$, as the spatial resolution decreases the ignition point
is pushed deeper into the star.  At 8 \msun, most of the models have
converge around 0.1-0.2 \msun \ ignition point, except the lowest
resolution model with has a center ignition. Models with
$\var=10^{-5}$ show little variation as \mesh \ changes, while as \var
\ increases in size the \mesh term becomes more significant.  
These studies suggest our baseline values for  \mesh \ and \var \ 
for off-center ignition are well within the convergence envelope.
All models converge on a 9\msun \ star having center ignition.

At \rot=0.5 (Figure \ref{fig:converge_grid} bottom), for the 7 \msun \ case our baseline parameters
agree with the highest resolution model, for the lack of
ignition. However, for all other values of \var \ and \mesh \ there is
considerable spread in ignition points. For the 8\msun \ case the
results have clustered around 0.05-2.0\msun, except for $\var=10^{-3}$
case, where the results have a spread of 0.5\msun. All models agree
again the 9\msun \ case has a central ignition.

Overall, our baseline models agree within $\approx$ 0.1\msun \ of the
highest resolution models we ran, for the ignition point. As the
rotation rate of the star increases we can see that the choice of
resolution terms becomes more significant and that there is a larger
spread in possible values. Changing the temporal resolution has the
most effect on the initial location of the flame.  The choice of
spatial resolution becomes more significant only as the temporal
resolution decreases.  Thus our choice of baseline parameters appear
to be a good compromise in terms of precision of results and
computational effort, decreasing \var \ increases the computational
time by a similar amount, while decreasing \mesh increases the memory
requirements for the model. However, they also show a necessary
requirement for carbon flame modelers to look critically at their
choice of model resolution \citep{timmes_1994_aa,ritossa_1996_aa,siess_2006_aa,
doherty_2010_aa,denissenkov_2013_ab,chen_2014_aa}.

\section{Discussion}
\label{sec:discussion}

We have investigated the detailed and global properties of carbon
burning in SAGB stars with 2755 stellar evolution models.  These
models consumed 200,000 core-hours (roughly 3 days per model) and
yielded over 2 TB of decimated data (a limited number of MESA profiles were stored).
To our knowledge this represents the largest block of compute
resources used for a MESA survey to date. We note that every model
ran from the pre-main-sequence to the end of carbon burning (if carbon
ignition was achieved) without failure and without intervention.

With these models, the location of first ignition whether off-center
or central, the quenching location of the carbon burning flames and
flashes, the angular frequency of the carbon core, and the carbon core
mass have been surveyed as a function of the ZAMS mass, initial
rotation rate, the magnitude of various mixing parameters such as
convective overshoot, semiconvection, thermohaline and angular
momentum transport. We now compare our results to previous efforts
and discuss methods for calibrating the \overshoot\ parameter within
a given overshoot implementation.

\citet{2013_georgy_aa} found that rotation of a 9 $M_{\odot}$ model can increase the lifetime 
spent on the MS compared to 
that of a non-rotating model. This increase in MS lifetime is caused by
rotational mixing supplying a sustained amount of fresh hydrogen into the convective core.
They include modifications to 
the stellar structure equations 
due to centrifugal acceleration described by \citet{kippenhahn_1970_aa,endal_1976_aa}, 
assuming the angular velocity is constant on isobars \citep{zahn_1992_aa}.
\citet{2013_georgy_aa} also adopt an instantaneous method of overshoot 
with $d_{\rm{over}}/H_{\rm{p}} =0.10$ applied to the H- and He- burning boundaries.
For a non-rotating 9 \msun\ model at the end of core He burning they find a 
convective core to total mass ratio of $M_{\rm{cc}}/M_{\rm{tot}} \approx 0.10$ while our corresponding
rotating model yields a larger value of  $M_{\rm{cc}}/M_{\rm{tot}} \approx 0.15$.
We find a more modest difference between the non-rotating and rotating 8 \msun\ model 
of $M_{\rm{cc}}/M_{\rm{tot}} \approx 0.14$ and $M_{\rm{cc}}/M_{\rm{tot}} \approx 0.15$, respectively, 
as shown in Figure~\ref{fig:HR}.

While our MESA models use a similar implementation for rotation, 
our calculations differ from \citet{2013_georgy_aa} in that
we include the effects magnetic torques which aid in significantly inhibiting the spin up of the
convective core of the star during its evolution. For example, we find for an 8 \msun\ ZAMS model with 
\rot\ $\approx 0.5$, an angular velocity at the center of the core of 
$\rm{log_{10}} \ \omega_{center} \approx -3.4 \ \rm{rad \ s^{-1}}$ at the start of 
carbon ignition, compared to a more rapid value of 
$\rm{log_{10}} \ \omega_{center} \approx -0.9 \ \rm{rad \ s^{-1}}$ 
when internal magnetic fields are neglected.
Magnetic field torques that inhibit spin up of the stellar model results 
in less massive convective cores due to the less efficient rotational mixing. 
We also find that magnetic torques can account for 
the less drastic shift in luminosity on the HRD for an \hbox{8.0 \msun} ZAMS model with 
\rot $\approx$ 0.2 (See Figure~\ref{fig:HR}), contrary to the larger differences 
in the rotating model HRD tracks shown in \citet{georgy_2012_aa}.

For their stars past 2DU, \citet{doherty_2015_aa} found the CO core mass
for a 7 \msun\ ZAMS model to be \hbox{$\approx$ 0.8 \msun} and increases to 1.375 \msun\ for their 9.5 \msun\ ZAMS models.
In mild contrast, our models predict the CO core mass for a 7 \msun\ ZAMS model to be $1.05$ \msun\ and 
the highest mass star to produce a Chandrasekhar core to be a 9 \msun\ ZAMS, assuming an overshoot of \overshoot=0.016.
These differences are likely due to the treatment of the convective boundaries,
with \citet{doherty_2015_aa} using a search for convective neutrality rather than a convective-decay prescription,
leading to differences in the size of the $^4$He and CO core masses.

\citet{siess_2006_aa} found for models without overshoot that the ZAMS
mass range which ignite carbon off-center is 9$-$11.3 \msun.
This is comparable to our $f_{\rm{ov}}=0.0$, rotating models (Figure~\ref{fig:mass_over}),
which yield a value of \hbox{$M_{\rm{up}} \approx 8.8 \ M_{\odot}$} and $M_{\rm{mas}}$ $>$ 11$ \ M_{\odot}$.

\citet{denissenkov_2013_ab} found for a 9.5 \msun\ ZAMS mass with no
overshoot or thermohaline mixing, an off-center ignition mass of 0.665
\msun \ with the flame proceeding to the center, consistent with our
results. With \overshoot=0.007 they found carbon ignites off-center
but the flames and flashes do not reach the center. In contrast, we
find in this case the model star undergoes a central ignition.
We speculate this difference is due to \citet{denissenkov_2013_ab} 
only including overshoot once the CO core has formed, where we include it from
the \hbox{pre-MS} onwards. Thus,
the CO core in \citet{denissenkov_2013_ab}, \overshoot=0.007 model will be 
smaller than our models, and hence the ignition occurring off-center.

While our models are not completely comparable to those of
\citet{jones_2013_aa}, who use \overshoot=0.014 except at the base
of a burning region where they use \overshoot=0.007, they find a
8.2 \msun \ model ignites off-center while models with
$M_{\rm{zams}}>$ 8.8 \msun \ centrally ignite.  We find that for models
with 0.007 $\le$ \overshoot $\le$ 0.014 that a 8.2 \msun \ will ignite
off-center while only models with $M_{\rm{zams}}>$ 9.4 \msun \ will
always centrally ignite.

Arguably the biggest uncertainty in stellar models is the treatment of
convection. The overshooting parameter in particular, regardless of
how it is implemented within a specific numerical instrument,
critically influences all outputs of stellar evolution
\citep[e.g.][]{maeder_1975_aa,maeder_1976_aa}. Figure
\ref{fig:mass_over} in particular demonstrates that the properties of
carbon burning in SAGB models is not an exception, especially the
range of ZAMS masses that experience off-center ignition. Testing on a small number of models
suggests that the most significant location for overshoot is in 
regions of He burning,
followed by \review{carbon} burning. Regions with H burning 
or no burning show little difference
in ignition location with respect to changes in overshoot. 
The effect of convective overshoot on the stellar models 
considered in this work are in agreement with previous 
work by \citet{siess_2007_aa} who showed that for $f_{\rm{ov}}=0.016$ 
applied at the edge of the convective boundary, M$_{\rm{up}}$ can 
transition from 8.90$\pm$ 0.10 $M_{\odot}$ to 7.25$\pm$ 0.25 
$M_{\odot}$ for $Z=Z_{\odot}$. We find a similar transition where
$M_{\rm{up}} \approx 8.8 M_{\odot}$ for our rotating, \hbox{$f_{\rm{ov}}$ = 0.0 model}, 
which shifts to a value of M$_{\rm{up}} \approx$ 7.2 $M_{\odot}$ for $f_{\rm{ov}}$= 0.016 
(See Figure \ref{fig:mass_over}). \citet{gil-pons_2007_aa} 
found similar results upon investigating a grid of zero metallicity stars 
with $f_{\rm{ov}} = 0.12$ using an instantaneous overshooting formalism 
\citep{herwig_1997_aa}, contrary to the diffusive approach used in 
this work. They find a value of \hbox{$M_{\rm{up}} \approx$ 6.0 \msun}\ and 
$M_{\rm{mas}} \approx$ 7.8 \msun. The adoption of instantaneous overshooting,
as well as $Z \ll Z_{\odot}$ are likely to contribute to the modest 
discrepancy in values of $M_{\rm{up}}$ and $M_{\rm{mas}}$.

Traditionally the value of the overshooting parameter for a given
overshooting model is calibrated by fitting isochrones against the
width of the terminal age main sequence in color-magnitude diagrams,
or the surface abundances, of young and intermediate age clusters
\citep[e.g.,][]{maeder_1976_aa,maeder_1981_aa,mermilliod_1986_aa,schaller_1992_aa,herwig_2000_aa, vandenberg_2006_aa, kamath_2012_aa}.  
Photometry and spectroscopy of binary systems offer another avenue for
calibration of overshooting because these measurements can provide the
radii, effective temperatures, and masses. In addition both components
of the binary need to lie on the same isochrone and fit their
respective evolutionary tracks
\citep{schroder_1997_aa,pols_1997_aa,ribas_2000_aa,claret_2007_aa,meng_2014_aa,stancliffe_2015_aa}
High-precision high-cadence space photometry from the {\it CoRoT} and
{\it Kepler} missions opens up a newer method for calibration of
overshooting and other mixing processes in stellar interiors
\citep{neiner_2012_aa,montalban_2013_aa,tkachenko_2014_aa,guenther_2014_aa,aerts_2015_aa}.

MESA implements the time-dependent treatment of convective overshoot
mixing of \citet{herwig_2000_aa} with the traditional calibration method
leading to \overshoot=0.016.  It is unknown if this value of
\overshoot \ in this specific overshoot model applies to masses other
than the ones used for calibration, is consistent with values derived
from binary systems or asteroseismology, or if it applies to advanced burning
stages of stellar evolution. However, we have shown that for a dense grid of 
SAGB models \review{taken to the end of carbon burning}, utilizing 
our adopted baseline parameters, values of $M_{\rm{up}} \sim 7.0 \msun\ $ and
 $M_{\rm{mas}} \sim$ 8.4 \msun\,  are nearly independent of initial of ZAMS
 rotational values of $\rot \sim$ 0.0 - 0.5. While our SAGB models have been evolved 
 from the pre-MS phase through the end of \review{carbon} burning, for models whom do not
  ignite carbon, and those that ignite carbon off-center, the initial rotational rate
 may play a larger role in the final rotational rates of the WD that will eventually be born.
For a given ZAMS mass and overshoot
parameterization, we suggest that strong claims of carbon burning
quenching at an appreciable distance from the center to yield hybrid
CO + ONeNa white dwarfs should be viewed with caution.

\acknowledgements 
The authors thank Pavel Denissenkov for sharing his MESA inlists
and Lars Bildsten for detailed discussions.
We also thank the SPIDER collaboration for insightful discussions:  
K. Augustson, 
M. Browning,
M. Cantiello, 
J. Fuller,
R. Orvedahl, 
B. Paxton,
J. Toomre, 
R. Townsend, and
E. Zweibel.
Finally, we thank the participants of the 2014 MESA Summer School
for experimenting with some of the SAGB models:
L. Arcavi,
W. Ball,
E. Bauer,
P. Beck,
J. Blumenkopf,
J. Brown,
T. Ceillier,
D. Clausen,
R. Connolly,
J. Goldstein,
A. Lauer,
E. Leiner,
J. McKeever,
B. Mulligan,
A. Nagy,
A. Ordasi,
J. Ostrowski,
M. Renzo,
V. Schmid,
W. Strickland,
T. Sukhbold,
M. Sun,
S. Triana,
S. Valenti,
T. Van Reeth,
J. Vos,
M. Vuckovic,
D. Wilcox, and
M. Windju.
This project was supported by NASA under TCAN grant NNX14AB53G, by NSF
under SI$^2$ grant 1339600, and by NSF under PHY 08-022648 for the
Physics Frontier Center ``Joint Institute for Nuclear Astrophysics - Center for 
the Evolution of the Elements'' (JINA-CEE).
C.E.F. acknowledges partial support from Arizona State University
under the 2014 CLAS Undergraduate Summer Enrichment Award. The computing 
resources for the grid of MESA models was provided by the ASU Advanced
Computing Center.

% Bibliography goes here

\bibliographystyle{apj}

\bibliography{ms}

\end{document}